\documentclass[a4paper,onecolumn,11pt,accepted=2025-04-15]{quantumarticle}
\pdfoutput=1
\usepackage[utf8]{inputenc}
\usepackage[english]{babel}
\usepackage[T1]{fontenc}

\usepackage[pagebackref=true, colorlinks, linkcolor = magenta!77!black, anchorcolor = blue, citecolor = green!70!black]{hyperref}

\usepackage{amsmath,amssymb,amsthm,mathtools,mathrsfs,marvosym}

\usepackage{caption}

\usepackage{pifont}% http://ctan.org/pkg/pifont

\usepackage{authblk}
\usepackage{csquotes}
\usepackage{bbm}
\usepackage{lscape}
\usepackage{etex} 
\usepackage[table]{xcolor}

\usepackage{tikz}
\usetikzlibrary{arrows.meta}
\usetikzlibrary{decorations.pathmorphing,shapes}
\usetikzlibrary{calc}
\usepackage{comment}
\usepackage{adjustbox}
\usepackage{scalerel,stackengine}%Used for reallywidehat
\usepackage{stmaryrd}
\usepackage{fancyhdr}
\usepackage{soul}
\usepackage{dsfont}%this is for mathbb{1} but you write mathds{1}
\usepackage[normalem]{ulem}
\usepackage{longtable}
\usepackage[bottom]{footmisc}%this forces footnotes to appear on the bottom of page
\usepackage{graphicx}
\usepackage{caption}
\usepackage{subcaption}
\usepackage{pictexwd, dcpic}
\usepackage{float}
\usepackage{enumerate}
\usepackage[all]{xy} %This is to make higher categorical stuff
\usetikzlibrary{circuits.ee.IEC}

\usepackage{lipsum}

\newtheorem{theorem}[equation]{Theorem}
\newtheorem{lemma}[equation]{Lemma}
\newtheorem{proposition}[equation]{Proposition}
\newtheorem{corollary}[equation]{Corollary}
\theoremstyle{definition}
\newtheorem{definition}[equation]{Definition}
\newtheorem{construction}[equation]{Construction}
\newtheorem{question}[equation]{Question}

\newtheorem{problem}[equation]{Problem}
\newtheorem{example}[equation]{Example}
\newtheorem{exercise}[equation]{Exercise}
\newtheorem*{answer}{Answer}
\newtheorem*{solution}{Solution}
\newtheorem{remark}[equation]{Remark}

\newtheorem{notation}[equation]{Notation}
\newtheorem{noterm}[equation]{Notation and Terminology}

\newcommand\define[1]{\emph{\textbf{#1}}}

\newcommand{\be}{\begin{equation}}
\newcommand{\ee}{\end{equation}}
\def\ba{\begin{align}} %previously this was ``array''
\def\ea{\end{align}}
\newcommand{\bea}{\begin{eqnarray}}
\newcommand{\eea}{\end{eqnarray}}
\newcommand{\bx}{\begin{example}}
\newcommand{\ex}{\end{example}}
\newcommand{\bex}{\begin{exercise}}
\newcommand{\eex}{\end{exercise}}
\newcommand{\ban}{\begin{answer}}
\newcommand{\ean}{\end{answer}}
\newcommand{\bt}{\begin{theorem}}
\newcommand{\et}{\end{theorem}}
\newcommand{\bc}{\begin{corollary}}
\newcommand{\ec}{\end{corollary}}
\newcommand{\blem}{\begin{lemma}}
\newcommand{\elem}{\end{lemma}}
\newcommand{\bp}{\begin{problem}}
\newcommand{\ep}{\end{problem}}
\newcommand{\bn}{\begin{proposition}}
\newcommand{\en}{\end{proposition}}
\newcommand{\bd}{\begin{definition}}
\newcommand{\ed}{\end{definition}}
\newcommand{\bcon}{\begin{construction}}
\newcommand{\econ}{\end{construction}}
\newcommand{\bq}{\begin{question}}
\newcommand{\eq}{\end{question}}
\newcommand{\bprf}{\begin{proof}}
\newcommand{\eprf}{\end{proof}}
\newcommand{\br}{\begin{remark}}
\newcommand{\er}{\end{remark}}
\newcommand{\bs}{\begin{solution}}
\newcommand{\es}{\end{solution}}
\newcommand{\beqs}{\begin{eqnarray}}
\newcommand{\eeqs}{\end{eqnarray}}
\newcommand{\bnt}{\begin{noterm}}
\newcommand{\ent}{\end{noterm}}
\newcommand{\bnot}{\begin{notation}}
\newcommand{\enot}{\end{notation}}

\newcommand{\id}{\mathrm{id}}

\newcommand{\lra}{\longrightarrow}

\newcommand{\tr}{{\rm tr} }

\newcommand{\<}{\langle}
\renewcommand{\>}{\rangle}

\def\F{{{\mathcal{F}}}}

\def\R{{{\mathbb R}}}
\def\C{{{\mathbb C}}}

\def\N{{{\mathbb N}}}

\def\invexcl{\rotatebox[origin=c]{180}{$!$}}
\newcommand{\bloom}{\operatorname{\invexcl}}

\newcommand{\tao}{\text{\Yinyang}}

\def\VA{\mathcal{A}}
\def\VB{\mathcal{B}}
\def\VC{\mathcal{C}}

\def\M{\mathbb{M}}
\def\E{\mathscr{E}}
\def\F{\mathscr{F}}
\def\J{\mathscr{J}}

\def\mA{{{\mathcal{A}}}}

\newcommand{\stoch}{\;\xy0;/r.25pc/:(-3,0)*{}="1";(3,0)*{}="2";{\ar@{~>}"1";"2"|(1.06){\hole}};\endxy\!}

\newcounter{sarrow}
\newcommand\xstoch[1]{%
\stepcounter{sarrow}%
\mathrel{\begin{tikzpicture}[baseline= {( $ (current bounding box.south) + (0,-0.1ex) $ )}]
\node[inner sep=.5ex] (\thesarrow) {\;$\scriptstyle #1$\;};
\path[draw,{<[scale=1.5,width=3,length=2]}-,decorate,
  decoration={snake,amplitude=0.3mm,segment length=2.1mm,pre=lineto,pre length=1pt}] 
    (\thesarrow.south east) -- (\thesarrow.south west);
\end{tikzpicture}}%
}

\newcounter{sqarrow}

\begin{document}

\title{Quantum dynamics as a pseudo-density matrix}

\author{James Fullwood}
\affiliation{School of Mathematics and Statistics, Hainan University, 58 Renmin Ave., 570228, Haikou, Hainan Province, China}
\orcid{0000-0002-2445-2701}
\maketitle

\begin{abstract}
While in relativity theory space evolves over time into a single entity known as spacetime, quantum theory lacks a standard notion of how to encapsulate the dynamical evolution of a quantum state into a single "state over time". Recently it was emphasized in the work of Fitzsimons, Jones and Vedral that if such a state over time is to encode not only spatial but also temporal correlations which exist within a quantum dynamical process, then it should be represented not by a density matrix, but rather, by a \emph{pseudo-density matrix}. A pseudo-density matrix is a hermitian matrix of unit trace whose marginals are density matrices, and in this work, we make use a factorization system for quantum channels together with the combinatorial structure of the associahedron to show there is a well-defined pseudo-density matrix associated with a quantum system which is to evolve according to a finite sequence of quantum channels. We then view such a pseudo-density matrix as a quantum analog of a local patch of spacetime, and we make an in-depth mathematical analysis of such quantum dynamical pseudo-density matrices and the properties they satisfy. We also show how to explicitly extract quantum dynamics from a given pseudo-density matrix, thus solving an open problem posed in the literature. 
\end{abstract}

\vspace{-7mm}
\tableofcontents

%%%%%%%%%%%%%%
\section{Introduction}
%%%%%%%%%%%%%%

In 1907 Hermann Minkowski showed that the work of Maxwell, Lorentz and Einstein could be viewed geometrically as a 4-dimensional theory of \emph{spacetime} \cite{MKWSKI}, after which he boldly predicted that space by itself, and time by itself, were then "doomed to fade away in mere shadows". Minkowski's prediction was indeed correct, and ever since we have only furthered our understanding of the inextricable connection between space and time. By 1927 quantum theory had been established through the revolutionary work Bohr, Heisenberg, Schr\"odinger, Born and others, where the fundamental view was that at the sub-atomic realm, reality was appropriately described by "quantum states" evolving in time. While debates have raged through the ages regarding the ontological status of a quantum state, the one thing that most can agree upon is that at a fundamental level a quantum state consists of \emph{information}, and there is an emerging viewpoint --- famously coined by Wheeler as "it from bit" \cite{Wheeler_1989}--- that such quantum information is the basis of our physical reality. But while our macroscopic view of the world has blossomed from Minkowski's unification of space and time, for the most part there has not been an analogous unification of information and time in quantum theory, where we seemed to have left the insights of Minkowski behind. In particular, while in relativity theory space evolves in time to form a single mathematical object called spacetime, quantum theory lacks a standard notion of how to encapsulate the global evolution of a quantum state over successive instances of time into a single entity, or rather, as a "state over time". In this work, we then aim to make progress toward filling in the "?" in the following analogy:
\begin{eqnarray*}
&\text{\underline{General Relativity}}:\quad \, \text{space + time}&=\,\text{ spacetime} \\
&\text{\underline{Quantum Theory}}: \quad \, \text{state + time}&=\hspace{2mm}\text{?} 
\end{eqnarray*}  

To summarize our approach to such an analogy, first consider a local patch $S$ of spacetime, which one may view as a fibration $\varphi:S\to [t_0,t_1]$ of spatial slices over an interval of time $[t_0,t_1]$. As in topological quantum field theories, we may view $S$ as a \emph{cobordism} \cite{Atiyah88,Ba06}, i.e., as representing the \emph{process} of the spatial slice $S_0=\varphi^{-1}(t_0)$ evolving over time into the spatial slice $S_1=\varphi^{-1}(t_1)$. As such, in general relativity, the local objects encoding evolution over time \emph{are of the same class of entity} as the objects which are evolving over time, as $S$, $S_0$ and $S_1$ are all manifolds. However, a crucial distinction between the process manifold $S$ and the spatial slices $S_0$ and $S_1$, is that while $S_0$ and $S_1$ are of Euclidean signature, the signature of the process manifold $S$ picks up a negative sign as it extends over time, thus resulting in a Lorentzian signature. 

For our quantum analogy of such a local patch of spacetime, the spatial manifolds $S_0$ and $S_1$ are replaced by density operators $\rho_0\in \VA_0$ and $\rho_1\in \VA_1$, where $\VA_0$ and $\VA_1$ denote the algebras of operators on the Hilbert spaces associated with timelike separated quantum systems at times $t=t_0$ and $t=t_1>t_0$, respectively. The spacetime manifold $S$ is then replaced by a bipartite operator $\psi(\rho,\E)\in \VA_0\otimes \VA_1$ which we refer to as a \emph{quantum state over time}, where $\rho=\rho_0$ and $\E:\VA_0\to \VA_1$ is a quantum channel such that $\rho_1=\E(\rho)$. In accordance with our spacetime analogy, we require the states $\rho$ and $\E(\rho)$ to be the reduced density matrices of the state over time $\psi(\rho,\E)$ with respect to tracing out $\VA_1$ and $\VA_0$ respectively. Moreover, we also require that one should be able to extract the channel $\E$ from knowledge of the quantum state over time $\psi(\rho,\E)$.   

In Ref.~\cite{HHPBS17}, a further list of desiderata one would expect from such a quantum state over time was put forth, and various proposals for such a quantum state over time were analyzed (for more on such proposals as well as other formulations of dynamical quantum states, see Refs.~\cite{Le06,Le07,LeSp13,Cotler2018,OCB12,Ohya1983,Ts22,Woot87,MaCh22,GuoZ1,Jia_2024}). It was then shown in Ref.~\cite{FuPa22} that if we let $\mathscr{J}[\E]=\sum_{i,j}|i\rangle\langle j|\otimes \E(|j\rangle\langle i|)$ (which is the partial transpose of the \emph{Choi matrix} of $\E$), then the quantum state over time $\psi(\rho,\E)$ given by
\be \label{SMTXBLM67}
\psi (\rho,\E) =\frac{1}{2}\Big((\rho\otimes \mathds{1})\mathscr{J}[\E]+\mathscr{J}[\E](\rho\otimes \mathds{1})\Big)
\ee
satisfies the full list of desiderata from Ref.~\cite{HHPBS17}. Following this, it was shown in Refs.~\cite{LiNg23,PFBC23} that the quantum state over time $\psi(\rho,\E)$ as given by \eqref{SMTXBLM67} is in fact uniquely characterized by the desiderata put forth in Ref.~\cite{HHPBS17}, and as such, one may view $\psi(\rho,\E)$ as given by \eqref{SMTXBLM67} as \emph{the} quantum state over time associated with the pair $(\rho,\E)$. 

We note however that while the quantum state over time $\psi(\rho,\E)$ is hermitian and of unit trace, it is not positive in general, and as such, it is not a quantum state in the traditional sense. While this may come across at first glance as a defect of the construction, quantum states over time have played in central role in the theory of quantum Bayesian inversion and retrodiction~\cite{PaRuBayes,PaRu19,FuPa22a,PaFu24,BHKK_2024,LeSp13}, and as argued in Ref.~\cite{FJV15}, the negative eigenvalues of a quantum state over time serve as a witness to quantum correlations which imply causation. Moreover, the appearance of negative eigenvalues in the spectrum a quantum state over time is in direct accordance with our spacetime analogy. Indeed, as mentioned earlier, while the spatial slices $S_0$ and $S_1$ have metrics of Euclidean signature, the metric of the local patch of spacetime $S$ picks up a negative sign as it extends over time. On the quantum side, the positivity of the quantum states $\rho$ and $\E(\rho)$ are analogous to the spatial slices having metrics of Euclidean signature, while the non-positivity of the state over time $\psi(\rho,\E)$ is a quantum analog of the minus sign appearing in Lorentzian signature of the spacetime manifold $S$.

One desideratum for quantum states over time appearing in Ref.~\cite{HHPBS17} is a property referred to as \emph{associativity}, which allows one to unambiguously associate a 2-step state over time $\psi(\rho,\E,\F) \in \VA_0\otimes \VA_1\otimes \VA_2$ with the evolution of a state $\rho$ according to a 2-chain 
\be \label{2CXNXs57}
\VA_0\overset{\E}\lra \VA_1\overset{\mathscr{F}}\lra \VA_2\, ,
\ee
where $\E$ and $\F$ are quantum channels. The associativity property is most easily formulated in terms of the mapping $\rho\mapsto \psi(\rho,\E)$, which we refer to as the \emph{bloom map} associated with the channel $\E$. Denoting the bloom map associated with a channel $\E$ by $\bloom(\E):\VA\to  \VA\otimes \VB$, associativity is the requirement that for every 2-chain \eqref{2CXNXs57},
\be \label{ASXS917}
\bloom\left(\bloom(\mathscr{F})\circ \mathscr{E}\right)=\bloom(\mathscr{F}\circ \text{tr})\circ \bloom(\mathscr{E})\, ,
\ee
where "tr" is a blanket notation for partial trace, which in this context is a mapping of the form $\text{tr}:\VA\otimes \VB\to \VB$. The formulation of the associative identity \eqref{ASXS917} comes from the fact that for every state $\rho\in \VA$, the operators $\bloom\left(\bloom(\mathscr{F})\circ \mathscr{E}\right)(\rho)$ and $(\bloom(\mathscr{F}\circ \text{tr})\circ \bloom(\mathscr{E}))(\rho)$ both yield unit trace Hermitian operators in $\VA\otimes \VB\otimes \VC$ such that tracing over $\VC$ and $\VA$ yields the quantum states over time $\psi(\rho,\E)$ and $\psi(\E(\rho),\F)$, respectively. If the associative identity \eqref{NCNX796} then holds, we are ensured that these two operators are in fact equal, thus yielding a well-defined notion of a state over time associated with 2-step processes. The "associative" terminology then stems from the fact that each side of the associativity identity \eqref{ASXS917} may be naturally identified with a distinct parenthezation of the matrix algebra $\VA_0\otimes \VA_1\otimes \VA_2$, thus \eqref{ASXS917} is analogous to the equation $(\VA_0\otimes \VA_1)\otimes \VA_2=\VA_0\otimes (\VA_1\otimes \VA_2)$.   

Although it is straightforward to show that the quantum state over time construction given by \eqref{SMTXBLM67} indeed satisfies associativity, showing that it inductively extends to $n$-step processes for arbitrary $n>0$ is a mathematically demanding problem, leaving open the question of whether or not $n$-step quantum states over time are well-defined. In particular, if an initial state $\rho$ is to evolve according to an $n$-chain
 \be \label{NCNX796}
\VA_0\overset{\E_1}\lra \VA_1\lra \cdots \lra \VA_{n-1}\overset{\E_n}\lra \VA_n\, ,
\ee
then there are an $n$th Catalan number $c_n=\frac{1}{n+1}\binom{2n}{n}$ of ways to parenthesize the matrix algebra $\VA_0\otimes \cdots \otimes \VA_n$, each of which may be associated with a distinct construction of an $n$-step state over time $\psi(\rho,\E_1,...,\E_n)\in \VA_0\otimes \cdots \otimes \VA_n$. In this work, we prove that for all $n>0$, the $c_n$ different constructions of such an $n$-step state over time are all in fact equal (see item \ref{STXFXT81872} of Theorem~\ref{STXFXT8187}). For our proof, we show that just as in the case of 2-chains, each of the $c_n=\frac{1}{n+1}\binom{2n}{n}$ constructions of an $n$-step state over time may be written in terms of the bloom maps associated with each of the channels in the $n$-chain. We then associate each of the $c_n$ mappings yielding quantum states over time with a vertex of the associahedron $K_{n+1}$ (which is an $n-1$-dimensional convex polytope encoding the combinatorics of parenthezations of strings of $n+1$ letters), and show that any two maps connected by an edge of the associahedron may be obtained from one another by a single application of the associativity identity \eqref{ASXS917}. Since the underlying graph of the associahedron is connected, this implies that any two constructions of $n$-step states over time may be transformed into one another by successive applications of the the associativity identity \eqref{ASXS917}, thus yielding a well-defined quantum state over time $\psi(\rho,\E_1,...,\E_n)$ associated with $n$-step processes. The idea behind our proof in the case of 3-chains is illustrated in Figure~\ref{fig:ASFH}.

\begin{figure}
\hspace{0cm}
\centering
\xy0;/r.25pc/:
(0,15)*+{\bloom\left(\bloom\left(\bloom(\mathscr{E}_3)\circ \mathscr{E}_2\right)\circ \mathscr{E}_1\right)}="0";
(-35,0)*+{\bloom\left(\bloom(\mathscr{E}_3)\circ \mathscr{E}_2\circ \text{tr}\right)\circ \bloom(\mathscr{E}_1)}="1";
(-25,-25)*+{\bloom\left(\mathscr{E}_3\circ \text{tr}\right)\circ \bloom\left(\mathscr{E}_2\circ \text{tr}\right)\circ \bloom(\mathscr{E}_1)}="2";
(25,-25)*+{\bloom\left(\mathscr{E}_3\circ \text{tr}\right)\circ \bloom(\bloom(\mathscr{E}_2)\circ \mathscr{E}_1)}="3";
(35,0)*+{\bloom\left(\bloom(\mathscr{E}_3\circ \text{tr})\circ \bloom(\mathscr{E}_2)\circ \mathscr{E}_1\right)}="4";
{\ar@{-} "0";"1"_{}};
{\ar@{-}"1";"2"_{}};
{\ar@{-}"3";"2"_{}};
{\ar@{-}"4";"3"_{}};
{\ar@{-}"0";"4"^{}};
\endxy
\caption{\textbf{The case of 3-chains}. The associahedron $K_4$ is a pentagon, and the diagram above illustrates a labeling of the vertices of the associahedron $K_4$ by mappings which yield quantum states over time associated with 3-chains. Note that each edge in the pentagon links mappings that may be obtained from one another by a single application of the associativity identity \eqref{ASXS917}.}
\label{fig:ASFH}
\end{figure}

As a consequence of our results, we show in Example~\ref{PDOEXP71} that in the case of dynamically evolving systems of qubits, the associated quantum state over time $\psi(\rho,\E_1,...,\E_n)$ coincides with the coarse-grained pseudo-density matrix (PDM) associated with the system, which was first introduced in \cite{FJV15} to treat both spatial and temporal correlations in quantum theory on equal footing. It then follows that coarse-grained PDMs may be viewed as quantum states over time, and conversely, quantum states over time may be viewed as a generalization of PDMs to arbitrary finite dimensional quantum systems evolving under arbitrary $n$-chains. As PDMs have received significant attention in the recent literature~\cite{Marletto_2021,ZPTGVF18,Pisar19,Marletto_2020,Marletto_2019,Liu_2024,Liu_2024X,LQDV,FuPa24,song23,jia2023,FZ_24,Utagi_2021}, we hope that the connection between quantum states over time and PDMs established in this work will serve as bridge linking various perspectives on the role of temporal quantum correlations in our understanding of quantum information and its relationship with spacetime. Finally, we prove that given a unit-trace hermitian element $\tau\in \VA_0\otimes \cdots \otimes \VA_n$ satisfying some technical conditions, one may explicitly extract a dynamical process $(\rho,\E_1,...,\E_n)$ such that $\psi(\rho,\E_1,...,\E_n)=\tau$ (see Theorem~\ref{MSXAX71}), thus establishing an explicit one-to-one correspondence between $n$-step states over time and a subclass of unit trace hermitian matrices. This partially solves an open question recently posed in \cite{jia2023}, where it is stated "Another interesting and closely relevant open question is, for a given PDM, how to find a quantum process to realize it.".   

{\bf Acknowledgments.} 
This work is supported by the Hainan University startup fund for the project "Spacetime from quantum information", and is also supported in part by the Blaumann Foundation. We thank Arthur J. Parzygnat and Francesco Buscemi for many useful discussions. 

%%%%%%%%%%%%%%%%%%%%%%%%%%%%%%
\section{Preliminaries} \label{S2}
%%%%%%%%%%%%%%%%%%%%%%%%%%%%%%
In this section we provide the basic definitions, notation and terminology which will be used throughout.

\bd
Let $X$ be a finite set. A function $p:X\to \R$ will be referred to as a \define{quasi-probability distribution} if and only if $\sum_{x\in X}p(x)=1$. In such a case, $p(x)\in \R$ will be denoted by $p_x$ for all $x\in X$. If $p_x\in [0,1]$ for all $x\in X$, then $p$ will be referred to as a \define{probability distribution}.  
\ed

\bd
Let $X$ and $Y$ be finite sets. A \define{stochastic} map $f:X\xstoch{} Y$ consists of the assignment of a probability distribution $f_x:Y\to [0,1]$ for every $x\in X$. In such a case, $(f_x)_y$ will be denoted by $f_{yx}$ for all $x\in X$ and $y\in Y$, which is interpreted as the probability of $y$ given $x$. A stochastic map $f:X\xstoch{} Y$ together with a prior distribution $p$ on its set of inputs $X$ will be denoted by $(p,f)$.  The set of stochastic maps from $X$ to $Y$ will be denoted by $\bold{Stoch}(X,Y)$.
\ed

\br
 A stochastic map $f:X\xstoch{} Y$ is also commonly referred to as a \define{Markov kernel}, or a \define{discrete memoryless channel}. 
\er

\bnt
Given a natural number $m\in \N$, the set of $m\times m$ matrices with complex entries will be denoted by $\M_m$, and will be referred to as a \define{matrix algebra}. As the matrix algebra $\M_1$ is simply the complex numbers, it will be denoted by $\C$. The matrix units in $\M_m$ will be denoted by $E_{ij}^{(m)}$ (or simply $E_{ij}$ if $m$ is clear from the context), and for every $\rho\in \M_m$, $\rho^{\dag}\in \M_m$ denotes the conjugate-transpose of $\rho$. Given a finite set $X$, a direct sum $\bigoplus_{x\in X}\M_{m_x}$ will be referred to as a \define{multi-matrix algebra}, whose multiplication and addition are defined component-wise. If $\VA$ and $\VB$ are multi-matrix algebras, then the vector space of all linear maps from $\VA$ to $\VB$ will be denoted by $\text{Hom}(\VA,\VB)$. The \define{trace} of an element $A=\bigoplus_{x\in X}A_{x}\in \bigoplus_{x\in X}\M_{m_x}$ is the complex number $\text{tr}(A)$ given by $\text{tr}(A)=\sum_{x\in X}\tr(A_{x})$ (where $\text{tr}(*)$ is the usual trace on matrices), and the \define{dagger} of $A$ is the element $A^{\dag}\in \bigoplus_{x\in X}\M_{m_x}$ given by $A^{\dag}=\bigoplus_{x\in X}A_{x}^{\dag}$. Given $\mathscr{E}\in \text{Hom}(\VA,\VB)$ with $\VA$ and $\VB$ multi-matrix algebras, we let $\mathscr{E}^*\in \text{Hom}(\VB,\VA)$ denote the \define{Hilbert--Schmidt dual} (or \define{adjoint}) of $\mathscr{E}$, which is uniquely determined by the condition
\[
\text{tr}\left(\mathscr{E}(A)^{\dag}B\right)=\text{tr}\left(A^{\dag}\mathscr{E}^*(B)\right)
\] 
for all $A\in \VA$ and $B\in \VB$. The identity map between algebras will be denoted by $\id$, while the unit element in an algebra will be denoted by $\mathds{1}$ (subscripts such as $\id_{\mA}$ and $\mathds{1}_{\mA}$ will be used if deemed necessary).
\ent

\br
Every finite-dimensional $C^*$-algebra is isomorphic to a multi-matrix algebra \cite{Fa01}. 
\er

\bnot
Given a finite set $X$,  the multi-matrix algebra $\bigoplus_{x\in X}\C$ is canonically isomorphic to algebra $\C^X$ of complex-valued functions on $X$. As such, we will write an element $\rho\in \bigoplus_{x\in X}\C\cong \C^X$ as 
\[
\rho=\sum_{x\in X}\rho_x\delta_x\in \C^X,
\]
where $\delta_x\in \C^X$ is the \define{Dirac-delta} of $x$, which is the function taking the value 1 at $x$ and 0 otherwise. In such a case, $\rho_x\in \C$ will be referred to as the \define{$x$-component} of $\rho$ for all $x\in X$.
\enot

\bd
Let $X$ be a finite set and let $\VA=\bigoplus_{x\in X}\M_{m_x}$ be a multi-matrix algebra. An element $A=\bigoplus_{x\in X} A_x\in \VA$ is said to be 
\begin{itemize}
\item
\define{self-adjoint} if and only if $A_x^{\dag}=A_x$ for all $x\in X$.
\item
\define{positive} if and only if $A_x\in \M_{m_x}$ is self-adjoint and has non-negative eigenvalues for all $x\in X$.
\item
a \define{state} if and only if $A$ is positive and of unit trace. If $X$ contains only one element, then a state $A$ will often be referred to as a \define{density matrix}. The set of all states on $\VA$ will be denoted by $\mathcal{S}(\VA)$.
\end{itemize}
\ed

\br
If $X$ is a finite set and $\VA=\C^X$, then a state $\rho\in \mathcal{S}(\C^X)$ is of the form $\rho=\sum_{x\in X}\rho_x\delta_x$ with $\rho_x\geq 0$ and $\sum_{x\in X}\rho_x=1$. As such, any state on $\C^X$ may be identified with a probability distribution on $X$.
\er

\bd
Let $\VA=\bigoplus_{x\in X}\M_{m_x}$ and $\VB=\bigoplus_{y\in Y}\M_{n_y}$. The \define{tensor product} of $\VA$ and $\VB$ is the multi-matrix algebra $\VA\otimes \VB$ given by
\[
\VA\otimes \VB=\bigoplus_{(x,y)\in X\times Y}\M_{m_x}\otimes \M_{n_y},
\] 
where $\M_{m_x}\otimes \M_{n_y}$ is the usual tensor product of matrix algebras. Given elements $\bigoplus_{x\in X}A_x\in \VA$ and $\bigoplus_{y\in Y}B_y\in \VB$, the element $(\bigoplus_{x\in X}A_x)\otimes (\bigoplus_{y\in Y}B_y)\in \VA\otimes \VB$ is the multi-matrix given by
\[
\Big(\bigoplus_{x\in X}A_x\Big)\otimes \Big(\bigoplus_{y\in Y}B_y\Big)=\bigoplus_{(x,y)\in X\times Y}A_x\otimes B_y,
\]
where $A_x\otimes B_y$ is the usual tensor product of matrices. Given maps $\mathscr{E}\in \text{Hom}(\VA ,\VA')$ and $\mathscr{F}\in \text{Hom}(\VB,\VB')$, then $\mathscr{E}\otimes \mathscr{F}\in \text{Hom}(\VA\otimes \VB,\VA'\otimes \VB')$ is the map corresponding to the linear extension of the assignment $(\mathscr{E}\otimes \mathscr{F})(A\otimes B)=\mathscr{E}(A)\otimes \mathscr{F}(B)$.
\ed

\bd
Given multi-matrix algebras $\VA=\bigoplus_{x\in X}\M_{m_x}$ and $\VB=\bigoplus_{y\in Y}\M_{m_y}$, an element $\E\in \text{Hom}(\VA,\VB)$ consists of elements $\E_{yx}\in \text{Hom}(\M_{m_x},\M_{n_y})$ such that 
\[
\E\Big(\bigoplus_{x\in X}\rho_x\Big)=\bigoplus_{y\in Y}\left(\sum_{x\in X}\E_{yx}(\rho_x)\right).
\]
In such a case, the $\E_{yx}$ will be referred to as the \define{component functions of $\E$}.
\ed

\bd
Let $\VA$ and $\VB$ be multi-matrix algebras. A map $\mathscr{E}\in \text{Hom}(\VA,\VB)$ is said to be
\begin{itemize}
\item
\define{$\dag$-preserving} if and only if $\mathscr{E}(A)^{\dag}=\mathscr{E}(A^{\dag})$ for all $A\in \VA$.
\item
\define{trace-preserving} if and only if $\text{tr}(\mathscr{E}(A))=\text{tr}(A)$ for all $A\in \VA$.
\item
\define{positive} if and only if $\mathscr{E}(A)$ is positive whenever $A\in \VA$ is positive.
\item
\define{completely positive} if and only if $\mathscr{E}\otimes \id_{\VC}:\VA\otimes \VC\to \VB\otimes \VC$ is positive for every multi-matrix algebra $\VC$.
\end{itemize} 
\ed

\bnot
Let $(\VA,\VB)$ be pair of multi-matrix algebras. The subset of $\text{Hom}(\VA,\VB)$ consisting of completely positive, trace-preserving maps will be denoted by $\bold{CPTP}(\VA,\VB)$, while the subset of  $\text{Hom}(\VA,\VB)$ consisting of trace-preserving maps will be denoted by $\bold{TP}(\VA,\VB)$. An element of $\bold{CPTP}(\VA,\VB)$ with $\VA$ and $\VB$ matrix algebras will often be referred to as a \define{quantum channel}.
\enot

\bd
Let $X$ and $Y$ be finite sets. An element $\E\in \bold{CPTP}(\C^X,\C^Y)$ will be referred to as a \define{classical channel}.  In such a case, we let $\mathscr{E}_{yx}\in [0,1]$ denote the elements such that
\[
\mathscr{E}(\delta_x)=\sum_{y\in Y}\mathscr{E}_{yx}\delta_y.
\]
In such a case, the elements $\mathscr{E}_{yx}$ will be referred to as the \define{conditional probabilities} associated with $\mathscr{E}$.
\ed

\bd
Given a pair $(\VA,\VB)$ of multi-matrix algebras, an element $(\rho,\mathscr{E})\in \mathcal{S}(\VA)\times \bold{CPTP}(\VA,\VB)$ will be referred to as a \define{process}, and the set of processes $\mathcal{S}(\VA)\times \bold{CPTP}(\VA,\VB)$ will be denoted by $\mathscr{P}(\VA,\VB)$. The subset of  $\mathscr{P}(\VA,\VB)$ consisting of processes $(\rho,\mathscr{E})$ with $\rho$ invertible will be denoted by $\mathscr{P}_{+}(\VA,\VB)$. When $\VA=\C^X$ and $\VB=\C^Y$ for finite sets $X$ and $Y$, then $(\rho,\mathscr{E})\in \mathscr{P}(\C^X,\C^Y)$ will be referred to as a \define{classical process}.
\ed

\bd
Let $(\VA,\VB)$ be a pair of multi-matrix algebras, and let $\mathscr{E}\in \text{Hom}(\VA,\VB)$. The \define{channel state} of $\mathscr{E}$ is the element $\mathscr{J}[\mathscr{E}]\in \VA\otimes \VB$ given by
\[
\mathscr{J}[\mathscr{E}]=(\text{id}_{\VA}\otimes \mathscr{E})(\mu^*(\mathds{1})),
\]
where $\mu^*:\VA\to \VA\otimes \VA$ is the Hilbert-Schmidt dual of the multiplication map $\mu:\VA\otimes \VA\to \VA$. 
\ed

\br
Given a pair $(\VA,\VB)$ of multi-matrix algebras, the map $\text{Hom}(\VA,\VB)\lra \VA\otimes \VB$ given by $\mathscr{E}\longmapsto \mathscr{J}[\mathscr{E}]$ is a linear isomorphism, which we refer to as the \define{Jamio{\l}kowski isomorphism} \cite{Jam72}.  In this work we will also make heavy use of the the inverse $\mathscr{J}^{-1}:\VA\otimes \VB\lra \text{Hom}(\VA,\VB)$ of the Jamio\l kowski isomorphism, which is given by
\[
\mathscr{J}^{-1}(\tau)(\rho)=\text{tr}_{\VA}((\rho\otimes \mathds{1})\tau).
\]
\er

\bd
If $\VA$, $\VB$ and $\VC$ are matrix algebras, there exists an associator isomorphism $\VA\otimes (\VB\otimes \VC)\lra (\VA\otimes \VB)\otimes \VC$ which then allows one to define tensor products of of a finite number of matrix algebras iteratively. We can then extend such a constriction to multi-matrix algebras as follows. Let $\{\VA_x\}_{x\in X}$ be a collection of multi-matrix algebras indexed by a finite set $X$, where $\VA_x=\bigoplus_{\lambda_x\in \Lambda_x}\VA_{\lambda_x}$. The \define{tensor product} of the algebras $\VA_x$ is the multi-matrix algebra $\bigotimes_{x\in X}\VA_x$ indexed by the set $\Gamma=\prod_{x\in X}\Lambda_x$ given by
\[
\bigotimes_{x\in X}\VA_x=\bigoplus_{(\lambda_x)_{x\in X}\in \Gamma}\Big(\bigotimes _{x\in X}A_{\lambda_x}\Big)\, .
\]
If the index set $X$ is of the form $X=\{0,1,...,n\}$, then we will write $\bigotimes_{x\in X}\VA_x$ as $\VA_0\otimes \cdots \otimes \VA_n$.
\ed

\bd
Let $\VA=\VA_0\otimes \cdots \otimes\VA_n$ be a tensor product of multi-matrix algebras. The \define{$i$th partial trace} for $i\in \{0,...,n\}$ is the map $\text{tr}_{i}:\VA\lra \VA_i$ given by the linear extension of the assignment
\[
\text{tr}_{i}(A_0\otimes \cdots \otimes A_n)=\text{tr}\left(A_0\otimes \cdots \otimes \widehat{A_i} \otimes \cdots \otimes A_n\right)A_i\in \VA_i,
\]
where $\widehat{A_i}$ denotes the empty matrix for all $i\in \{0,...,n\}$. 
\ed

\bnot
If $\VC=\VA\otimes \VB$, then the partial trace maps $\text{tr}_1:\VA\otimes \VB\lra \VA$ and $\text{tr}_2:\VA\otimes \VB\lra \VB$ will be denoted by $\text{tr}_{\VB}$ and $\text{tr}_{\VA}$ respectively. If $\VA=\C^X$ and $\VB=\C^Y$, then the partial trace maps $\text{tr}_{\C^X}:\C^X\otimes \C^Y\lra \C^Y$ and $\text{tr}_{\C^Y}:\C^X\otimes \C^Y\lra \C^X$ will be denoted by $\text{tr}_X$ and $\text{tr}_Y$. If $\VA=\VA_0\otimes \cdots \otimes \VA_n$, then we will frequently make use of the $n$th partial trace $\text{tr}_n:\VA\lra \VA_n$, and as such, we will often denote $\text{tr}_n$ simply by $\text{tr}$.
\enot

\bd \label{PDOXS787}
Let $X$ be a finite set and let $\VA=\bigoplus_{x\in X}\M_{m_x}$ be a multi-matrix algebra. A self-adjoint element $\tau\in \VA$ is said to be a \define{pseudo-density operator} with respect to the factorization $\VA=\VA_0\otimes \cdots \otimes \VA_n$ if and only if for all $i\in \{0,...,n\}$ we have
\[
\text{tr}_{i}(\tau)\in \mathcal{S}(\VA_i).
\]
If $\VA$ is a matrix algebra (i.e., when $X$ consists of a single element), then a pseudo-density operator $\tau\in \VA$ with respect to the factorization $\VA=\VA_0\otimes \cdots \otimes \VA_n$ will be referred to as a \define{pseudo-density matrix}. The set of all pseudo-density operators with respect to the factorization $\VA=\VA_0\otimes \cdots \otimes \VA_n$ will be denoted by $\mathscr{T}(\VA_0\otimes \cdots\otimes \VA_n)$.
\ed

\br
A pseudo-density operator with respect to the trivial factorization $\VA=\VA_0$ is simply a state, so that $\mathscr{T}(\VA_0)=\mathcal{S}(\VA_0)$.
\er

\br
Since all of the marginals of a pseudo-density operator are of unit trace, it follows that a pseudo-density operator is necessarily of unit trace as well.  
\er

%%%%%%%%%%%%%%%%%%%%%%%%%%%%%%%%%%
\section{Classical probability as CPTP dynamics} \label{S3}
%%%%%%%%%%%%%%%%%%%%%%%%%%%%%%%%%%
In this section we recall how classical probability may be recast in terms of CPTP maps between multi-matrix algebras. In particular, we will show how the joint distribution associated with a classical channel $\E:\C^X\lra \C^Y$ arises from a unique factorization of the channel of the form $\E=\tr_{X}\circ \bloom_{\E}$, where $\bloom_{\E}:\C^X\lra \C^{X\times Y}\cong \C^X\otimes \C^Y$ is a map referred to as the \emph{bloom} of $\E$. If $\rho\in \C^X$ is a state, then the element $\bloom_{\E}(\rho)$ is the joint distribution associated with the classical process $(\rho,\E)$, which in a more dynamical language we refer to as the \emph{state over time} associated with $(\rho,\E)$. We will then re-write the classical state over time $\bloom_{\E}(\rho)$ in a way that is valid for any \emph{quantum} process $(\rho,\E)\in \mathscr{P}(\VA,\VB)$ with $\VA$ and $\VB$ arbitrary multi-matrix algebras, thus paving the way for a quantum generalization of the classical state over time $\bloom_{\E}(\rho)$.

The following proposition shows how classical stochastic maps may be recast in terms of CPTP dynamics.

\bn
Let $(X,Y)$ be a pair of finite sets, and let $\bold{Q}:\bold{Stoch}(X,Y)\lra \bold{CPTP}(\C^X,\C^Y)$ be the map given by
\[
\bold{Q}(f)(\delta_x)=\sum_{y\in Y}f_{yx}\delta_y.
\] 
Then $\bold{Q}$ is a continuous bijection.
\en

\bprf
This follows form the fact that stochastic maps $f:X\xstoch{} Y$ and CPTP maps $\E:\C^X\lra \C^Y$ are completely determined by their associated conditional probabilities $f_{yx}$ and $\mathscr{E}_{yx}$.
\eprf

The next proposition shows that factorizations of classical channels are essentially unique, which we will use to characterize joint distributions associated with stochastic dynamics.

\bn \label{CPXQX191}
Let $(X,Y)$ be a pair of finite sets, let $\mathscr{E}\in \bold{CPTP}(\C^X,\C^Y)$, and let $\bloom_{\mathscr{E}}\in \bold{CPTP}(\C^X, \C^{X}\otimes \C^{Y})$ be such that
$\mathscr{E}=\emph{tr}_X\circ \bloom_{\mathscr{E}}$ and $\id_{\C^X}=\emph{tr}_Y\circ \bloom_{\mathscr{E}}$. Then $\bloom_{\mathscr{E}}$ is the linear map given by
\be \label{BLMXN87}
\bloom_{\mathscr{E}}\Big(\sum_{x\in X}\rho_x\delta_x\Big)=\sum_{(x,y)\in X\times Y}\rho_x\mathscr{E}_{yx}(\delta_x\otimes \delta_y),
\ee
where $\mathscr{E}_{yx}$ are the conditional probabilities associated with the map $\mathscr{E}$.
\en

\bprf
We show that given $x_0\in X$, we have
\[
\bloom_{\mathscr{E}}(\delta_{x_0})=\sum_{(x,y)\in X\times Y}\delta_{xx_0}\mathscr{E}_{yx}(\delta_x\otimes \delta_y).
\]
The result then follows as $\bloom_{\mathscr{E}}$ is assumed to be linear and Dirac deltas form a basis of $\C^X$. So let $x_0\in X$ be arbitrary, and let $\widetilde{\mathscr{E}}_{(x,y)x_{0}}$ be the conditional probabilities associated with the map $\bloom_{\mathscr{E}}$, so that
\[
\bloom_{\mathscr{E}}(\delta_{x_0})=\sum_{(x,y)\in X\times Y}\widetilde{\mathscr{E}}_{(x,y)x_0}(\delta_x\otimes \delta_y) .
\]
From the conditions $\mathscr{E}=\text{tr}_X\circ \bloom_{\mathscr{E}}$ and $\id_{\C^X}=\text{tr}_Y\circ \bloom_{\mathscr{E}}$ we then have
\[
\sum_{y\in Y}\mathscr{E}_{yx_0}\delta_y=\mathscr{E}(\delta_{x_0})=\text{tr}_X( \bloom_{\mathscr{E}}(\delta_{x_0}))=\sum_{y\in Y}\Big(\sum_{x\in X}\widetilde{\mathscr{E}}_{(x,y)x_0}\Big)\delta_y
\]
and
\[
\sum_{x\in X}\delta_{xx_0}\delta_x=\delta_{x_0}=\text{tr}_Y\left(\bloom_{\mathscr{E}}(\delta_{x_0})\right)=\sum_{x\in X}\Big(\sum_{y\in Y}\widetilde{\mathscr{E}}_{(x,y)x_0}\Big)\delta_{x},
\]
where $\delta_{xx_0}$ is the Kronecker delta. For all $y\in Y$ we then have 
\[
\sum_{x\in X}\widetilde{\mathscr{E}}_{(x,y)x_0}=\mathscr{E}_{yx_0},
\]
and for all $x\in X$ we have 
\[
\sum_{y\in Y}\widetilde{\mathscr{E}}_{(x,y)x_0}=\delta_{xx_0}.
\]
It then follows that for all $(x,y)\in X\times Y$ we have $\widetilde{\mathscr{E}}_{(x,y)x_0}=\delta_{xx_0}\mathscr{E}_{yx}$, thus
\[
\bloom_{\mathscr{E}}(\delta_{x_0})=\sum_{(x,y)\in X\times Y}\widetilde{\mathscr{E}}_{(x,y)x_0}(\delta_x\otimes \delta_y)=\sum_{(x,y)\in X\times Y}\delta_{xx_0}\mathscr{E}_{yx}(\delta_x\otimes \delta_y),
\]
as desired.
\eprf

\bd
Let $(X,Y)$ be a pair of finite sets and let $(\rho,\mathscr{E})\in \mathscr{P}(\C^X,\C^Y)$ be a classical process.
\begin{itemize}
\item
The map $\bloom_{\mathscr{E}}\in \bold{CPTP}(\C^X, \C^{X}\otimes \C^{Y})$ defined by \eqref{BLMXN87} will be referred to as the \define{bloom} of $\mathscr{E}$.
\item
The element $\bloom_{\mathscr{E}}(\rho)\in \mathcal{S}(\C^X\otimes \C^Y)$ will be referred to as the \define{state over time} associated with the classical process $(\rho,\mathscr{E})$.
\item
The map on classical processes given by $(\rho,\E)\longmapsto \bloom_{\E}(\rho)$ will be referred to as the \define{classical state over time function}. 
\end{itemize}
\ed

\begin{figure}
\centering
\begin{tabular}{m{7cm} m{8cm}}
\includegraphics[width=6cm]{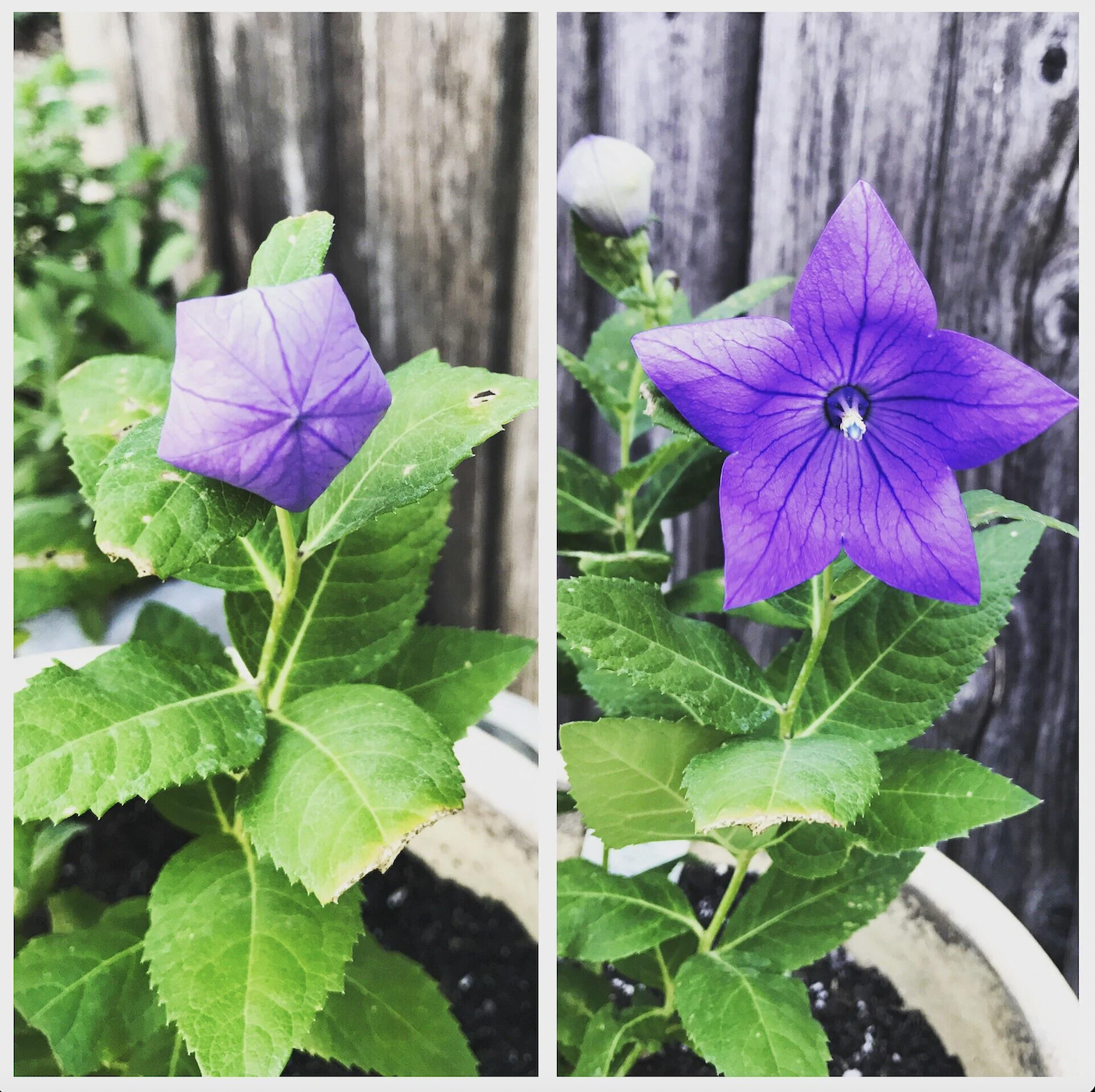}
& 
\includegraphics[width=8.5cm]{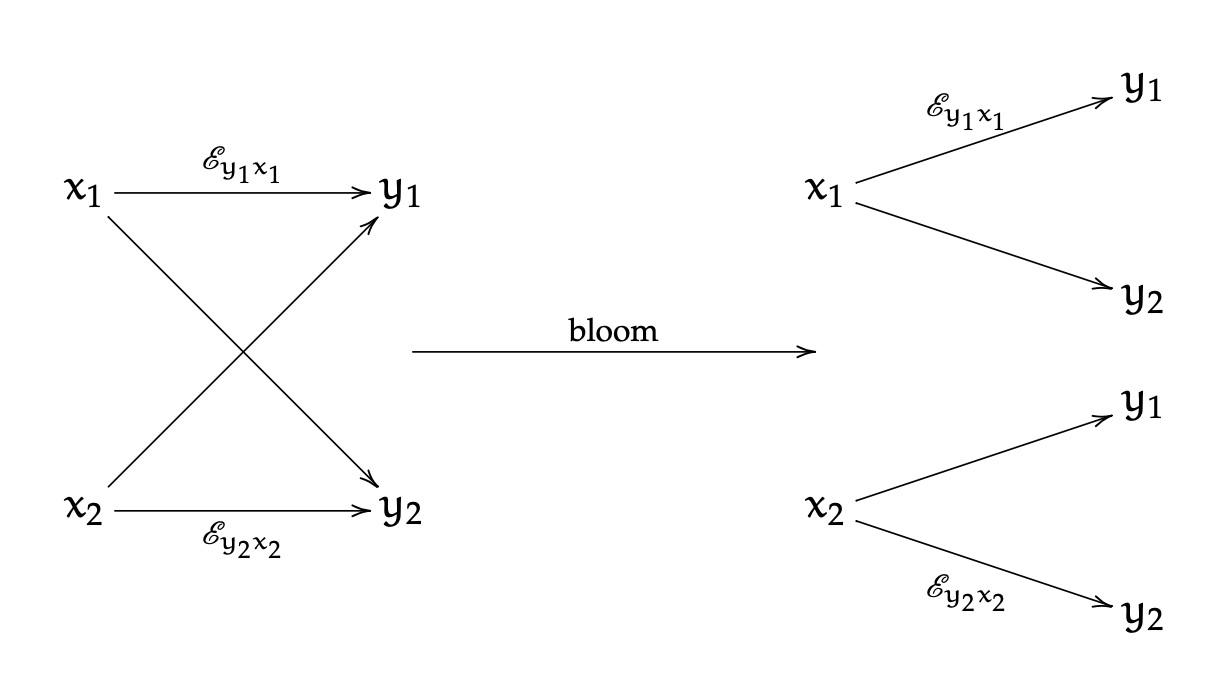}
\end{tabular}
\caption{The bloom map "opens up" the classical channel $\E$.}
\label{fig:SAFH}
\end{figure}

\br
Given a pair of finite sets $(X,Y)$ and a classical process $(\rho,\mathscr{E})\in \mathscr{P}(\C^X,\C^Y)$, if we identify $\mathscr{E}_{yx}$ with conditional probabilities $\mathbb{P}(y|x)$, $\rho\in \mathcal{S}(\C^X)$ with a probability distribution $\mathbb{P}(x)$ on $X$, and $\bloom_{\mathscr{E}}(\rho)$ with a joint probability distribution $\mathbb{P}(x,y)$ on $X\times Y$, then the equation 
\[
\bloom_{\mathscr{E}}(\rho)_{(x,y)}=\rho_x\mathscr{E}_{yx}
\]
may be re-written as
\be \label{CDXPX791}
\mathbb{P}(x,y)=\mathbb{P}(x)\mathbb{P}(y|x),
\ee
which is the classical equation relating a joint distribution with a conditional distribution and a prior. As such, Proposition~\ref{CPXQX191} may be interpreted as saying that the joint distribution $\mathbb{P}(x,y)$ associated with a classical process $(\rho,\mathscr{E})\in \mathscr{P}(\C^X,\C^Y)$ arises as an intermediary step in a canonical factorization $\E=\text{tr}_X\circ \bloom_{\E}$ of the associated channel. This observation will be crucial for generalizing joint distributions to the quantum domain.
\er

Next we show that in the classical domain, joint distributions are in a bijective correspondence with classical processes.

\bnot
Given a pair $(X,Y)$ of finite sets, let $\mathcal{S}^{X}_{+}(\C^X\otimes \C^Y)\subset \mathcal{S}(\C^X\otimes \C^Y)$ denote the subset given by
\[
\mathcal{S}^{X}_{+}(\C^X\otimes \C^Y)=\left\{\tau\in \mathcal{S}(\C^X\otimes \C^Y) \hspace{1mm}|\hspace{1mm} \text{$\text{tr}_Y(\tau)$ is invertible}\right\}.
\]
\enot

\bn \label{STXDNX797}
Let $(X,Y)$ be a pair of finite sets, and let 
\[
\mathfrak{S}:\mathscr{P}_{+}(\C^X,\C^Y)\lra \mathcal{S}^{X}_{+}(\C^{X}\otimes \C^{Y})
\]
be the map given by $\mathfrak{S}(\rho,\mathscr{E})=\bloom_{\mathscr{E}}(\rho)$. Then $\mathfrak{S}$ is a bijection, whose inverse is determined by the condition
\[
\mathfrak{S}^{-1}(\tau)=(\rho,\mathscr{E})\implies \mathscr{E}_{yx}=\frac{\tau_{(x,y)}}{\rho_x} \quad \quad \forall (x,y)\in X\times Y.
\]
\en

\bprf
Let $(X,Y)$ be a pair of finite sets, and let $\mathfrak{C}:\mathcal{S}^{X}_{+}(\C^{X}\otimes \C^{Y})\lra \mathscr{P}_{+}(\C^X,\C^Y)$ be the map uniquely determined by the condition
\[
\mathfrak{C}(\tau)=(\rho,\mathscr{E})\implies \mathscr{E}_{yx}=\frac{\tau_{(x,y)}}{\rho_x} \quad \quad \forall (x,y)\in X\times Y.
\]
We will show $\mathfrak{C}\circ \mathfrak{S}=\text{id}$ and $\mathfrak{S}\circ \mathfrak{C}=\text{id}$. For the former case, let $(\rho,\mathscr{E})\in \mathscr{P}_{+}(\C^X,\C^Y)$, and let $(\widetilde{\rho},\widetilde{\mathscr{E}})=\mathfrak{C}\left(\mathfrak{S}(\rho,\mathscr{E})\right)$. By the definition of $\mathfrak{C}$ we have
\begin{eqnarray*}
\mathfrak{C}(\tau)=(\rho,\mathscr{E})&\implies& \mathscr{E}_{yx}=\frac{\tau_{(x,y)}}{\rho_x} \quad \quad \hspace{2.51cm} \forall (x,y)\in X\times Y \\
&\implies&\sum_{y\in Y}\tau_{(x,y)}=\sum_{y\in Y}\rho_x\mathscr{E}_{yx}=\rho_x \quad \quad \forall x\in X \\
&\implies& \text{tr}_{Y}(\tau)=\rho,
\end{eqnarray*}
thus
\[
\widetilde{\rho}=\text{tr}_Y\left(\mathfrak{S}(\rho,\mathscr{E})\right)=\text{tr}_Y\left(\bloom_{\mathscr{E}}(\rho)\right)=\rho.
\]
Moreover, by definition of $\mathfrak{C}$, for all $(x,y)\in X\times Y$ we have
\[
\widetilde{\mathscr{E}}_{yx}=\frac{\bloom_{\mathscr{E}}(\rho)_{(x,y)}}{\widetilde{\rho}_x}=\frac{\rho_x\mathscr{E}_{yx}}{\rho_x}=\mathscr{E}_{yx} \implies \widetilde{\mathscr{E}}=\mathscr{E},
\]
thus $\mathfrak{C}\circ \mathfrak{S}=\text{id}$. 

Now let $\tau\in \mathcal{S}^{X}_{+}(\C^X\otimes \C^Y)$, let $(\rho,\mathscr{E})=\mathfrak{C}(\tau)$, and let $\widetilde{\tau}=\mathfrak{S}(\mathfrak{C}(\tau))$. Then for all $(x,y)\in X\times Y$ we have
\[
\widetilde{\tau}_{(x,y)}=\mathfrak{S}(\rho,\mathscr{E})_{(x,y)}=\rho_x\mathscr{E}_{yx}=\rho_x\left(\frac{\tau_{(x,y)}}{\rho_x}\right)=\tau_{(x,y)} \implies \widetilde{\tau}=\tau,
\]
thus $\mathfrak{S}\circ \mathfrak{C}=\text{id}$, as desired.
\eprf

\br
In light of Proposition~\ref{STXDNX797}, it follows that a state $\tau\in \mathcal{S}^{X}_{+}(\C^{X}\otimes \C^{Y})$ may either be viewed as a joint distribution associated with two space-like separated random variables $X$ and $Y$ occurring in parallel, or as a \emph{state over time} $\bloom_{\mathscr{E}}(\rho)$ associated with the process $(\rho,\E)=\mathfrak{C}(\tau)\in \mathscr{P}_{+}(\C^X,\C^Y)$. These two equivalent viewpoints are what we refer to as the \define{static-dynamic duality} for classical joint states in $\mathcal{S}^{X}_{+}(\C^{X}\otimes \C^{Y})$. Another way to think of the static-dynamic duality for classical joint states, is that classical probability does not distinguish between temporal correlations and spatial correlations, revealing a certain symmetry between space and time for classical random variables. For quantum systems however it turns out that such a symmetry between space and time \emph{does not} exist, as there exists quantum processes for which temporal correlations may not be viewed as spatial correlations and vice-versa \cite{FJV15}. As such, the static-dynamic duality for classical joint states does not generalize to the quantum domain, and we will see that this is manifested in the fact that quantum states over time are not positive in general. This fact will play a crucial role in the general theory of states over time for quantum processes.
\er

In the next proposition, we show how the state over time $\bloom_{\mathscr{E}}(\rho)$ associated with a classical process $(\rho,\mathscr{E})\in \mathscr{P}(\C^X,\C^Y)$ may be re-written in a way that is valid for all quantum processes, which we will use in the next section for a quantum generalization of states over time.

\bn \label{PXPSN683}
Let $(X,Y)$ be a pair of finite sets, and let $\mathscr{E}\in \bold{CPTP}(\C^X,\C^Y)$. Then 
\be \label{IQBX981}
\bloom_{\mathscr{E}}(\rho)=(\rho\otimes \mathds{1})\mathscr{J}[\mathscr{E}]
\ee
for all $\rho\in \C^X$. In particular, if $\rho\in \mathcal{S}(\C^X)$, then $(\rho\otimes \mathds{1})\mathscr{J}[\mathscr{E}]$ is the state over time associated with the classical process $(\rho,\mathscr{E})$.
\en

\bprf
Let $\mu:\C^X\otimes \C^X\lra \C^X$ denote the multiplication map. From the definition of the Hilbert-Schmidt dual we have
\[
\mu^*(\mathds{1})=\sum_{x\in X}\delta_{x}\otimes \delta_x \hspace{0.25mm},
\]
thus
\begin{eqnarray*}
\mathscr{J}[\mathscr{E}]&=&\left(\text{id}_{\C^X}\otimes \mathscr{E}\right)(\mu^*(\mathds{1}))=\left(\text{id}_{\C^X}\otimes \mathscr{E}\right)\Big(\sum_{x\in X}\delta_{x}\otimes \delta_x\Big)=\sum_{x\in X}\delta_{x}\otimes \mathscr{E}(\delta_x) \\
&=&\sum_{x\in X}\delta_{x}\otimes \Big(\sum_{y\in Y}\mathscr{E}_{yx}\delta_y\Big)=\sum_{(x,y)\in X\times Y}\mathscr{E}_{yx}(\delta_x\otimes \delta_y). 
\end{eqnarray*}
And since for all $\rho\in \C^X$ we have
\[
\rho\otimes \mathds{1}=\Big(\sum_{x\in X}\rho_x\delta_x\Big)\otimes \Big(\sum_{y\in Y}\delta_y\Big)=\sum_{(x,y)\in X\times Y}\rho_x(\delta_x\otimes \delta_y),
\]
it follows that
\begin{eqnarray*}
(\rho\otimes \mathds{1})\mathscr{J}[\mathscr{E}]&=&\Big(\sum_{(x,y)\in X\times Y}\rho_x(\delta_x\otimes \delta_y)\Big)\Big(\sum_{(x,y)\in X\times Y}\mathscr{E}_{yx}(\delta_x\otimes \delta_y)\Big) \\
&=&\sum_{(x,y)\in X\times Y}\rho_x\mathscr{E}_{yx}(\delta_x\otimes \delta_y) \\
&=&\bloom_{\mathscr{E}}(\rho),
\end{eqnarray*}
as desired.
\eprf

\br
While up to now the left-hand side of equation \eqref{IQBX981} is only defined for classical processes, the right-hand side is defined for \emph{any} quantum process $(\rho,\mathscr{E})$. Moreover, since for classical processes we have $\left[\mathscr{J}[\mathscr{E}],(\rho\otimes \mathds{1})\right]=0$, the right-hand side of equation \eqref{IQBX981} coincides with 
\be \label{QSXT837}
\lambda (\rho\otimes \mathds{1})\mathscr{J}[\mathscr{E}]+(1-\lambda)\mathscr{J}[\mathscr{E}](\rho\otimes \mathds{1})
\ee
for all $\lambda \in \C$, which is also well-defined for any quantum process $(\rho,\mathscr{E})$. However when $\left[(\rho\otimes \mathds{1}),\mathscr{J}[\mathscr{E}]\right]\neq 0$ and $\lambda\neq 1$ the expression \eqref{QSXT837} is distinct from $(\rho\otimes \mathds{1})\J[\E]$, thus in the quantum domain, \eqref{QSXT837} yields a parametric family of states over time associated with a general quantum process $(\rho,\E)$ which generalizes the classical state over time $\bloom_{\mathscr{E}}(\rho)$. Another crucial point, is that while the multi-matrix \eqref{QSXT837} is of unit trace for all $\lambda \in \C$, in the quantum domain it is rarely positive (even for $\lambda\in \R$), and is only guaranteed to be self-adjoint for $\lambda=1/2$.  As such, when generalizing states over time to the quantum domain, we will relax the positivity condition and only require states over time to be self-adjoint elements of unit trace. In fact, as first pointed out in \cite{FJV15}, it is the negative eigenvalues of a quantum state over time which act as a witness to causal correlations which exist in the associated process, and are a feature of the theory rather than a defect.
\er

%%%%%%%%%%%%%%%%%%%%%%
\section{In quantum bloom} \label{S4}
%%%%%%%%%%%%%%%%%%%%%%

The classical bloom map $\bloom_{\mathscr{E}}$ from Proposition~\ref{CPXQX191} may be viewed as the output of a mapping that associates every pair $(X,Y)$ of finite sets with a function
\be \label{CLSXCL947}
\bloom_{(*)}:\bold{CPTP}(\C^X,\C^Y)\lra \bold{CPTP}(\C^X,\C^X\otimes \C^Y)
\ee
such that $\text{tr}_X\circ \bloom_{\mathscr{E}}=\mathscr{E}$ and $\text{tr}_Y\circ \bloom_{\mathscr{E}}=\id_{\C^X}$ for all $\mathscr{E}\in \bold{CPTP}(\C^X,\C^Y)$. Moreover, the bloom map $\bloom_{(*)}$ naturally extends to a map on all of $\text{Hom}(\C^X,\C^Y)$, which together with the partial trace provides a factorization system on $\text{Hom}(\C^X,\C^Y)$ which yields the state over time $\bloom_{\mathscr{E}}(\rho)$ associated with every classical process $(\rho,\mathscr{E})\in \mathscr{P}(\C^X,\C^Y)$. In this section, we extend the classical bloom map \eqref{CLSXCL947} to the quantum domain. Though there are various such extensions, the extension referred to as the \emph{symmetric bloom} yields a state over time which is always self-adjoint, and will be the focus of our work.

\bd
A \define{bloom map} associates every pair $(\VA,\VB)$ of multi-matrix algebras with a map $\bloom:\bold{TP}(\VA,\VB)\lra \bold{TP}(\VA,\VA\otimes \VB)$ such that $\text{tr}_{\VA}\circ \bloom(\mathscr{E})=\mathscr{E}$ and $\text{tr}_{\VB}\circ \bloom(\mathscr{E})=\text{id}_{\VA}$ for all $\mathscr{E}\in \bold{TP}(\VA,\VB)$. 
\ed

\bd
If $(\rho,\mathscr{E})\in \mathscr{P}(\VA,\VB)$ is a quantum process, and $\bloom$ is a bloom map, then $\bloom(\mathscr{E})(\rho)$ will be referred to as  the \define{state over time} associated with the process $(\rho,\mathscr{E})$ and the bloom map $\bloom$, and will be denoted by $\bloom(\rho,\E)$. The map on quantum processes given by
\[
(\rho,\E)\longmapsto \bloom(\rho,\E)
\]
will then be referred to as the \define{state over time function} associated with $\bloom$.
\ed

\bd
Given a bloom map $\bloom$, the factorization $\mathscr{E}=\text{tr}_{\VA}\circ \bloom(\E)$ will be referred to as the \define{bloom-shriek factorization} of $\E\in \bold{TP}(\VA,\VB)$. 
\ed

\br
An analogue of bloom-shriek factorization for free gs-monoidal categories appears in \cite{fritz2023free} under the name \emph{bloom-circuitry factorization}.
\er

\bx
Let $(\VA,\VB)$ be a pair of multi-matrix algebras, and let 
\[
\bloom:\bold{TP}(\VA,\VB)\lra \bold{TP}(\VA,\VA\otimes \VB)
\]
be the map given by $\bloom(\mathscr{E})(\rho)=(\rho\otimes \mathds{1})\mathscr{J}[\mathscr{E}]$. When  $\VA$ is a matrix algebra it was shown in \cite{PaRuBayes} that $\mu_{\VA}^*(\mathds{1})=\sum_{i,j}E_{ij}\otimes E_{ji}$, where $E_{ij}$ are the matrix units in $\VA$. We then have
\begin{eqnarray*}
\bloom(\mathscr{E})(\rho)&=&(\rho\otimes \mathds{1})\mathscr{J}[\mathscr{E}]=(\rho\otimes \mathds{1})(\id_{\VA}\otimes \E)\left(\mu_{\VA}^*(\mathds{1})\right) \\
&=&(\rho\otimes \mathds{1})(\id_{\VA}\otimes \E)\Big(\sum_{i,j}E_{ij}\otimes E_{ji}\Big)=(\rho\otimes \mathds{1})\Big(\sum_{i,j}E_{ij}\otimes \E(E_{ji})\Big) \\
&=&\sum_{i,j}\rho E_{ij}\otimes \E(E_{ji}),
\end{eqnarray*}
from which it follows that
\[
\text{tr}_{\VA}\left(\bloom(\mathscr{E})(\rho)\right)=\sum_{i,j}\text{tr}(\rho E_{ij})\E(E_{ji})=\sum_{i,j}\rho_{ji}\E(E_{ji})=\E(\rho),
\]
and
\[
\text{tr}_{\VB}\left(\bloom(\mathscr{E})(\rho)\right)=\sum_{i,j}\text{tr}(\E(E_{ji}))\rho E_{ij}=\sum_{i,j}\delta_{ji}\rho E_{ij}=\sum_{i}\rho E_{ii}=\rho,
\]
thus $\bloom$ defines a bloom map in this case. When $\VA=\bigoplus_{x\in X}\M_{m_x}$ and $\VB=\bigoplus_{y\in Y}\M_{n_y}$ are both multi-matrix algebras, then 
\[
\rho=\bigoplus_{x\in X}\rho_x \quad \& \quad \J[\E]=\bigoplus_{(x,y)\in X\times Y}\J[\E_{yx}],
\]
where $\E_{yx}$ are the component functions of $\E\in \text{Hom}(\VA,\VB)$. We then have
\[
\bloom(\E)(\rho)=(\rho\otimes \mathds{1})\J[\E]=\bigoplus_{(x,y)\in X\times Y}(\rho_x\otimes \mathds{1})\J[\E_{yx}],
\] 
thus the above arguments in the case when $\VA$ is a matrix algebra may be applied component-wise to deduce $\text{tr}_{\VA}\circ \bloom(\mathscr{E})=\mathscr{E}$ and $\text{tr}_{\VB}\circ \bloom(\mathscr{E})=\text{id}_{\VA}$, thus $\bloom$ defines a bloom map. Moreover, it follows from Proposition~\ref{PXPSN683} that this bloom map recovers the classical state over time $\bloom_{\E}(\rho)$ associated with classical processes $(\rho,\E)\in \mathscr{P}(\C^X,\C^Y)$.
\ex

In light of the previous example, the following definition provides three examples of bloom maps. 

\bd
Let $(\VA,\VB)$ be a pair of multi-matrix algebras.
\begin{itemize} 
\item
The \define{right bloom} is the map $\bloom_{R}:\bold{TP}(\VA,\VB)\lra \bold{TP}(\VA,\VA\otimes \VB)$ given by $\mathscr{E}\longmapsto \bloom_{R}(\mathscr{E})$, where 
\[
\bloom_{R}(\mathscr{E})(\rho)=(\rho\otimes \mathds{1})\mathscr{J}[\mathscr{E}]
\]
for all $\rho\in \VA$.
\item
The \define{left bloom} is the map $\bloom_{L}:\bold{TP}(\VA,\VB)\lra \bold{TP}(\VA,\VA\otimes \VB)$ given by $\mathscr{E}\longmapsto \bloom_{L}(\mathscr{E})$, where 
\[
\bloom_{L}(\mathscr{E})(\rho)=\mathscr{J}[\mathscr{E}](\rho\otimes \mathds{1})
\]
for all $\rho\in \VA$.
\item
The \define{symmetric bloom} is the map $\tao:\bold{TP}(\VA,\VB)\lra \bold{TP}(\VA,\VA\otimes \VB)$ given by $\mathscr{E}\longmapsto \tao(\mathscr{E})$, where
\[
\tao(\mathscr{E})(\rho)=\frac{1}{2}\left((\rho\otimes \mathds{1})\mathscr{J}[\mathscr{E}]+\mathscr{J}[\mathscr{E}](\rho\otimes \mathds{1})\right)
\]
for all $\rho\in \VA$.
\end{itemize}
\ed

\br
The symmetric bloom was introduced in \cite{FuPa22}, where it was proved that the associated state over time function $(\rho,\E)\longmapsto \tao(\rho,\E):= \tao(\E)(\rho)$ satisfies a list of axioms set forth in \cite{HHPBS17}. The list of axioms include hermiticity of $\tao(\rho,\E)$, bilinearity of $\tao(*,*)$, a classical limit axiom, and an associativity axiom which ensures that $\tao$ extends in a well-defined way to $n$-step processes $(\rho,\E_1,...,\E_n)$. In Refs.\cite{LiNg23,PFBC23} it is shown that the symmetric bloom is in fact the \emph{only} state over time function satisfying the aforementioned list of axioms. Moreover, for dynamically evolving systems of qubits, the state over time $\tao(\rho,\E)$ coincides with the \emph{pseudo-density matrix} associated with coarse-grained Pauli observables~\cite{Marletto_2021,ZPTGVF18,Pisar19,Marletto_2020,Marletto_2019,Liu_2024,Liu_2024X,LQDV,FuPa24,song23,jia2023,FZ_24,Utagi_2021}.
\er

The following statement yields an important property of the symmetric bloom which will be useful for our purposes.

\bn \label{HXT1987}
The $(\VA,\VB)$ be a pair of multi-matrix algebras, and let $\mathscr{E}\in \bold{TP}(\VA,\VB)$. Then the following statements are equivalent.
\begin{enumerate}[i.] 
\item \label{HXT1}
$\mathscr{E}$ is $\dag$-preserving.
\item \label{HXT2}
 $\mathscr{J}[\mathscr{E}]$ is self-adjoint.
\item \label{HXT3}
$\emph{\Yinyang}(\mathscr{E})$ is $\dag$-preserving.
\end{enumerate}
\en

\bprf
We prove the statement in the case of matrix algebras. 

\underline{\ref{HXT1}$\implies$ \ref{HXT2}}: Let $\mu:\VA\otimes \VA\lra \VA$ denote the multiplication map. It follows from \cite{PaRuBayes} that
\[
\mu^*(\mathds{1})=\sum_{i,j}E_{ij}\otimes E_{ji},
\]
thus
\[
\mathscr{J}[\mathscr{E}]=\left(\id_{\VA}\otimes \mathscr{E}\right)\mu^*(\mathds{1}) =\left(\id_{\VA}\otimes \mathscr{E}\right)\Big(\sum_{i,j}E_{ij}\otimes E_{ji}\Big) =\sum_{i,j}E_{ij}\otimes \mathscr{E}(E_{ji}).
\]
We then have
\begin{eqnarray*}
\mathscr{J}[\mathscr{E}]^{\dag}&=&\Big(\sum_{i,j}E_{ij}\otimes \mathscr{E}(E_{ji})\Big)^{\dag} =\sum_{i,j}E_{ij}^{\dag}\otimes \mathscr{E}(E_{ji})^{\dag} \\
&=&\sum_{i,j}E_{ji}\otimes \mathscr{E}(E_{ji}^{\dag})=\sum_{i,j}E_{ji}\otimes \mathscr{E}(E_{ij}) \\
&=&\mathscr{J}[\mathscr{E}],
\end{eqnarray*}
as desired.

\underline{\ref{HXT2}$\implies$ \ref{HXT3}}: Let $\rho\in \VA$. Then
\begin{eqnarray*}
\tao(\mathscr{E})(\rho)^{\dag}&=&\frac{1}{2}\left((\rho\otimes \mathds{1})\mathscr{J}[\mathscr{E}]+\mathscr{J}[\mathscr{E}](\rho\otimes \mathds{1})\right)^{\dag} \\
&=&\frac{1}{2}\left(\mathscr{J}[\mathscr{E}]^{\dag}(\rho\otimes \mathds{1})^{\dag}+(\rho\otimes \mathds{1})^{\dag}\mathscr{J}[\mathscr{E}]^{\dag}\right) \\
&=&\frac{1}{2}\left(\mathscr{J}[\mathscr{E}](\rho^{\dag}\otimes \mathds{1})+(\rho^{\dag}\otimes \mathds{1})\mathscr{J}[\mathscr{E}]\right) \\
&=&\tao(\mathscr{E})(\rho^{\dag}),
\end{eqnarray*}
thus $\tao(\mathscr{E})$ is $\dag$-preserving.

\underline{\ref{HXT3}$\implies$ \ref{HXT1}}: Since 
\[
\text{tr}_{\VA}(A\otimes B)^{\dag}=\left(\text{tr}(A)B\right)^{\dag}=\overline{\text{tr}(A)}B^{\dag}=\text{tr}(A^{\dag})B^{\dag}=\text{tr}_{\VA}(A^{\dag}\otimes B^{\dag})=\text{tr}_{\VA}\left((A\otimes B)^{\dag}\right),
\]
it follows that $\text{tr}_{\VA}$ is $\dag$-preserving. For all $\rho\in A$ we then have
\[
\mathscr{E}(\rho)^{\dag}=(\text{tr}_{\VA}\circ \tao(\mathscr{E}))(\rho)^{\dag}=\text{tr}_{\VA}\left(\tao(\mathscr{E})(\rho)\right)^{\dag}=\text{tr}_{\VA}\left(\tao(\mathscr{E})(\rho)^{\dag}\right)=\text{tr}_{\VA}\left(\tao(\mathscr{E})(\rho^{\dag})\right)=\mathscr{E}(\rho^{\dag}),
\]
as desired.
\eprf

\bd
A bloom map $\bloom$ is said to be
\itemize
\item
\define{classically reducible} if and only if $\left[(\rho\otimes \mathds{1}),\mathscr{J}[\mathscr{E}]\right]=0\implies \bloom(\mathscr{E})(\rho)=(\rho\otimes \mathds{1})\mathscr{J}[\mathscr{E}]$.
\item
\define{hermitian} if and only if $\bloom(\mathscr{E})$ is $\dag$-preserving whenever $\mathscr{E}$ is $\dag$-preserving.
\ed

\br
By Proposition~\ref{HXT1987} it follows that the symmetric bloom is hermitian, and it follows directly from its definition that the symmetric bloom is classically reducible. These are two key properties one should expect from a quantum bloom map if it is to be viewed as  generalization of the classical bloom. While we have reason to believe that the symmetric bloom is the only bloom map which is both hermitian and classically reducible, we have yet to prove such a result.  
\er

We now generalize Proposition~\ref{STXDNX797} using the symmetric bloom. In particular, given a pseudo-density operator on $\tau \in \VA\otimes \VB$, we derive an explicit formula for a process $(\rho,\E)$ such that $\tau=\tao(\rho,\E)$. For this, we recall that $\mathscr{T}(\VA\otimes \VB)$ denotes the set of pseudo-density operators with respect to the tensor factorization $\VA_0=\VA\otimes \VB$ (see Definition~\ref{PDOXS787}).

\blem \label{LXMA11309}
Let $\tau\in \mathscr{T}(\VA\otimes \VB)$ be such that $\emph{tr}_{\VB}(\tau)$ is invertible. Then there exists a unique element $X_{\tau}\in \VA\otimes \VB$ such that
\be \label{XTAU97}
(\emph{tr}_{\VB}(\tau)\otimes \mathds{1})X_{\tau}+X_{\tau}(\emph{tr}_{\VB}(\tau)\otimes \mathds{1})=2\tau.
\ee
\elem

\bprf
We prove the statement in the case of matrix algebras, where we will use the fact from linear algebra that if $A,B,C\in \M_n$, then the Sylvester equation $AX+XB=C$ has a unique solution $X\in \M_n$ if and only if $A$ and $-B$ have disjoint spectra~\cite{Datta}. So let $\tau \in \mathscr{T}(\VA\otimes \VB)$ be such that $\text{tr}_{\VB}(\tau)$ is invertible. Since $\text{tr}_{\VB}(\tau)$ is a state which is invertible, it follows that the eigenvalues of $\text{tr}_{\VB}(\tau)$ are strictly positive, thus the spectra of $\text{tr}_{\VB}(\tau)\otimes \mathds{1}$ and $-(\text{tr}_{\VB}(\tau)\otimes \mathds{1})$ are disjoint. It then follows that the Sylvester equation
\[
(\text{tr}_{\VB}(\tau)\otimes \mathds{1})X+X(\text{tr}_{\VB}(\tau)\otimes \mathds{1})=2\tau
\]
has a unique solution $X_{\tau}\in \VA\otimes \VB$, thus proving the statement.
\eprf

\bnot
Given a pair $(\VA,\VB)$ of multi-matrix algebras, we let 
\be \label{PDMXT671}
\mathscr{T}_{*}(\VA\otimes \VB)\subset \mathscr{T}^{\VA}_{+}(\VA\otimes \VB)\subset \mathscr{T}(\VA\otimes \VB)
\ee
denote the subsets given by
\[
\mathscr{T}^{\VA}_{+}(\VA \otimes \VB)=\left\{\tau\in \mathscr{T}(\VA\otimes \VB) \hspace{1mm}| \hspace{1mm} \text{tr}_{\VB}(\tau) \hspace{1mm} \text{is invertible} \right\},
\]
and
\[
\mathscr{T}_{*}(\VA\otimes \VB)=\left\{\tau\in \mathscr{T}^{\VA}_{+}(\VA\otimes \VB) \hspace{1mm}| \hspace{1mm} \mathscr{J}^{-1}(X_{\tau})\in \bold{CPTP}(\VA,\VB) \right\},
\]
where $X_{\tau}\in \VA\otimes \VB$ is the unique element satisfying \eqref{XTAU97}.
\enot

\blem \label{TXGDX687}
Let $(\VA,\VB)$ be a pair of multi-matrix algebras. Then $\tao(\rho,\mathscr{E})\in \mathscr{T}_{*}(\VA\otimes \VB)$ for all $(\rho,\mathscr{E})\in \mathscr{P}_{+}(\VA,\VB)$.
\elem

\bprf
Let $(\rho,\mathscr{E})\in \mathscr{P}_{+}(\VA,\VB)$ and let $\tau=\tao(\rho,\mathscr{E})$. To show $\tau\in \mathscr{T}_{*}(\VA\otimes \VB)$ we need to show that $\tau$ is a pseudo-density operator such that $\text{tr}_{\VB}(\tau)$ is invertible and $\mathscr{J}^{-1}(X_{\tau})$ is CPTP.  To show $\tau$ is a pseudo-density operator we first show $\tau$ is self-adjoint. Now since $\mathscr{E}$ is CPTP it is $\dag$-preserving thus $\tao(\mathscr{E})$ is $\dag$-preserving by Proposition~\ref{HXT1987}. We then have
\[
\tau^{\dag}=\tao(\rho,\mathscr{E})^{\dag}=\tao(\mathscr{E})(\rho)^{\dag}=\tao(\mathscr{E})(\rho^{\dag})=\tao(\mathscr{E})(\rho)=\tau,
\]
thus $\tau$ is self-adjoint. And since the marginals of $\tau$ are $\rho$ and $\mathscr{E}(\rho)$ with $\rho$ invertible (since $(\rho,\mathscr{E})\in \mathscr{P}_{+}(\VA,\VB)$), it follows that $\tau \in \mathscr{T}^{\VA}_{+}(\VA\otimes \VB)$. Moreover, since $\rho=\text{tr}_{\VB}(\tau)$ and
\[
\tau=\frac{1}{2}\left((\rho\otimes \mathds{1})\mathscr{J}[\mathscr{E}]+\mathscr{J}[\mathscr{E}](\rho\otimes \mathds{1})\right),
\]
it follows that $X_{\tau}=\mathscr{J}[\mathscr{E}]$ and $\mathscr{J}^{-1}(X_{\tau})=\mathscr{E}\in \bold{CPTP}(\VA,\VB)$, thus $\tau\in \mathscr{T}_{*}(\VA\otimes \VB)$, as desired.
\eprf

\bt \label{EXT987189}
Let $(\VA,\VB)$ be a pair of multi-matrix algebras, and let $\mathfrak{S}:\mathscr{P}_{+}(\VA,\VB)\lra \mathscr{T}_{*}(\VA\otimes \VB)$ be the map given by $\mathfrak{S}(\rho,\mathscr{E})=\tao(\rho,\mathscr{E})$. Then $\mathfrak{S}$ is a bijection, whose inverse is given by
\be \label{EXT987}
\mathfrak{S}^{-1}(\tau)=\left(\emph{tr}_{\VB}(\tau),\mathscr{J}^{-1}(X_{\tau})\right),
\ee
where $X_{\tau}\in \VA\otimes \VB$ is the unique element satisfying \eqref{XTAU97}.
\et

\bprf
Let $\mathfrak{C}:\mathscr{T}_{*}(\VA\otimes \VB)\lra \mathscr{P}_{+}(\VA,\VB)$ be the map given by $\mathfrak{C}(\tau)=(\text{tr}_{\VB}(\tau),\mathscr{J}^{-1}(X_{\tau}))$. We now show $\mathfrak{C}\circ \mathfrak{S}=\id$ and $\mathfrak{S}\circ \mathfrak{C}=\id$. So let $(\rho,\mathscr{E})\in \mathscr{P}_{+}(\VA,\VB)$ and let $\tau=\mathfrak{S}(\rho,\mathscr{E})$, so that $\rho=\text{tr}_{\VB}(\tau)$ and
\[
\tau=\frac{1}{2}\left((\rho\otimes \mathds{1})\mathscr{J}[\mathscr{E}]+\mathscr{J}[\mathscr{E}](\rho\otimes \mathds{1})\right) \implies \mathscr{J}[\mathscr{E}]=X_{\tau}\implies \mathscr{E}=\mathscr{J}^{-1}(X_{\tau}).
\]
We then have
\[
(\mathfrak{C}\circ \mathfrak{S})(\rho,\mathscr{E})=\mathfrak{C}(\tau)=(\text{tr}_{\VB}(\tau),\mathscr{J}^{-1}(X_{\tau}))=(\rho,\mathscr{E}),
\]
thus $\mathfrak{C}\circ \mathfrak{S}=\id$. Now let $\tau\in \mathscr{T}_{*}(\VA\otimes \VB)$, so that
\[
\frac{1}{2}\left(\text{tr}_{\VB}(\tau)\otimes \mathds{1})X_{\tau}+X_{\tau}(\text{tr}_{\VB}(\tau)\otimes \mathds{1}\right)=\tau.
\]
We then have
\begin{eqnarray*}
(\mathfrak{S}\circ \mathfrak{C})(\tau)&=&\mathfrak{S}\left(\mathfrak{C}(\tau)\right) \\
&=&\mathfrak{S}(\text{tr}_{\VB}(\tau),\mathscr{J}^{-1}(X_{\tau})) \\
&=&\frac{1}{2}\left((\text{tr}_{\VB}(\tau)\otimes \mathds{1})\mathscr{J}[\mathscr{J}^{-1}(X_{\tau})]+\mathscr{J}[\mathscr{J}^{-1}(X_{\tau})](\text{tr}_{\VB}(\tau)\otimes \mathds{1})\right) \\
&=&\frac{1}{2}\left(\text{tr}_{\VB}(\tau)\otimes \mathds{1})X_{\tau}+X_{\tau}(\text{tr}_{\VB}(\tau)\otimes \mathds{1}\right) \\
&=&\tau,
\end{eqnarray*}
thus $\mathfrak{S}\circ \mathfrak{C}=\id$, as desired.
\eprf

\br
As pointed out in Ref.~\cite{song23}, when $\text{tr}_{\VB}(\tau)$ is not invertible there exists infinitely many solutions to \eqref{XTAU97}. It then follows that Theorem~\ref{EXT987189} establishes the largest subset of $\VA\otimes \VB$ for which their is a one-to-one correspondence with quantum processes of the form $(\rho,\E)$.
\er

%%%%%%%%%%%%%%%%%%%%
\section{Blooming 2-chains} \label{S5}
%%%%%%%%%%%%%%%%%%%%
Our next goal is to extend bloom maps to $n$-chains, but first we need to consider the case $n=2$. In this section we will show is that if a bloom map satisfies an associativity condition on $2$-chains, then it naturally extends to a unique bloom map on $2$-chains. We then show that the right bloom, left bloom and symmetric bloom are all associative, and as such, yield well-defined state over time functions for 2-step processes. In such a case we also prove an explicit formula for the symmetric bloom state over time function, and show that is satisfies a certain compositionality property. This will provide us with the mathematical foundation for using bloom maps to define explicit states over time associated with the dynamics of $n$-step quantum processes for all $n>0$. 

\bd
Given a triple $(\VA,\VB,\VC)$ of multi-matrix algebras, a \define{$2$-chain} consists of a pair $(\mathscr{E},\mathscr{F})\in \bold{TP}(\VA,\VB)\times \bold{TP}(\VB,\VC)$. The set of all such $2$-chains will be denoted by $\bold{TP}(\VA,\VB,\VC)$.
\ed 

\bx[Blooming 2-chains] \label{B2CX77}
Let $\VA\overset{\mathscr{E}}\longrightarrow \VB\overset{\mathscr{F}}\longrightarrow \VC$ be a 2-chain, let $\bloom$ be a bloom map, and consider the following diagram.
\be \label{DGX3SPX872}
\xy0;/r.25pc/:
(-25,-12.5)*+{\VA}="0";
(0,-12.5)*+{\VB}="1";
(25,-12.5)*+{\VC}="2";
(-12.5,5)*+{\VA\otimes \VB}="4";
(12.5,5)*+{\VB\otimes \VC}="5";
{\ar"0";"1"_{\mathscr{E}}};
{\ar"1";"2"_{\mathscr{F}}};
{\ar"0";"4"^{\bloom}};
{\ar"4";"1"^{\text{tr}}};
{\ar"1";"5"^{\bloom}};
{\ar"5";"2"^{\text{tr}}};
\endxy
\ee
Using only maps in the above diagram, there are two maps one may construct in $\bold{TP}(\VA,\VA\otimes \VB\otimes \VC)$, namely, $\bloom\left(\bloom(\mathscr{F})\circ \mathscr{E}\right)$ and $\bloom(\mathscr{F}\circ \text{tr})\circ \bloom(\mathscr{E})$. As the map $\bloom\left(\bloom(\mathscr{F})\circ \mathscr{E}\right)$ has codomain $\VA\otimes (\VB\otimes \VC)$ and the map $\bloom(\mathscr{F}\circ \text{tr})\circ \bloom(\mathscr{E})$ has codomain $(\VA\otimes \VB)\otimes \VC$, there is a bijection between 2-blooms constructed from the above diagram and parenthezations of the multi-matrix algebra $\VA\otimes \VB\otimes \VC$. It is then a natural to question whether or not the two maps $\bloom\left(\bloom(\mathscr{F})\circ \mathscr{E}\right)$ and $\bloom(\mathscr{F}\circ \text{tr})\circ \bloom(\mathscr{E})$ are equal. If the two maps are in fact equal, then the bloom map $\bloom$ naturally extends to a unique bloom map on 2-chains. Such considerations then motivate the following definition.
\ex

\bd
A bloom map $\bloom$ is said to be \define{associative} if and only if for every 2-chain 
\[
\VA\overset{\mathscr{E}}\longrightarrow \VB\overset{\mathscr{F}}\longrightarrow \VC
\]
we have
\be \label{ASXS91}
\bloom\left(\bloom(\mathscr{F})\circ \mathscr{E}\right)=\bloom(\mathscr{F}\circ \text{tr})\circ \bloom(\mathscr{E}).
\ee
\ed

It turns out that the right bloom, left bloom and symmetric bloom are indeed associative, but before giving the proof, we will first prove the following lemma, which will also be useful later on. 

\blem \label{CPXLX731}
Let $(\VA,\VB,\VC)$ be a triple of matrix algebras, and let $(\mathscr{E},\mathscr{F})\in \bold{TP}(\VA,\VB,\VC)$ be a 2-chain. Then the following statements hold. 
\begin{enumerate}[i.]
\item \label{CPXLX1}
$\mathscr{J}[\bloom_{R}(\mathscr{F})\circ \mathscr{E}]=\left(\mathscr{J}[\mathscr{E}]\otimes \mathds{1}\right)\left(\mathds{1}\otimes \mathscr{J}[\mathscr{F}]\right)$ 
\item \label{CPXLX2}
$\mathscr{J}[\bloom_{L}(\mathscr{F})\circ \mathscr{E}]=\left(\mathds{1}\otimes \mathscr{J}[\mathscr{F}]\right)\left(\mathscr{J}[\mathscr{E}]\otimes \mathds{1}\right)$ 
\item \label{CPXLX3}
$\mathscr{J}[\F\circ \tr]=\mathds{1}\otimes \mathscr{J}[\F]$
\item \label{CPXLX5}
$\bloom_{L}(\bloom_{R}(\mathscr{F})\circ \mathscr{E})=\bloom_{R}(\mathscr{F}\circ \emph{tr})\circ \bloom_{L}(\mathscr{E})$
\item \label{CPXLX4}
$\bloom_{R}(\bloom_{L}(\mathscr{F})\circ \mathscr{E})=\bloom_{L}(\mathscr{F}\circ \emph{tr})\circ \bloom_{R}(\mathscr{E})$
\end{enumerate}
\elem

\bprf
\underline{Item \ref{CPXLX1}}: Indeed, for all $\rho\in \VA$ we have
\begin{eqnarray*}
\mathscr{J}[\bloom_{R}(\mathscr{F})\circ \mathscr{E}]&=&\sum_{i,j} E_{ij}^{\VA}\otimes \left(\bloom_{R}(\mathscr{F})\circ \mathscr{E}\right)(E_{ji}^{\VA}) \\
&=&\sum_{i,j} E_{ij}^{\VA}\otimes \bloom_{R}(\mathscr{F})\left(\mathscr{E}(E_{ji}^{\VA})\right) \\
&=&\sum_{i,j} E_{ij}^{\VA}\otimes \left((\mathscr{E}(E_{ji}^{\VA})\otimes \mathds{1})\mathscr{J}[\mathscr{F}]\right) \\
&=&\sum_{i,j} E_{ij}^{\VA}\otimes \left((\mathscr{E}(E_{ji}^{\VA})\otimes \mathds{1})\Big(\sum_{k,l}E_{kl}^{\VB}\otimes \mathscr{F}(E_{lk}^{\VB})\Big)\right) \\
&=&\sum_{i,j,k,l} E_{ij}^{\VA}\otimes \mathscr{E}(E_{ji}^{\VA})E_{kl}^{\VB}\otimes \mathscr{F}(E_{lk}^{\VB}) \\
&=&\Big(\sum_{i,j}E^{\VA}_{ij}\otimes \mathscr{E}(E^{\VA}_{ji})\otimes \mathds{1}\Big)\Big(\sum_{k,l}\mathds{1}\otimes E^{\VB}_{kl}\otimes \mathscr{F}(E^{\VB}_{lk})\Big) \\
&=&\left(\mathscr{J}[\mathscr{E}]\otimes \mathds{1}\right)\left(\mathds{1}\otimes \mathscr{J}[\mathscr{F}]\right),
\end{eqnarray*}
as desired.

\underline{Item \ref{CPXLX2}}: The proof is similar to that of item \ref{CPXLX1}.

\underline{Item \ref{CPXLX3}}: Indeed,
\begin{eqnarray*}
\mathscr{J}[\F\circ \tr] &=&\sum_{r,s,k,l}\left(E_{rs}^{\VA}\otimes E_{kl}^{\VB}\right)\otimes \left((\mathscr{F}\circ \text{tr})(E_{sr}^{\VA}\otimes E_{lk}^{\VB})\right) \\
&=&\sum_{r,s,k,l}\left(E_{rs}^{\VA}\otimes E_{kl}^{\VB}\right)\otimes \left(\delta_{sr}\mathscr{F}(E_{lk}^{\VB})\right) \\
&=&\sum_{r,k,l}\left(E_{rr}^{\VA}\otimes E_{kl}^{\VB}\right)\otimes \mathscr{F}(E_{lk}^{\VB}) \\
&=&\sum_{k,l}\left(\mathds{1}\otimes E_{kl}^{\VB}\right)\otimes \mathscr{F}(E_{lk}^{\VB}) \\
&=&\mathds{1}\otimes \mathscr{J}[\F],
\end{eqnarray*}
as desired.

\underline{Item \ref{CPXLX5}}: Indeed, for all $\rho\in \VA$ we have
\begin{eqnarray*}
\left(\bloom_{R}(\F\circ \tr)\circ \bloom_{L}(\E)\right)(\rho)&=&\bloom_{R}(\mathscr{F}\circ \text{tr})\left((\bloom_{L}(\mathscr{E})(\rho)\right) \\
&=&\left(\bloom_{L}(\mathscr{E})(\rho)\otimes \mathds{1}\right)\mathscr{J}[\F\circ \tr] \\
&=&\left(\mathscr{J}[\E]\otimes \mathds{1}\right)\left(\rho \otimes \mathds{1}\otimes \mathds{1}\right)\left(\mathds{1}\otimes \mathscr{J}[\F]\right) \\
&=&\left(\mathscr{J}[\E]\otimes \mathds{1}\right)\left(\mathds{1}\otimes \mathscr{J}[\F]\right)\left(\rho \otimes \mathds{1}\otimes \mathds{1}\right) \\
&=&\mathscr{J}[\bloom_{R}(\mathscr{F})\circ \mathscr{E}](\rho\otimes \mathds{1}) \\
&=&\bloom_{L}(\bloom_{R}(\mathscr{F})\circ \mathscr{E})(\rho),
\end{eqnarray*}
where the third and fifth equalities follows from items \ref{CPXLX3} and \ref{CPXLX1}. It then follows that $\bloom_{L}(\bloom_{R}(\mathscr{F})\circ \mathscr{E})=\bloom_{R}(\mathscr{F}\circ \tr)\circ \bloom_{L}(\mathscr{E})$, as desired.

\underline{Item \ref{CPXLX4}}: The proof is similar to that of item \ref{CPXLX5}.
\eprf

\bn \label{ASXTVX676}
The right bloom, left bloom and symmetric bloom are all associative.
\en

\bprf
\underline{Right bloom}: Indeed, for all $\rho\in \VA$ we have
\begin{eqnarray*}
\left(\bloom_{R}(\mathscr{F}\circ \text{tr})\circ \bloom_{R}(\mathscr{E})\right)(\rho)&=&\bloom_{R}(\mathscr{F}\circ \text{tr})\left((\bloom_{R}(\mathscr{E})(\rho)\right) \\
&=&\left(\bloom_{R}(\mathscr{E})(\rho)\otimes \mathds{1}\right)\mathscr{J}[\F\circ \tr] \\
&=&(\rho\otimes \mathds{1})\left(\mathscr{J}[\mathscr{E}]\otimes \mathds{1}\right)\left(\mathds{1}\otimes \mathscr{J}[\mathscr{F}]\right) \\
&=&(\rho\otimes \mathds{1})\mathscr{J}[\bloom_{R}(\mathscr{F})\circ \mathscr{E}] \\
&=&\bloom_{R}(\bloom_{R}(\mathscr{F})\circ \mathscr{E})(\rho),
\end{eqnarray*}
where the third and fourth equalities follow from items \ref{CPXLX3} and  \ref{CPXLX1} of Lemma~\ref{CPXLX731}. It then follows that $\bloom_{R}\left(\bloom_{R}(\mathscr{F})\circ \mathscr{E}\right)=\bloom_{R}(\mathscr{F}\circ \text{tr})\circ \bloom_{R}(\mathscr{E})$, as desired. 

\underline{Left bloom}: The proof is similar to that for the right bloom.

\underline{Symmetric bloom}: We first compute
\begin{eqnarray*}
\tao(\tao(\mathscr{F})\circ \mathscr{E})&=&\frac{1}{2}\left(\bloom_{R}\left(\frac{1}{2}\left(\bloom_{R}(\mathscr{F})+\bloom_{L}(\mathscr{F})\right)\circ \mathscr{E}\right)+\bloom_{L}\left(\frac{1}{2}\left(\bloom_{R}(\mathscr{F})+\bloom_{L}(\mathscr{F})\right)\circ \mathscr{E}\right)\right) \\
&=&\frac{1}{4}\left(\bloom_{R}\left(\bloom_{R}(\mathscr{F})\circ \mathscr{E}\right)+\bloom_{L}\left(\bloom_{L}(\mathscr{F})\circ \mathscr{E}\right)+\bloom_{R}\left(\bloom_{L}(\mathscr{F})\circ \mathscr{E}\right)+\bloom_{L}\left(\bloom_{R}(\mathscr{F})\circ \mathscr{E}\right)\right).
\end{eqnarray*}
Now let $\aleph=\tao(\mathscr{F}\circ \text{tr})\circ \tao(\mathscr{E})$. Then
\begin{eqnarray*}
\aleph&=&\tao(\mathscr{F}\circ \text{tr})\circ \tao(\mathscr{E}) \\
&=&\frac{1}{2}\left(\bloom_{R}(\mathscr{F}\circ \text{tr})+\bloom_{L}(\mathscr{F}\circ \text{tr})\right)\circ \frac{1}{2}\left(\bloom_{R}(\mathscr{E})+\bloom_{L}(\mathscr{E})\right) \\
&=&\frac{1}{4}\left(\bloom_{R}(\mathscr{F}\circ \text{tr})\circ \bloom_{R}(\mathscr{E})+\bloom_{L}(\mathscr{F}\circ \text{tr})\circ \bloom_{L}(\mathscr{E})+\bloom_{R}(\mathscr{F}\circ \text{tr})\circ \bloom_{L}(\mathscr{E})+\bloom_{L}(\mathscr{F}\circ \text{tr})\circ \bloom_{R}(\mathscr{E})\right) \\
&=&\frac{1}{4}\left(\bloom_{R}\left(\bloom_{R}(\mathscr{F})\circ \mathscr{E}\right)+\bloom_{L}\left(\bloom_{L}(\mathscr{F})\circ \mathscr{E}\right)+\bloom_{R}\left(\bloom_{L}(\mathscr{F})\circ \mathscr{E}\right)+\bloom_{L}\left(\bloom_{R}(\mathscr{F})\circ \mathscr{E}\right)\right) \\
&=&\tao(\tao(\mathscr{F})\circ \mathscr{E}),
\end{eqnarray*}
where the second-to-last equality follows from the associativity of $\bloom_{R}$, the associativity of $\bloom_{L}$, and items \ref{CPXLX4} and \ref{CPXLX5} of Lemma~\ref{CPXLX731}.
\eprf

Associative bloom maps yield well-defined state over time functions on 2-step processes, as we now show.

\bd
Given a triple $(\VA,\VB,\C)$ of multi-matrix algebras, an element $(\rho,\mathscr{E},\F)\in \mathcal{S}(\VA)\times \bold{CPTP}(\VA,\VB)\times \bold{CPTP}(\VB,\VC)$ will be referred to as a \define{2-step process}, and the set of 2-step processes $\mathcal{S}(\VA)\times \bold{CPTP}(\VA,\VB)\times \bold{CPTP}(\VB,\VC)$ will be denoted by $\mathscr{P}(\VA,\VB,\VC)$. The subset of  $\mathscr{P}(\VA,\VB,\VC)$ consisting of processes $(\rho,\mathscr{E},\F)$ with $\rho$ invertible will be denoted by $\mathscr{P}_{+}(\VA,\VB,\VC)$.
\ed

\bd
Suppose $\bloom$ is associative. If $(\rho,\mathscr{E},\F)\in \mathscr{P}(\VA,\VB,\VC)$ is a 2-step process, then
\[
\bloom\left(\bloom(\mathscr{F})\circ \mathscr{E}\right)(\rho)=\left(\bloom(\mathscr{F}\circ \text{tr})\circ \bloom(\mathscr{E})\right)(\rho)\in \VA\otimes \VB\otimes \VC
\]
will be referred to as  the \define{state over time} associated with the 2-step process $(\rho,\mathscr{E},\F)$ and the bloom map $\bloom$, and will be denoted by $\bloom(\rho,\E,\F)$. The map on 2-step quantum processes given by
\[
(\rho,\E,\F)\longmapsto \bloom(\rho,\E,\F)
\]
will then be referred to as the \define{2-step state over time function} associated with $\bloom$.
\ed

The next result shows that the state over time $\bloom(\rho,\E,\F)$ has the expected marginals.

\bn \label{PXPNAXS771}
Suppose $\bloom$ is associative, and let $(\rho,\E,\F)\in \mathscr{P}(\VA,\VB,\VC)$ be a 2-step process. Then the following statements hold.
\begin{enumerate}[i.]
\item \label{PXPNAXS7711}
$\emph{tr}_{\VC}\left(\bloom(\rho,\E,\F)\right)=\bloom(\rho,\E)$ 
\item \label{PXPNAXS7712}
$\emph{tr}_{\VA}\left(\bloom(\rho,\E,\F)\right)=\bloom(\E(\rho),\F)$ 
\item \label{PXPNAXS7714}
$\emph{tr}_{\VB\otimes \VC}\left(\bloom(\rho,\E,\F)\right)=\rho$ 
\item \label{PXPNAXS7715}
$\emph{tr}_{\VA\otimes \VC}\left(\bloom(\rho,\E,\F)\right)=\E(\rho)$ 
\item \label{PXPNAXS7716}
$\emph{tr}_{\VA\otimes \VB}\left(\bloom(\rho,\E,\F)\right)=\F(\E(\rho))$ 
\end{enumerate}
\en

\bprf
Associativity is used in the proofs by identifying $\bloom(\rho,\E,\F)$ with either $\left(\bloom(\F\circ \text{tr})\circ \bloom(\E)\right)(\rho)$ or $\bloom(\bloom(\F)\circ \E)(\rho)$.

\underline{Item \ref{PXPNAXS7711}}:
\begin{eqnarray*}
\text{tr}_{\VC}\left(\bloom(\rho,\E,\F)\right)&=&\text{tr}_{\VC}\left(\bloom(\F\circ \text{tr}) \left(\bloom(\E)(\rho)\right)\right) \\
&=&\bloom(\E)(\rho) \\
&=&\bloom(\rho,\E),
\end{eqnarray*}
as desired.

\underline{Item \ref{PXPNAXS7712}}:
\begin{eqnarray*}
\text{tr}_{\VA}\left(\bloom(\rho,\E,\F)\right)&=&\text{tr}_{\VA}\left(\bloom(\bloom(\F)\circ \E)(\rho)\right) \\
&=&(\bloom(\F)\circ \E)(\rho) \\
&=&\bloom(\F)(\E(\rho)) \\
&=&\bloom(\E(\rho),\F),
\end{eqnarray*}
as desired.

\underline{Item \ref{PXPNAXS7714}}: Indeed,
\begin{eqnarray*}
\text{tr}_{\VB\otimes \VC}\left(\bloom(\rho,\E,\F)\right)&=&\text{tr}_{\VB}\left( \text{tr}_{\VC}\left(\bloom(\rho,\E,\F)\right)\right) \\
&=&\text{tr}_{\VB}\left(\bloom(\rho,\E)\right) \\
&=&\rho,
\end{eqnarray*}
where the second equality follows from item \ref{PXPNAXS7711}.

\underline{Item \ref{PXPNAXS7715}}: Indeed,
\begin{eqnarray*}
\text{tr}_{\VA\otimes \VC}\left(\bloom(\rho,\E,\F)\right)&=&\text{tr}_{\VA}\left( \text{tr}_{\VC}\left(\bloom(\rho,\E,\F)\right)\right) \\
&=&\text{tr}_{\VA}\left(\bloom(\rho,\E)\right) \\
&=&\E(\rho),
\end{eqnarray*}
where the second equality follows from item \ref{PXPNAXS7711}.

\underline{Item \ref{PXPNAXS7716}}: Indeed,
\begin{eqnarray*}
\text{tr}_{\VA\otimes \VB}\left(\bloom(\rho,\E,\F)\right)&=&\text{tr}_{\VB}\left(\text{tr}_{\VA}\left(\bloom(\rho,\E,\F)\right)\right) \\
&=&\text{tr}_{\VB}\left(\bloom(\E(\rho),\F)\right) \\
&=&(\F\circ \E)(\rho),
\end{eqnarray*}
where the second equality follows from item \ref{PXPNAXS7712}.
\eprf

We now derive a formula for the symmetric bloom state over time $\tao(\rho,\E,\F)$ in terms of $\rho$, $\J[\E]$ and $\J[\F]$.

\bd
Let $\VA$ be a multi-matrix algebra. The \define{normalized Jordan product} is the map $\bold{Jor}:\VA\times \VA\lra \VA$ given by
\[
\bold{Jor}\left(A,B\right)=\frac{1}{2}\left(AB+BA\right).
\]
\ed

\blem
Let $(\E,\F)\in \bold{TP}(\VA,\VB,\VC)$ be a 2-chain. Then
\[
\mathscr{J}[\tao(\mathscr{F})\circ \mathscr{E}]=\bold{Jor}\left(\mathscr{J}[\E]\otimes \mathds{1},\mathds{1}\otimes \mathscr{J}[\F]\right).
\]
\elem 

\bprf
 Indeed,
\begin{eqnarray*}
\mathscr{J}[\tao(\mathscr{F})\circ \mathscr{E}]&=&\mathscr{J}\left[\frac{1}{2}(\bloom_{R}(\mathscr{F})+\bloom_{L}(\mathscr{F}))\circ \mathscr{E}\right] \\
&=&\frac{1}{2}\mathscr{J}\left[\bloom_{R}(\mathscr{F})\circ \mathscr{E}+\bloom_{L}(\mathscr{F})\circ \mathscr{E}\right] \\
&=&\frac{1}{2}\left(\mathscr{J}\left[\bloom_{R}(\mathscr{F})\circ \mathscr{E}]+\mathscr{J}[\bloom_{L}(\mathscr{F})\circ \mathscr{E}\right]\right) \\
&=&\frac{1}{2}\left(\left(\mathscr{J}[\mathscr{E}]\otimes \mathds{1}\right)\left(\mathds{1}\otimes \mathscr{J}[\mathscr{F}]\right)+\left(\mathds{1}\otimes \mathscr{J}[\mathscr{F}]\right)\left(\mathscr{J}[\mathscr{E}]\otimes \mathds{1}\right)\right) \\
&=&\bold{Jor}\left(\mathscr{J}[\E]\otimes \mathds{1},\mathds{1}\otimes \mathscr{J}[\F]\right),
\end{eqnarray*}
where the second to last equation follows from items \ref{CPXLX1} and \ref{CPXLX2} of Lemma~\ref{CPXLX731}.
\eprf

\bn
Let $(\rho,\E,\F)\in \mathscr{P}(\VA,\VB,\VC)$ be a 2-step process. Then
\be \label{2STPFXM91}
\tao(\rho,\E,\F)=\bold{Jor}\left(\rho\otimes \mathds{1},\bold{Jor}(\mathscr{J}[\E]\otimes \mathds{1},\mathds{1}\otimes \mathscr{J}[\F])\right).
\ee
\en

\bprf
Indeed,
\begin{eqnarray*}
\mathscr{J}[\tao(\F)\circ \E])&=&\mathscr{J}\left[\frac{1}{2}\left(\bloom_{R}(\F)+\bloom_{L}(\F)\right)\circ \E\right] \\
&=&\frac{1}{2}\left(\mathscr{J}[\bloom_{R}(\F)\circ \E]+\bloom_{L}(\F)\circ \E]\right) \\
&=&\frac{1}{2}\left((\mathscr{J}[\E]\otimes \mathds{1})(\mathds{1}\otimes \mathscr{J}[\F])+(\mathds{1}\otimes \mathscr{J}[\F])(\mathscr{J}[\E]\otimes \mathds{1})\right) \\
&=&\bold{Jor}\left(\mathscr{J}[\E]\otimes \mathds{1},\mathds{1}\otimes \mathscr{J}[\F]\right), \\
\end{eqnarray*}
where the third equality follows from items \ref{CPXLX5} and \ref{CPXLX4} Lemma~\ref{CPXLX731}. We then have
\begin{eqnarray*}
\tao(\rho,\E,\F)&=&\tao\left(\tao(\mathscr{F})\circ \mathscr{E}\right)(\rho) \\
&=&\bold{Jor}(\rho\otimes \mathds{1},\mathscr{J}[\tao(\F)\circ \E]) \\
&=&\bold{Jor}(\rho\otimes \mathds{1},\bold{Jor}\left(\mathscr{J}[\E]\otimes \mathds{1},\mathds{1}\otimes \mathscr{J}[\F]\right)), \\
\end{eqnarray*}
as desired.
\eprf

To conclude the section, we prove a result which will be useful later on.

\bn[Compositionality] 
Let $(\rho,\E,\F)\in \mathscr{P}(\VA,\VB,\VC)$ be a two-step process. Then
\be \label{CMPXALTX67}
\emph{tr}_{\VB}\left(\tao(\rho,\E,\F)\right)=\tao(\rho,\F\circ \E).
\ee
\en

\bprf
It follows by direct computation that 
\[
\bold{Jor}\left(\J[\E]\otimes \mathds{1},\mathds{1}\otimes \J[\F]\right)=\sum_{i,j,k,l}E_{ij}^{\VA}\otimes \bold{Jor}\left(\E(E_{ji}^{\VA}),E_{kl}^{\VB}\right)\otimes \F(E_{lk}^{\VB}),
\]
thus
\begin{eqnarray*}
\tr_{\VB}\left((\rho\otimes \mathds{1})\bold{Jor}\left(\J[\E]\otimes \mathds{1},\mathds{1}\otimes \J[\F]\right)\right)&=&\tr_{\VB}\left(\sum_{i,j,k,l}\rho E_{ij}^{\VA}\otimes \bold{Jor}\left(\E(E_{ji}^{\VA}),E_{kl}^{\VB}\right)\otimes \F(E_{lk}^{\VB})\right) \\
&=&\sum_{i,j,k,l}\rho E_{ij}^{\VA}\otimes \left(\E(E_{ji}^{\VA})E_{kl}^{\VB} \F(E_{lk}^{\VB})\right) \\
&=&\sum_{i,j,k,l}\rho E_{ij}^{\VA}\otimes \left(\E(E_{ji}^{\VA})_{lk} \F(E_{lk}^{\VB})\right) \\
&=&(\rho\otimes \mathds{1})\sum_{i,j}E_{ij}^{\VA}\otimes (\F\circ \E)(E_{ji}^{\VA}) \\
&=&(\rho\otimes \mathds{1})\J[\F\circ \E].
\end{eqnarray*}
Similarly, we have
\[
\tr_{\VB}\left(\bold{Jor}\left(\J[\E]\otimes \mathds{1},\mathds{1}\otimes \J[\F]\right)(\rho\otimes \mathds{1})\right)=\J[\F\circ \E](\rho\otimes \mathds{1}),
\]
thus
\begin{eqnarray*}
\tr_{\VB}\left(\tao(\rho,\E,\F)\right)&\overset{\eqref{2STPFXM91}}=&\tr_{\VB}\left(\bold{Jor}\left(\rho\otimes \mathds{1},\bold{Jor}(\mathscr{J}[\E]\otimes \mathds{1},\mathds{1}\otimes \mathscr{J}[\F])\right)\right) \\
&=&\bold{Jor}\left(\rho\otimes \mathds{1},\J[\F\circ \E]\right) \\
&=&\tao(\rho,\F\circ \E),
\end{eqnarray*}
as desired.
\eprf

%%%%%%%%%%%%%%%%%%%%%%
\section{Blooming $n$-chains} \label{S6}
%%%%%%%%%%%%%%%%%%%%%%
In this section we show how associative bloom maps naturally extend to bloom maps for arbitrary $n$-chains.

\bd
Given an $(n+1)$-tuple $(\VA_0,...,\VA_n)$ of multi-matrix algebras, an \define{$n$-chain} consists of a  $n$-tuple 
\[
(\mathscr{E}_1,...,\mathscr{E}_n)\in \bold{TP}(\VA_0,\VA_1)\times \cdots \times \bold{TP}(\VA_{n-1},\VA_n).
\]
The set of all such $n$-chains will be denoted by $\bold{TP}(\VA_0,...,\VA_n)$.
\ed 

\bd \label{NXBLXA87}
An \define{$n$-bloom} associates every $(n+1)$-tuple $(\VA_0,...,\VA_n)$ of multi-matrix algebras with a map $\bloom:\bold{TP}(\VA_0,...,\VA_n)\lra \bold{TP}(\VA_0,\VA_0\otimes \cdots \otimes \VA_n)$ such that
\be \label{BLXCDX9177}
\text{tr}_{i}\circ \bloom(\mathscr{E}_1,...,\mathscr{E}_n)=
\begin{cases}
\mathscr{E}_i\circ \cdots \circ \mathscr{E}_1 \quad \quad \text{for $i\in \{1,...,n\}$} \\
\id_{\VA_0} \quad \quad \hspace{1.23cm} \text{for $i=0$}
\end{cases}
\ee
for all $(\mathscr{E}_1,...,\mathscr{E}_n)\in \bold{TP}(\VA_0,...,\VA_n)$.
\ed

\bx[Blooming 3-chains]
Let $\bloom$ be a bloom map, and let 
\[
\VA_{0}\overset{\mathscr{E}_1}\longrightarrow \VA_{1}\overset{\mathscr{E}_2}\longrightarrow \VA_{2}\overset{\mathscr{E}_3}\longrightarrow \VA_3
\]
be a 3-chain, which after incorporating bloom-shriek factorizations yields the following diagram:
\be \label{DGX3SPX87}
\xy0;/r.25pc/:
(-25,-12.5)*+{\VA_0}="0";
(0,-12.5)*+{\VA_1}="1";
(25,-12.5)*+{\VA_2}="2";
(50,-12.5)*+{\VA_3}="3";
(-12.5,5)*+{\VA_0\otimes \VA_1}="4";
(12.5,5)*+{\VA_1\otimes \VA_2}="5";
(37.5,5)*+{\VA_2\otimes \VA_3}="6";
{\ar"0";"1"_{\mathscr{E}_1}};
{\ar"1";"2"_{\mathscr{E}_2}};
{\ar"2";"3"_{\mathscr{E}_2}};
{\ar"0";"4"^{\bloom}};
{\ar"4";"1"^{\text{tr}}};
{\ar"1";"5"^{\bloom}};
{\ar"5";"2"^{\text{tr}}};
{\ar"2";"6"^{\bloom}};
{\ar"6";"3"^{\text{tr}}};
\endxy
\ee
Similar to the case of 2-chains, there are then five 3-blooms in $\bold{TP}(\VA_0,\VA_0\otimes \VA_1 \otimes \VA_2\otimes \VA_3)$ one may construct from the above diagram \eqref{DGX3SPX87}, each of which corresponds to a parenthezation of  $\VA_0\otimes \VA_1\otimes \VA_2\otimes \VA_3$.

For example, the 3-bloom associated with the parenthezation $\VA_0\otimes (\VA_1\otimes (\VA_2\otimes \VA_3))$ is constructed as follows. We first associate a map with what is inside the outermost parentheses of $\VA_0\otimes (\VA_1\otimes (\VA_2\otimes \VA_3))$, namely, $\VA_1\otimes (\VA_2\otimes \VA_3)$, which is a parenthezation of the multi-matrix algebra $\VA_1\otimes \VA_2\otimes \VA_3$. As we've already seen in the case of 2-chains, the 2-bloom associated with $\VA_1\otimes (\VA_2\otimes \VA_3)$ is then
\[
\bloom\left(\bloom(\mathscr{E}_3)\circ \mathscr{E}_2\right):\VA_1\longrightarrow \VA_1\otimes (\VA_2\otimes \VA_3),
\]
and then post-composing with $\mathscr{E}_1$ yields
\[
\bloom\left(\bloom(\mathscr{E}_3)\circ \mathscr{E}_2\right)\circ \mathscr{E}_1:\VA_0\longrightarrow \VA_1\otimes (\VA_2\otimes \VA_3),
\]
so that taking a final bloom yields 
\[
\bloom\left(\bloom\left(\bloom(\mathscr{E}_3)\circ \mathscr{E}_2\right)\circ \mathscr{E}_1\right):\VA_0\longrightarrow \VA_0\otimes (\VA_1\otimes (\VA_2\otimes \VA_3)),
\]
which is the desired 3-bloom. The 3-blooms associated with the other 4 parenthezations of $\VA_0\otimes \VA_1\otimes \VA_2\otimes \VA_3$ may then be obtained by considering the pentagon where each vertex corresponds to one of the 5 parenthezations of $\VA_0\otimes \VA_1\otimes \VA_2\otimes \VA_3$, and a directed edge is drawn between two vertices if and only if they are related by a single application of the associator transformation $\VA\otimes (\VB\otimes \VC)\longmapsto (\VA\otimes \VB)\otimes \VC$:
\[
\xy0;/r.25pc/:
(0,25)*+{\VA_0\otimes (\VA_1\otimes (\VA_2\otimes \VA_3))}="0";
(-55,0)*+{(\VA_0\otimes \VA_1)\otimes (\VA_2\otimes \VA_3)}="1";
(-30,-35)*+{((\VA_0\otimes \VA_1)\otimes \VA_2)\otimes \VA_3}="2";
(30,-35)*+{(\VA_0\otimes (\VA_1\otimes \VA_2))\otimes \VA_3}="3";
(55,0)*+{\VA_0\otimes ((\VA_1\otimes \VA_2)\otimes \VA_3)}="4";
{\ar"0";"1"_{}};
{\ar"1";"2"_{}};
{\ar"3";"2"_{}};
{\ar"4";"3"_{}};
{\ar"0";"4"^{}};
\endxy
\]
The associated 3-blooms for the other 4 parenthezations of $\VA_0\otimes \VA_1\otimes \VA_2\otimes \VA_3$ can then be obtained from $\bloom\left(\bloom\left(\bloom(\mathscr{E}_3)\circ \mathscr{E}_2\right)\circ \mathscr{E}_1\right)$ by applying corresponding associator transformations of maps $\bloom\left(\bloom(\mathscr{F})\circ \mathscr{E}\right)\longmapsto \bloom(\mathscr{F}\circ \text{tr})\circ \bloom(\mathscr{E})$ along the pentagon:
\[
\xy0;/r.25pc/:
(0,25)*+{\bloom\left(\bloom\left(\bloom(\mathscr{E}_3)\circ \mathscr{E}_2\right)\circ \mathscr{E}_1\right)}="0";
(-55,0)*+{\bloom\left(\bloom(\mathscr{E}_3)\circ \mathscr{E}_2\circ \text{tr}\right)\circ \bloom(\mathscr{E}_1)}="1";
(-30,-35)*+{\bloom\left(\mathscr{E}_3\circ \text{tr}\right)\circ \bloom\left(\mathscr{E}_2\circ \text{tr}\right)\circ \bloom(\mathscr{E}_1)}="2";
(30,-35)*+{\bloom\left(\mathscr{E}_3\circ \text{tr}\right)\circ \bloom(\bloom(\mathscr{E}_2)\circ \mathscr{E}_1)}="3";
(55,0)*+{\bloom\left(\bloom(\mathscr{E}_3\circ \text{tr})\circ \bloom(\mathscr{E}_2)\circ \mathscr{E}_1\right)}="4";
{\ar"0";"1"_{}};
{\ar"1";"2"_{}};
{\ar"3";"2"_{}};
{\ar"4";"3"_{}};
{\ar"0";"4"^{}};
\endxy
\]

This procedure then yields the following 3-blooms associated with each parenthezation of $\VA_0\otimes \VA_1\otimes \VA_2\otimes \VA_3$:

\begin{enumerate}[i.]
\item \label{BMX1}
$\underline{\VA_0\otimes (\VA_1\otimes (\VA_2\otimes \VA_3))}: \hspace{2mm}\bloom\left(\bloom\left(\bloom(\mathscr{E}_3)\circ \mathscr{E}_2\right)\circ \mathscr{E}_1\right)$ 
\item \label{BMX2}
$\underline{(\VA_0\otimes \VA_1)\otimes (\VA_2\otimes \VA_3)}: \hspace{2mm}\bloom\left(\bloom(\mathscr{E}_3)\circ \mathscr{E}_2\circ \text{tr}\right)\circ \bloom(\mathscr{E}_1)$
\item \label{BMX3}
$\underline{((\VA_0\otimes \VA_1)\otimes \VA_2)\otimes \VA_3}: \hspace{2mm}\bloom\left(\mathscr{E}_3\circ \text{tr}\right)\circ \bloom\left(\mathscr{E}_2\circ \text{tr}\right)\circ \bloom(\mathscr{E}_1)$
\item \label{BMX4}
$\underline{\VA_0\otimes ((\VA_1\otimes \VA_2)\otimes \VA_3)}: \hspace{2mm}\bloom\left(\bloom(\mathscr{E}_3\circ \text{tr})\circ \bloom(\mathscr{E}_2)\circ \mathscr{E}_1\right)$
\item \label{BMX5}
$\underline{(\VA_0\otimes (\VA_1\otimes \VA_2))\otimes \VA_3}: \hspace{2mm} \bloom\left(\mathscr{E}_3\circ \text{tr}\right)\circ \bloom(\bloom(\mathscr{E}_2)\circ \mathscr{E}_1)$
\end{enumerate}
If $\bloom$ is assumed to be associative, then this implies that these five 3-blooms are all equal:

\underline{\ref{BMX1} $=$ \ref{BMX2}}: The statement follows from \eqref{ASXS91} with $\mathscr{E}=\mathscr{E}_1$ and $\mathscr{F}=\bloom(\mathscr{E}_3)\circ \mathscr{E}_2$.

\underline{\ref{BMX2} $=$ \ref{BMX3}}: The statement follows from \eqref{ASXS91} with $\mathscr{E}=\mathscr{E}_2\circ \text{tr}$ and $\mathscr{F}=\mathscr{E}_3$.

\underline{\ref{BMX1} $=$ \ref{BMX4}}: The statement follows from \eqref{ASXS91} with $\mathscr{E}=\mathscr{E}_2$ and $\mathscr{F}=\mathscr{E}_3$.

\underline{\ref{BMX4} $=$ \ref{BMX5}}: The statement follows from \eqref{ASXS91} with $\mathscr{E}=\bloom(\mathscr{E}_2)\circ \mathscr{E}_1$ and $\mathscr{F}=\mathscr{E}_3\circ \text{tr}$.
\ex

\br
Note that when showing \ref{BMX4}$=$\ref{BMX5} we have set $\text{tr}\circ \text{tr}=\text{tr}$ as we are eliding the domains and codomains of the partial trace, which we will continue to do throughout.
\er

We now generalize the previous two examples to $n$-chains for all $n>1$. Let $\bloom$ be a bloom map, let $n>1$, and let 
\be \label{NXCXN57}
\VA_{0}\overset{\mathscr{E}_1}\longrightarrow \VA_{1} \longrightarrow \cdots \longrightarrow \VA_{n-1}\overset{\mathscr{E}_n}\longrightarrow \VA_{n}
\ee
be an $n$-chain. With every parenthezation $\vartheta(\VA_0,...,\VA_n)$ of the multi-matrix algebra $\VA_0\otimes \cdots \otimes \VA_n$ we associate a map
\[
\bloom_{\vartheta}(\mathscr{E}_1,...,\mathscr{E}_n):\VA_0\longrightarrow \vartheta(\VA_0,...,\VA_n)
\]
constructed from the bloom map $\bloom$, the maps $\mathscr{E}_i$ for $i=1,...,n$ and the partial trace $\text{tr}$. For the case $n=2$, we know from Example~\ref{B2CX77} that there are two parenthezations of $\VA_0\otimes \VA_1\otimes \VA_2$, namely,
\[
\vartheta(\VA_0,\VA_1,\VA_2)=(\VA_0\otimes \VA_1)\otimes \VA_3 \quad \& \quad
\widetilde{\vartheta}(\VA_0,\VA_1,\VA_2)=\VA_0\otimes (\VA_1\otimes \VA_2).
\]
and we let 
\[
\bloom_{\vartheta}(\mathscr{E}_1,\mathscr{E}_2)=\bloom(\mathscr{E}_2\circ \text{tr})\circ \bloom(\mathscr{E}_1) \quad \& \quad \bloom_{\widetilde{\vartheta}}(\mathscr{E}_1,\mathscr{E}_2)=\bloom(\bloom(\mathscr{E}_2)\circ \mathscr{E}_1).
\]
In such a case we have that $\bloom_{\vartheta}(\mathscr{E}_1,\mathscr{E}_2)$ may be obtained from $\bloom_{\widetilde{\vartheta}}(\mathscr{E}_1,\mathscr{E}_2)$ by a transformation of the form 
\[
\bloom(\bloom(\mathscr{F})\circ \mathscr{E})\longmapsto \bloom(\mathscr{F}\circ \text{tr})\circ \bloom(\mathscr{E})
\]
with $\mathscr{F}=\mathscr{E}_2$ and $\mathscr{E}=\mathscr{E}_1$. For general $n>2$ we will build up recursively from this case. 

So now let $n>2$ and let $\vartheta(\VA_0,...,\VA_n)$ be an arbitrary parenthezation of $\VA_0\otimes \cdots \otimes\VA_n$. It then follows that $\vartheta(\VA_0,...,\VA_n)$ is in one of the three forms:
\begin{eqnarray*}
I:&  \vartheta(\VA_0,...,\VA_n)&= \hspace{1mm} \VA_{0}\otimes \left(\omega(\VA_1,...,\VA_n)\right) \\
II:& \vartheta(\VA_0,...,\VA_n)&=\hspace{1mm} \left(\omega(\VA_0,...,\VA_{n-1})\right)\otimes \VA_n \\
III:& \vartheta(\VA_0,...,\VA_n)&=\hspace{1mm} \left(\omega(\VA_0,...,\VA_{k-1})\right)\otimes \left(\chi(\VA_k,...,\VA_{n})\right) 
\end{eqnarray*}
where $1<k<n$ and $\omega(\VA_1,...,\VA_n)$, $\omega(\VA_0,...,\VA_{n-1})$, $\omega(\VA_0,...,\VA_{k-1})$ and $\chi(\VA_k,...,\VA_{n})$ are the parenthezations of the multi-matrix algebras $\VA_1\otimes \cdots \otimes\VA_n$, $\VA_0\otimes \cdots \otimes\VA_{n-1}$, $\VA_0\otimes \cdots \otimes\VA_{k-1}$ and $\VA_k\otimes \cdots \otimes\VA_n$ for which the above equations hold.  In each of the three cases we define the associated $n$-blooms recursively
as follows.
\begin{eqnarray*}
I:&  \bloom_{\vartheta}(\mathscr{E}_1,...,\mathscr{E}_n)&=\hspace{2mm}\bloom\left(\bloom_{\omega}(\mathscr{E}_2,...,\mathscr{E}_n)\circ \hspace{0.75mm} \mathscr{E}_1\right) \\
II:& \bloom_{\vartheta}(\mathscr{E}_1,...,\mathscr{E}_n)&=\hspace{2mm}\bloom(\mathscr{E}_n\circ \text{tr})\circ \bloom_{\omega}(\mathscr{E}_1,...,\mathscr{E}_{n-1})  \\
III:& \bloom_{\vartheta}(\mathscr{E}_1,...,\mathscr{E}_n)&=\hspace{2mm}\bloom\left(\bloom_{\chi}(\mathscr{E}_{k+1},...,\mathscr{E}_n)\circ  \hspace{0.75mm}  \mathscr{E}_k\circ \text{tr}\right)\circ \bloom_{\omega}(\mathscr{E}_1,...,\mathscr{E}_{k-1})
\end{eqnarray*}
The next proposition shows that $\bloom_{\vartheta}$, when viewed as a map on $n$-chains, is an actual $n$-bloom, i.e., that $\bloom_{\vartheta}$ satisfies the condition \eqref{BLXCDX9177} in Definition~\ref{NXBLXA87}. 

\bn
Let $n>1$, let $(\VA_0,...,\VA_n)$ be an $(n+1)$-tuple of multi-matrix algebras, let $\vartheta(\VA_0,...,\VA_n)$ be a parenthezation of $\VA_0\otimes \cdots \otimes \VA_n$, and let $\bloom$ be a bloom map. Then $\bloom_{\vartheta}$ is an $n$-bloom, i.e.,
\be \label{BXCXDMA381}
\emph{tr}_{i}\circ \bloom_{\vartheta}(\mathscr{E}_1,...,\mathscr{E}_n)=
\begin{cases}
\mathscr{E}_i\circ \cdots \circ \mathscr{E}_1 \quad \quad \text{for $i\in \{1,...,n\}$} \\
\id_{\VA_0} \quad \quad \hspace{1.23cm} \text{for $i=0$}
\end{cases}
\ee
for all $(\mathscr{E}_1,...,\mathscr{E}_n)\in \bold{TP}(\VA_0,...,\VA_n)$.
\en

\bprf
We use induction on $n$. For $n=2$ the statement follows from Proposition~\ref{PXPNAXS771}. Now suppose the result holds for $n=m-1>2$, and let $\vartheta(\VA_0,...,\VA_{m})$ be a parenthezation of $\VA_0\otimes \cdots \otimes \VA_m$. We then consider the three cases as for $\vartheta(\VA_0,...,\VA_{m})$ as defined above.

\underline{Case I:} In this case we have $\vartheta(\VA_0,...,\VA_{m})=\VA_{0}\otimes \left(\omega(\VA_1,...,\VA_m)\right)$ and
\[
\bloom_{\vartheta}(\mathscr{E}_1,...,\mathscr{E}_m)=\bloom\left(\bloom_{\omega}(\mathscr{E}_2,...,\mathscr{E}_m)\circ \hspace{0.75mm} \mathscr{E}_1\right)
\]
for all $m$-chains $(\mathscr{E}_1,...,\mathscr{E}_m)$. For $i=0$ we then have
\[
\text{tr}_{0}\circ \bloom_{\vartheta}(\mathscr{E}_1,...,\mathscr{E}_n)=\text{tr}_{0}\circ \bloom\left(\bloom_{\omega}(\mathscr{E}_2,...,\mathscr{E}_m)\circ \hspace{0.75mm} \mathscr{E}_1\right)=\id_{\VA_0},
\]
where the last equality follows from the definition of a bloom map. As for $0<i\leq m$, we have
\begin{eqnarray*}
\text{tr}_{i}\circ \bloom_{\vartheta}(\mathscr{E}_1,...,\mathscr{E}_n)&=&\text{tr}_{\VA_{0}\otimes \cdots \otimes\widehat{\VA_{i}}\otimes \cdots \otimes\VA_{m}}\circ \bloom\left(\bloom_{\omega}(\mathscr{E}_2,...,\mathscr{E}_m)\circ \hspace{0.75mm} \mathscr{E}_1\right) \\
&=&\text{tr}_{\VA_{1}\otimes \cdots \otimes\widehat{\VA_{i}}\otimes \cdots \otimes\VA_{m}}\circ \text{tr}_{\VA_0}\circ \bloom\left(\bloom_{\omega}(\mathscr{E}_2,...,\mathscr{E}_m)\circ \hspace{0.75mm} \mathscr{E}_1\right) \\
&=&\text{tr}_{\VA_{1}\otimes \cdots \otimes\widehat{\VA_{i}}\otimes \cdots \otimes\VA_{m}}\circ \bloom_{\omega}(\mathscr{E}_2,...,\mathscr{E}_m)\circ \hspace{0.75mm} \mathscr{E}_1 \\
&\overset{\text{induction}}=&(\mathscr{E}_i\circ \cdots \circ \mathscr{E}_2)\circ \mathscr{E}_1 \\
&=&\mathscr{E}_i\circ \cdots \circ \mathscr{E}_1, \\
\end{eqnarray*}
thus $\bloom_{\vartheta}$ is an $m$-bloom.

\underline{Case II:} In this case we have $\vartheta(\VA_0,...,\VA_{m})=\left(\omega(\VA_0,...,\VA_{m-1})\right)\otimes \VA_m$ and
\[
\bloom_{\vartheta}(\mathscr{E}_1,...,\mathscr{E}_m)=\bloom(\mathscr{E}_m\circ \text{tr})\circ \bloom_{\omega}(\mathscr{E}_1,...,\mathscr{E}_{m-1})
\]
for all $m$-chains $(\mathscr{E}_1,...,\mathscr{E}_m)$, where $\text{tr}=\text{tr}_{\VA_0\otimes \cdots \otimes \VA_{m-2}}:\VA_0\otimes \cdots \otimes \VA_{m-1}\lra \VA_{m-1}$. For $i=m$ we then have
\begin{eqnarray*}
\text{tr}_{m}\circ \bloom_{\vartheta}(\mathscr{E}_1,...,\mathscr{E}_n)&=&\text{tr}_{\VA_{0}\otimes \cdots \otimes\VA_{m-1}}\circ \bloom(\mathscr{E}_m\circ \text{tr})\circ \bloom_{\omega}(\mathscr{E}_1,...,\mathscr{E}_{m-1}) \\
&=&\mathscr{E}_m\circ \left( \text{tr}\circ \bloom_{\omega}(\mathscr{E}_1,...,\mathscr{E}_{m-1})\right) \\
&\overset{\text{induction}}=&\mathscr{E}_m\circ \left(\mathscr{E}_{m-1}\circ \cdots \circ \mathscr{E}_1\right) \\
&=&\mathscr{E}_{m}\circ \cdots \circ \mathscr{E}_1.
\end{eqnarray*}
As for $0\leq i <m$, we have
\begin{eqnarray*}
\text{tr}_{i}\circ \bloom_{\vartheta}(\mathscr{E}_1,...,\mathscr{E}_n)&=&\text{tr}_{\VA_{0}\otimes \cdots \otimes\widehat{\VA_{i}}\otimes \cdots \otimes\VA_{m-1}}\circ \left(\text{tr}_{\VA_m}\circ \bloom(\mathscr{E}_m\circ \text{tr})\right)\circ \bloom_{\omega}(\mathscr{E}_1,...,\mathscr{E}_{m-1}) \\
&=&\text{tr}_{\VA_{0}\otimes \cdots \otimes\widehat{\VA_{i}}\otimes \cdots \otimes\VA_{m-1}}\circ \id_{\VA_0\otimes \cdots \otimes \VA_{m-1}}\circ \bloom_{\omega}(\mathscr{E}_1,...,\mathscr{E}_{m-1}) \\
&=&\text{tr}_{\VA_{0}\otimes \cdots \otimes\widehat{\VA_{i}}\otimes \cdots \otimes\VA_{m-1}}\circ \bloom_{\omega}(\mathscr{E}_1,...,\mathscr{E}_{m-1}) \\
&\overset{\text{induction}}=&
\begin{cases}
\mathscr{E}_i\circ \cdots \circ \mathscr{E}_1 \quad \quad \text{for $i\in \{1,...,m-1\}$} \\
\id_{\VA_0} \quad \quad \hspace{1.23cm} \text{for $i=0$},
\end{cases}
\end{eqnarray*}
thus $\bloom_{\vartheta}$ is an $m$-bloom.

\underline{Case III:} In this case we have $\vartheta(\VA_0,...,\VA_{m})=\left(\omega(\VA_0,...,\VA_{k-1})\right)\otimes \left(\chi(\VA_k,...,\VA_{m})\right)$ for some $1<k<m$ and
\[
\bloom_{\vartheta}(\mathscr{E}_1,...,\mathscr{E}_m)=\bloom\left(\bloom_{\chi}(\mathscr{E}_{k+1},...,\mathscr{E}_m)\circ  \hspace{0.75mm}  \mathscr{E}_k\circ \text{tr}\right)\circ \bloom_{\omega}(\mathscr{E}_1,...,\mathscr{E}_{k-1})
\]
for all $m$-chains $(\mathscr{E}_1,...,\mathscr{E}_m)$, where $\text{tr}=\text{tr}_{\VA_0\otimes \cdots \otimes \VA_{k-2}}:\VA_0\otimes \cdots \otimes \VA_{k-1}\lra \VA_{k-1}$. Now let $\aleph=\text{tr}_{i}\circ \bloom_{\vartheta}(\mathscr{E}_1,...,\mathscr{E}_n)$. Then for $0\leq i <k$ we have
\begin{eqnarray*}
\aleph&=&\text{tr}_{i}\circ \bloom_{\vartheta}(\mathscr{E}_1,...,\mathscr{E}_n) \\
&=&\text{tr}_{\VA_{0}\otimes \cdots \otimes\widehat{\VA_{i}}\otimes \cdots \otimes\VA_{m}}\circ \bloom\left(\bloom_{\chi}(\mathscr{E}_{k+1},...,\mathscr{E}_m)\circ  \hspace{0.75mm}  \mathscr{E}_k\circ \text{tr}\right)\circ \bloom_{\omega}(\mathscr{E}_1,...,\mathscr{E}_{k-1}) \\
&=&\text{tr}_{\VA_{0}\otimes \cdots \otimes\widehat{\VA_{i}}\otimes \cdots \otimes\VA_{k-1}}\circ \left(\text{tr}_{\VA_k\otimes \cdots \VA_{m}}\circ \bloom\left(\bloom_{\chi}(\mathscr{E}_{k+1},...,\mathscr{E}_m)\right)\circ  \hspace{0.75mm}  \mathscr{E}_k\circ \text{tr}\right)\circ \bloom_{\omega}(\mathscr{E}_1,...,\mathscr{E}_{k-1}) \\
&=&\text{tr}_{\VA_{0}\otimes \cdots \otimes\widehat{\VA_{i}}\otimes \cdots \otimes\VA_{k-1}}\circ \id_{\VA_0\otimes \cdots \otimes \VA_{k-1}}\circ \bloom_{\omega}(\mathscr{E}_1,...,\mathscr{E}_{k-1}) \\
&=&\text{tr}_{\VA_{0}\otimes \cdots \otimes\widehat{\VA_{i}}\otimes \cdots \otimes\VA_{k-1}}\circ \bloom_{\omega}(\mathscr{E}_1,...,\mathscr{E}_{k-1}) \\
&\overset{\text{induction}}=&
\begin{cases}
\mathscr{E}_i\circ \cdots \circ \mathscr{E}_1 \quad \quad \text{for $i\in \{1,...,k-1\}$} \\
\id_{\VA_0} \quad \quad \hspace{1.23cm} \text{for $i=0$}.
\end{cases}
\end{eqnarray*}
As for $k\leq i\leq m$, we have
\begin{eqnarray*}
\aleph&=&\text{tr}_{i}\circ \bloom_{\vartheta}(\mathscr{E}_1,...,\mathscr{E}_n) \\
&=&\text{tr}_{\VA_{0}\otimes \cdots \otimes\widehat{\VA_{i}}\otimes \cdots \otimes\VA_{m}}\circ \bloom\left(\bloom_{\chi}(\mathscr{E}_{k+1},...,\mathscr{E}_m)\circ  \hspace{0.75mm}  \mathscr{E}_k\circ \text{tr}\right)\circ \bloom_{\omega}(\mathscr{E}_1,...,\mathscr{E}_{k-1}) \\
&=&\text{tr}_{\VA_{k}\otimes \cdots \otimes\widehat{\VA_{i}}\otimes \cdots \otimes\VA_{m}}\circ \left(\text{tr}_{\VA_0\otimes \cdots \otimes \VA_{k-1}}\circ \bloom\left(\bloom_{\chi}(\mathscr{E}_{k+1},...,\mathscr{E}_m)\circ  \hspace{0.75mm}  \mathscr{E}_k\circ \text{tr}\right)\right)\circ \bloom_{\omega}(\mathscr{E}_1,...,\mathscr{E}_{k-1}) \\
&=&\text{tr}_{\VA_{k}\otimes \cdots \otimes\widehat{\VA_{i}}\otimes \cdots \otimes\VA_{m}}\circ \left(\bloom_{\chi}(\mathscr{E}_{k+1},...,\mathscr{E}_m)\circ  \hspace{0.75mm}  \mathscr{E}_k\circ \text{tr}\right)\circ \bloom_{\omega}(\mathscr{E}_1,...,\mathscr{E}_{k-1}) \\
&=&\left(\text{tr}_{\VA_{k}\otimes \cdots \otimes\widehat{\VA_{i}}\otimes \cdots \otimes\VA_{m}}\circ \bloom_{\chi}(\mathscr{E}_{k+1},...,\mathscr{E}_m)\right)\circ  \hspace{0.75mm}  \mathscr{E}_k\circ \left(\text{tr}_{\VA_0\otimes \cdots \otimes \VA_{k-2}}\circ \bloom_{\omega}(\mathscr{E}_1,...,\mathscr{E}_{k-1})\right) \\
&\overset{\text{induction}}=&
\left(\begin{cases}
\mathscr{E}_i\circ \cdots \circ \mathscr{E}_{k+1} \quad \quad \text{for $i\in \{k+1,...,m\}$} \\
\id_{\VA_k} \quad \quad \hspace{1.63cm} \text{for $i=k$}
\end{cases}\right)
\circ \mathscr{E}_k\circ \mathscr{E}_{k-1}\circ \cdots \circ \mathscr{E}_1 \\
&=&\mathscr{E}_i\circ \cdots \circ \mathscr{E}_1,
\end{eqnarray*}
thus $\bloom_{\vartheta}$ is an $m$-bloom, as desired. 
\eprf

\br
If $\vartheta(\VA_0,...,\VA_n)$ and $\widetilde{\vartheta}(\VA_0,...,\VA_n)$ are two parenthezations of $\VA_0\otimes \cdots \otimes \VA_n$ such that $\widetilde{\vartheta}(\VA_0,...,\VA_n)$ is obtained from $\vartheta(\VA_0,...,\VA_n)$ by an associator transformation of the form $\VA\otimes (\VB\otimes \VC)\longmapsto (\VA\otimes \VB)\otimes \VC$, then there exists a 2-chain
\[
\VA\overset{\mathscr{E}}\longrightarrow \VB\overset{\mathscr{F}}\longrightarrow \VC
\]
such that $\bloom_{\widetilde{\vartheta}}$ is obtained from $\bloom_{\vartheta}$ by an associator transformation at the level of \emph{maps} associated with the 2-chain $(\E,\mathscr{F})$, i.e., 
\[
\bloom_{\vartheta}=\bloom(\bloom(\mathscr{F})\circ \mathscr{E})\longmapsto \bloom(\mathscr{F}\circ \text{tr})\circ \bloom(\mathscr{E})=\bloom_{\widetilde{\vartheta}}.
\]
\er

The previous remark motivates the following definition.

\bd
Let $n>1$ and let $0<k< l\leq n$. The \define{associator relation} on the set of maps $\bold{TP}(\VA_{l-k},\VA_{l-k}\otimes \cdots \otimes \VA_{l})$ is the subset 
\[
\mathscr{A}\subset \bold{TP}(\VA_{l-k},\VA_{l-k}\otimes \cdots \otimes \VA_{l})\times \bold{TP}(\VA_{l-k},\VA_{l-k}\otimes \cdots \otimes \VA_{l})
\]
given by $(\psi,\varphi)\in \mathscr{A}$ if and only if $\varphi$ is obtained from $\psi$ from successive transformations of the form 
\[
\bloom(\bloom(\mathscr{F})\circ \mathscr{E})\longmapsto \bloom(\mathscr{F}\circ \text{tr})\circ \bloom(\mathscr{E}).
\]
In such a case, we use the notation $\psi\searrow \varphi$ to denote the fact that $(\psi,\varphi)\in \mathscr{A}$.
\ed

\blem \label{LXMMA8971}
Let $k>0$. Then $\bloom(\bloom(\mathscr{F}_1)\circ \cdots \circ \bloom(\mathscr{F}_k)\circ \mathscr{E})\searrow \bloom(\mathscr{F}_1\circ \emph{tr})\circ \cdots \bloom(\mathscr{F}_k\circ \emph{tr})\circ \bloom(\mathscr{E})$.
\elem

\bprf
We use induction on $k$. The case $k=1$ holds by definition of the associator relation, so now assume the result holds for $k=m-1$. Then
\begin{eqnarray*}
\bloom(\bloom(\mathscr{F}_1)\circ \cdots \circ \bloom(\mathscr{F}_{m-1})\circ \bloom(\mathscr{F}_m)\circ \mathscr{E})&\overset{\text{induction}}\searrow&  \bloom(\mathscr{F}_1\circ \text{tr})\circ \cdots \bloom(\mathscr{F}_{m-1}\circ \text{tr})\circ \bloom(\bloom(\mathscr{F}_m)\circ \mathscr{E}) \\
&\overset{\text{associator}}\searrow&\bloom(\mathscr{F}_1\circ \text{tr})\circ \cdots \bloom(\mathscr{F}_{m-1}\circ \text{tr})\circ \bloom(\mathscr{F}_m\circ \text{tr})\circ \bloom(\mathscr{E}),
\end{eqnarray*}
thus the result holds for $k=m$, as desired.
\eprf

\bn \label{PXPNRSX67}
Let $n>1$, and let $\vartheta(\VA_0,...,\VA_n)$ be a parenthezation of $\VA_0\otimes \cdots \otimes \VA_n$. Then
\[
\bloom_{\vartheta}(\mathscr{E}_1,...,\mathscr{E}_n)\searrow \bloom(\mathscr{E}_n\circ \emph{tr})\circ \bloom(\mathscr{E}_{n-1}\circ \emph{tr})\circ \cdots \circ \bloom(\mathscr{E}_2\circ \emph{tr})\circ \bloom(\mathscr{E}_1).
\]
\en

\bprf
We will consider each of the three cases given above for the definition of $\bloom_{\vartheta}$, and use strong induction on $n$. The case $n=2$ follows by definition, so now assume the result holds for all $m$ with $1<m<n$.

\underline{Case I}: Indeed,
\begin{eqnarray*}
\bloom_{\vartheta}(\mathscr{E}_1,...,\mathscr{E}_n)&=&\bloom(\bloom_{\omega}(\mathscr{E}_2,...,\mathscr{E}_n)\circ \hspace{0.75mm} \mathscr{E}_1) \\
&\overset{\text{induction}}\searrow& \bloom\left(\bloom(\mathscr{E}_n\circ \text{tr})\circ \cdots \circ \bloom(\mathscr{E}_3\circ \text{tr})\circ \bloom(\mathscr{E}_2)\circ \mathscr{E}_1\right) \\
&\overset{\text{Lemma~\ref{LXMMA8971}}}\searrow&\bloom(\mathscr{E}_n\circ \text{tr})\circ \bloom(\mathscr{E}_{n-1}\circ \text{tr})\circ \cdots \circ \bloom(\mathscr{E}_2\circ \text{tr})\circ \bloom(\mathscr{E}_1),
\end{eqnarray*}
where in the last line we note that we have repeatedly used the elision $\text{tr}\circ \text{tr}=\text{tr}$, thus the result holds.

\underline{Case II}: Indeed,
\begin{eqnarray*}
\bloom_{\vartheta}(\mathscr{E}_1,...,\mathscr{E}_n)&=&\bloom(\mathscr{E}_n\circ \text{tr})\circ \bloom_{\omega}(\mathscr{E}_1,...,\mathscr{E}_{n-1}) \\
&\overset{\text{induction}}\searrow&\bloom(\mathscr{E}_n\circ \text{tr})\circ \bloom(\mathscr{E}_{n-1}\circ \text{tr})\circ \cdots \circ \bloom(\mathscr{E}_2\circ \text{tr})\circ \bloom(\mathscr{E}_1),
\end{eqnarray*}
as desired.

\underline{Case III}: Let $\aleph=\bloom_{\vartheta}(\mathscr{E}_1,...,\mathscr{E}_n)$. Then
\begin{eqnarray*}
\aleph&=&\bloom_{\vartheta}(\mathscr{E}_1,...,\mathscr{E}_n) \\
&=&\bloom\left(\bloom_{\chi}(\mathscr{E}_{k+1},...,\mathscr{E}_n)\circ  \hspace{0.75mm}  \mathscr{E}_k\circ \text{tr}\right)\circ \bloom_{\omega}(\mathscr{E}_1,...,\mathscr{E}_{k-1}) \\
&\overset{\text{induction}}\searrow&\bloom\left(\bloom(\mathscr{E}_n\circ \text{tr})\circ \cdots \circ \bloom(\mathscr{E}_{k+1})\circ \mathscr{E}_{k}\circ \text{tr}\right)\circ \bloom(\mathscr{E}_{k-1}\circ \text{tr})\circ \cdots \circ \bloom(\mathscr{E}_2\circ \text{tr})\circ \bloom(\mathscr{E}_1) \\
&\overset{\text{Lemma~\ref{LXMMA8971}}}\searrow&\bloom(\mathscr{E}_n\circ \text{tr})\circ \cdots \circ \bloom(\mathscr{E}_{k+1}\circ \text{tr})\circ \bloom(\mathscr{E}_{k}\circ \text{tr})\circ \bloom(\mathscr{E}_{k-1}\circ \text{tr})\circ \cdots \circ \bloom(\mathscr{E}_2\circ \text{tr})\circ \bloom(\mathscr{E}_1),
\end{eqnarray*}
where again in the last line we note that we have repeatedly used the elision $\text{tr}\circ \text{tr}=\text{tr}$, thus the result holds.
\eprf

The following theorem follows directly from Proposition~\ref{PXPNRSX67}, and will form the basis for a definition of states over time associated with $n$-step processes, as we will see in the next section.

\bt \label{CXRSX9817}
If $\bloom$ is associative, then 
\[
\bloom_{\vartheta}(\mathscr{E}_1,...,\mathscr{E}_n)= \bloom(\mathscr{E}_n\circ \emph{tr})\circ \bloom(\mathscr{E}_{n-1}\circ \emph{tr})\circ \cdots \circ \bloom(\mathscr{E}_2\circ \emph{tr})\circ \bloom(\mathscr{E}_1)
\]
for every parenthezation $\vartheta(\VA_0,...,\VA_n)$ of $\VA_0\otimes \cdots \otimes \VA_n$.
\et

%%%%%%%%%%%%%%%%%%%%%%%%%%%%%%%%
\section{States over time for $n$-step processes} \label{S7}
%%%%%%%%%%%%%%%%%%%%%%%%%%%%%%%%
In this section we use bloom maps for $n$-chains to define states over time for arbitrary finite-step quantum processes. 

\bd
Given an $(n+1)$-tuple $(\VA_0,...,\VA_n)$ of multi-matrix algebras, an element
\[
\left(\rho,\mathscr{E}_1,...,\mathscr{E}_n\right)\in \mathcal{S}(\VA_0)\times \bold{CPTP}(\VA_0,\VA_1)\times \cdots \times \bold{CPTP}(\VA_{n-1},\VA_n)
\] 
will be referred to as an \define{$n$-step process}, and the set $\mathcal{S}(\VA_0)\times \bold{CPTP}(\VA_0,\VA_1)\times \cdots \times \bold{CPTP}(\VA_{n-1},\VA_n)$ of all such $n$-step processes will be denoted by $\mathscr{P}(\VA_0,...,\VA_n)$. Given an $n$-step process $\left(\rho,\mathscr{E}_1,...,\mathscr{E}_n\right)$, we let $\rho_0=\rho$ and $\rho_i=(\mathscr{E}_{i}\circ \cdots \circ \mathscr{E}_1)(\rho)$ for all $i\in \{1,...,n\}$.
\ed

\bd
An \define{$n$-step state over time function} associates every $(n+1)$-tuple $(\mathcal{A}_0,\ldots,\mathcal{A}_n)$ of multi-matrix algebras with a map 
\[
\psi:\mathscr{P}(\VA_0,...,\VA_n)\longrightarrow \mathscr{T}\left(\VA_0\otimes \cdots \otimes \VA_{n}\right)
\]
such that 
\be \label{TRCXND71}
\text{tr}_{i}\left(\psi(\rho,\mathscr{E}_{1},...,\mathscr{E}_n)\right)=\rho_{i},
\ee
for all $i\in \{0,...,n\}$. In such a case, the pseudo-density operator $\psi(\rho,\mathscr{E}_{1},...,\mathscr{E}_n)$ will be referred to as the \define{state over time} associated with the \define{$n$-step process} $(\rho,\mathscr{E}_{1},...,\mathscr{E}_n)$. For $n=1$ an $n$-step state over time function will be referred to simply as a \define{state over time function}.
\ed

\bt \label{STXFXT8187}
Let $\bloom$ be a bloom map which is hermitian, let $n>0$, and let $\psi$ be the map on $n$-step processes given by
\be \label{STXFXT81}
\psi(\rho,\mathscr{E}_1,...,\mathscr{E}_n)=(\bloom(\mathscr{E}_{n}\circ \emph{tr})\circ \cdots \circ \bloom(\mathscr{E}_2\circ \emph{tr})\circ \bloom(\mathscr{E}_1))(\rho).
\ee
Then the following statements hold.
\begin{enumerate}[i.]
\item \label{STXFXT81871}
The map $\psi$ is an $n$-step state over time function.
\item \label{STXFXT81872}
If $\bloom$ is also associative and $(\rho,\E_1,...,\E_n)\in \mathscr{P}(\VA_0,...,\VA_n)$, then $\psi(\rho,\mathscr{E}_1,...,\mathscr{E}_n)=\bloom_{\vartheta}(\E_1,...,\E_n)(\rho)$ for every parenthezation $\vartheta(\VA_0,...,\VA_n)$ of $\VA_0\otimes \cdots \otimes \VA_n$.
\end{enumerate} 
\et

\bprf
\underline{Item \ref{STXFXT81871}}: Let $(\VA_0,...,\VA_n)$ be an $(n+1)$-tuple of multi-matrix algebras, let $(\rho,\mathscr{E}_1,...,\mathscr{E}_n)\in \mathscr{P}(\VA_0,...,\VA_n)$ be an $n$-step process. Then for all $i\in \{0,...,n\}$ we have
\begin{eqnarray*}
\text{tr}_{i}\left(\psi(\rho,\mathscr{E}_{1},...,\mathscr{E}_n)\right)&=&\text{tr}_{\VA_{0}\otimes \cdots \otimes\widehat{\VA_{i}}\otimes \cdots \otimes\VA_{n}}\left((\bloom(\mathscr{E}_{n}\circ \text{tr})\circ \cdots \circ \bloom(\mathscr{E}_2\circ \text{tr})\circ \bloom(\mathscr{E}_1))(\rho)\right) \\
&\overset{\eqref{BXCXDMA381}}=&
\begin{cases}
(\mathscr{E}_i\circ \cdots \circ \mathscr{E}_1)(\rho) \quad \quad \text{for $i\in \{1,...,n\}$} \\
\id_{\VA_0}(\rho) \quad \quad \hspace{1.56cm} \text{for $i=0$}
\end{cases} \\
&=&\rho_i,
\end{eqnarray*}
thus $\psi$ satisfies condition \eqref{TRCXND71} of being a state over time function. 

To conclude the proof, we must show that $\psi(\rho,\mathscr{E}_1,...,\mathscr{E}_n)$ is self-adjoint, for which
we use induction on $n$. For $n=1$ we have $\psi(\rho,\mathscr{E}_1)=\bloom(\mathscr{E}_1)(\rho)$, and since $\mathscr{E}_1$ is CPTP it is $\dag$-preserving, thus $\bloom(\mathscr{E}_1)$ is $\dag$-preserving by the hermitian assumption on $\bloom$. We then have
\[
\bloom(\mathscr{E}_1)(\rho)^{\dag}=\bloom(\mathscr{E}_1)(\rho^{\dag})=\bloom(\mathscr{E}_1)(\rho),
\]
where the last equality follows from the fact that $\rho$ is a state (and so self-adjoint), thus $\bloom(\mathscr{E}_1)(\rho)$ is self-adjoint. Now assume the result holds for $n=m-1>1$, and let $\sigma=(\bloom(\mathscr{E}_{m-1}\circ \text{tr})\circ \cdots \circ \bloom(\mathscr{E}_2\circ \text{tr})\circ \bloom(\mathscr{E}_1))(\rho)$, so that 
\[
\psi(\rho,\mathscr{E}_1,...,\mathscr{E}_m)=\bloom(\mathscr{E}_m\circ \text{tr})(\sigma).
\]
Now since $\mathscr{E}_m$ and $\text{tr}$ are both completely positive, it follows that $\mathscr{E}_m\circ \text{tr}$ is $\dag$-preserving, thus by the hermitian assumption on $\bloom$ it follows that $\bloom(\mathscr{E}_m\circ \text{tr})$ is $\dag$-preserving as well. We then have
\[
\psi(\rho,\mathscr{E}_1,...,\mathscr{E}_m)^{\dag}=\bloom(\mathscr{E}_m\circ \text{tr})(\sigma)^{\dag}=\bloom(\mathscr{E}_m\circ \text{tr})(\sigma^{\dag})=\bloom(\mathscr{E}_m\circ \text{tr})(\sigma)=\psi(\rho,\mathscr{E}_1,...,\mathscr{E}_m),
\]
where the last equality follows from the fact that $\sigma$ is self-adjoint by the inductive hypothesis, thus $\psi(\rho,\mathscr{E}_1,...,\mathscr{E}_m)$ is self-adjoint. 

\underline{Item \ref{STXFXT81872}}: The statement follows from Theorem~\ref{CXRSX9817}.
\eprf

%%%%%%%%%%%%%%%%%%%%%%%%%%%%
\section{The $\tao$-function} \label{S8}
%%%%%%%%%%%%%%%%%%%%%%%%%%%%

Since the symmetric bloom $\tao$ is hermitian and associative, it follows from Theorem~\ref{STXFXT8187} that $\tao$ extends to a uniquely defined $n$-step state over time function for all $n\in \N$, which we refer to as the \emph{$\tao$-function}. In this section we will prove various properties the $\tao$-function satisfies, including multi-linearity, classical reducibility and a \emph{multi-marginal property} that will be used in the next section to solve an open problem from the literature. We also prove a general formula for the $\tao$-function evaluated on an $n$-step process $(\rho,\E_1,...,\E_n)$ in terms of $\rho$ and the channel states $\J[\E_1],...,\J[\E_n]$. To conclude the section, we show that in the case of a closed system of dynamically evolving qubits, the $\tao$-function recovers the pseudo-density matrix formalism first introduced by Fitzsimons, Jones and Vedral \cite{FJV15}.

\bd
The \define{$\tao$-function} is the function on all finite-step processes given by
\be \label{TAOFXTX71}
\tao(\rho,\mathscr{E}_1,...,\mathscr{E}_n)=(\tao(\mathscr{E}_{n}\circ \text{tr})\circ \cdots \circ \tao(\mathscr{E}_2\circ \text{tr})\circ \tao(\mathscr{E}_1))(\rho).
\ee
\ed

\br
Note that the $\tao$-function as given by \eqref{TAOFXTX71} is well-defined even if $\rho$ is not a state and even if $\E_{i}$ is not CPTP for any $i\in \{1,...,n\}$. As such, we may view the $\tao$-function as being defined on the set
\[
\VA_0\times \bold{TP}(\VA_0,...,\VA_n)
\]
for any $(n+1)$-tuple of multi-matrix algebras $(\VA_0,...,\VA_n)$. However, the element $\tao(\rho,\E_1,...,\E_n)$ will only be viewed as a state over time when $(\rho,\E_1,...,\E_n)\in \mathscr{P}(\VA_0,...,\VA_n)$.
\er

We now prove various properties the $\tao$-function satisfies, but first we introduce some notation.

\bnot
Let $\VA=\VA_0\otimes \cdots \otimes \VA_n$. Given $0\leq i_1<\cdots<i_m\leq n$, we let 
\[
\tr_{i_1\cdots i_{m}}:\VA\lra \VA_{i_1}\otimes \cdots \otimes \VA_{i_m}
\]
denote the partial trace map.
\enot

\bt \label{TAOFCTX931}
The $\tao$-function satisfies the following properties.
\begin{enumerate}[i.]
\item \label{TAOFCTX9311}
\emph{(Hermiticity)} $\tao(\rho,\E_1,...,\E_n)$ is self-adjoint of unit trace for all finite step processes $(\rho,\E_1,...,\E_n)$.
\item \label{TAOFCTX9312}
\emph{(Multi-Linearity)} The $\tao$-function is multi-linear, i.e., 
\be \label{MTXL1}
\tao(\lambda \rho+(1-\lambda)\sigma,\E_1,...,\E_n)=\lambda \tao(\rho,\E_1,...,\E_n)+(1-\lambda)\tao(\sigma,\E_1,...,\E_n)
\ee
and
\be \label{MTXL2}
\tao(\rho,\E_1,...,\lambda \E_i+(1-\lambda)\F_i,...,\E_n)=\lambda \tao(\rho,\E_1,...,\E_i,...,\E_n)+(1-\lambda)\tao(\rho,\E_n,...,\F_i,...,\E_n)
\ee
for all $i\in \{1,...,n\}$ and for all $\lambda \in \C$.
\item \label{TAOFCTX93123}
\emph{(Reduction)} $\tao(\rho,\E_1,...,\E_n)=\tao\left(\rho,\E_1,... ,\tao(\E_{i+1})\circ \E_i\hspace{0.35mm},\hspace{0.35mm}\E_{i+2}\circ \emph{tr},\hspace{0.35mm} \E_{i+3},...,\E_n\right)$ for all $i\in \{1,...,n-1\}$.
\item \label{TAOFCTX93147}
\emph{(The Multi-Marginal Property)} Let $0\leq i_1<\cdots<i_m\leq n$. Then
\be
\emph{tr}_{i_1\cdots i_m}\left(\tao(\rho,\E_1,...,\E_n)\right)=\tao(\rho_{i_1},\E_{i_2}\circ \cdots \circ \E_{i_1+1},...,\E_{i_m}\circ \cdots \circ \E_{i_{m-1}+1})
\ee 
\item \label{TAOFCTX9315}
\emph{(Classical Reducibility)} If $(\rho,\E_1,...,\E_n)\in \mathscr{P}(\C^{X_0},...,\C^{X_n})$ is a classical process, then
\[
\tao(\rho,\E_1,...,\E_n)=\bigoplus_{(x_0,...,x_n)\in X_0\times \cdots \times X_n}\mathbb{P}(x_n|x_{n-1})\cdots \mathbb{P}(x_1|x_0)\mathbb{P}(x_0),
\]
where $\mathbb{P}(x_0)$ is the classical distribution associated with $\rho$ and $\mathbb{P}(x_i|x_{i-1})$ are conditional probabilities associated with $\mathscr{J}[\E_i]$.
\end{enumerate}
\et

\bprf
\underline{Item \ref{TAOFCTX9311}}: The statement follows from item \ref{STXFXT81871} of Theorem~\ref{STXFXT8187}.

\underline{Item \ref{TAOFCTX9312}}: Equation \eqref{MTXL1} follows from the fact that $\tao(\E_n\circ \tr)\circ \cdots \circ \tao(\E_1)$ is a linear map, while equation \eqref{MTXL2} follows from the fact that 
\[
\tao((\lambda \E_i+(1-\lambda)\F_i)\circ \tr)=\lambda\tao(\E_i\circ \tr)+(1-\lambda)\tao(\F_i\circ \tr)
\]
for all $i\in \{1,...,n\}$ and all $\lambda\in \C$.

\underline{Item \ref{TAOFCTX93123}}: Let $i\in \{1,...,n-1\}$, and let $\aleph=\tao\left(\rho,\E_1,... ,\tao(\E_{i+1})\circ \E_i\hspace{0.35mm},\hspace{0.35mm}\E_{i+2}\circ \text{tr},\hspace{0.35mm} \E_{i+3},...,\E_n\right)$. Then
\begin{eqnarray*}
\aleph&=&\tao\left(\rho,\E_1,... ,\tao(\E_{i+1})\circ \E_i\hspace{0.35mm},\hspace{0.35mm}\E_{i+2}\circ \text{tr},\hspace{0.35mm} \E_{i+3},...,\E_n\right) \\
&=&\left(\tao(\E_n\circ \tr)\circ \cdots \circ \tao\left(\tao(\E_{i+1})\circ \E_i\circ \tr\right)\circ \cdots \circ \tao(\E_1)\right)(\rho) \\
&=&\left(\tao(\E_n\circ \tr)\circ \cdots \circ \tao(\E_{i+1}\circ \tr)\circ \tao(\E_i\circ \tr)\circ \cdots \circ \tao(\E_1)\right)(\rho) \\
&=&\tao(\rho,\E_1,...,\E_n), 
\end{eqnarray*}
where the third equality follows from the associativity of the symmetric bloom.

\underline{Item \ref{TAOFCTX93147}}: The proof relies on two lemmas, which we will prove after we finish proving the theorem. The case $m=1$ follows from item \ref{STXFXT81871} of Theorem~\ref{STXFXT8187}, so now suppose $m>1$. The partial trace map $\tr_{i_1\cdots i_m}$ may be written as 
\[
\tr_{i_1\cdots i_m}=\left(\tr_{\VA_{i_1+1}\otimes \cdots \otimes \VA_{i_2-1}}\circ \cdots \circ \tr_{\VA_{i_{m-1}+1}\otimes \cdots \otimes \VA_{i_m-1}}\right)\circ \tr_{\VA_0\otimes \cdots \otimes \VA_{{i_1}-1}\otimes \VA_{{i_m}+1}\otimes \cdots \otimes \VA_{n}},
\]
where if $i_j-i_{j-1}=1$, we set $\tr_{\VA_{i_{j-1}+1}\otimes \cdots \otimes \VA_{i_j-1}}=\id$. Now by items \ref{TAOFCTXTRNK47} and  \ref{TAOFCTXTRNK49} Lemma~\ref{TAOFCTXTRNK747} we have
\[
\tr_{\VA_0\otimes \cdots \otimes \VA_{{i_1}-1}\otimes \VA_{{i_m}+1}\otimes \cdots \otimes \VA_{n}}\left(\tao(\rho,\E_1,...,\E_n)\right)=\tao(\rho_{i_1},\E_{i_1+1},...,\E_{i_m})\in \VA_{i_1}\otimes \VA_{i_1+1}\otimes \cdots \otimes \VA_{i_m},
\] 
thus
\begin{eqnarray*}
\tr_{i_1\cdots i_m}\left(\tao(\rho,\E_1,...,\E_n)\right)&=&\tr_{\VA_{i_1+1}\otimes \cdots \otimes \VA_{i_2-1}}\circ \cdots \circ \tr_{\VA_{i_{m-1}+1}\otimes \cdots \otimes \VA_{i_m-1}}\left(\tao(\rho_{i_1},\E_{i_1+1},...,\E_{i_m})\right) \\
&=&\tao(\rho_{i_1},\E_{i_2}\circ \cdots \circ \E_{i_1+1},...,\E_{i_m}\circ \cdots \circ \E_{i_{m-1}+1}),
\end{eqnarray*}
where the last equality follows from Lemma~\ref{LMXHLLX767}.

\underline{Item \ref{TAOFCTX9315}}: In the proof of Proposition~\ref{PXPSN683} we established that if $\E:\C^X\lra \C^Y$ is a classical channel, then
\[
\J[\E]=\sum_{(x,y)\in X\times Y}\E_{yx}(\delta_x\otimes \delta_y),
\]
where we recall $\E_{yx}$ are the conditional probabilities associated with the classical channel $\E$. The statement then follows directly from the definition of the symmetric bloom.
\eprf

We now prove the lemmas needed for the proof of the multi-marginal property of the $\tao$-function (item \ref{TAOFCTX93147} of Theorem~\ref{TAOFCTX931}).

\blem \label{TAOFCTXTRNK747}
Let $(\rho,\E_1,...,\E_n)\in \mathscr{P}(\VA_0,...,\VA_n)$ be an $n$-step process. Then the following statements hold.
\begin{enumerate}[i.]
\item \label{TAOFCTXTRNK47}
$\emph{tr}_{\VA_0\otimes \cdots \otimes\VA_{i-1}}(\tao(\rho,\E_1,...,\E_n))=\tao(\rho_i,\E_{i+1},...,\E_{n})$ for all $i\in \{1,...,n\}$.
\item \label{TAOFCTXTRNK49}
$\emph{tr}_{\VA_{i+1}\otimes \cdots \otimes\VA_{n}}(\tao(\rho,\E_1,...,\E_n))=\tao(\rho,\E_{1},...,\E_{i})$ for all $i\in \{0,...,n-1\}$.
\end{enumerate}
\elem

\bprf
\underline{Item \ref{TAOFCTXTRNK47}}: We use induction on $i$. For $i=0$ we have
\begin{eqnarray*}
\tr_{\VA_0}\left(\tao(\rho,\E_1,...,\E_n)\right)&=&\tr_{\VA_0}\left(\left(\tao(\E_n\circ \tr)\circ \cdots \circ \tao(\E_2\circ \tr)\circ \tao(\E_1)\right)(\rho)\right) \\
&=&\tr_{\VA_0}\left(\tao\left(\tao(\E_n\circ \tr)\circ \cdots \circ \tao(\E_2)\circ \E_1\right)(\rho)\right) \\
&=&\left(\tao(\E_n\circ \tr)\circ \cdots \circ \tao(\E_2)\right)(\E_1(\rho)) \\
&=&\tao(\rho_1,\E_2,...,\E_n),
\end{eqnarray*}
where the second equality follows from Lemma~\ref{LXMMA8971} and the third equality follows from the definition of a bloom map, thus the result holds for $i=0$. Now suppose the result holds for $i=m-1$ with $m-1<n$. Then
\begin{eqnarray*}
\tr_{\VA_0\otimes \cdots \otimes \VA_{m-1}}\left(\tao(\rho,\E_1,...,\E_n)\right)&=&\tr_{\VA_{m-1}}\left(\tr_{\VA_0\otimes \cdots \otimes \VA_{m-2}}\left(\tao(\rho,\E_1,...,\E_n)\right)\right) \\
&=&\tr_{\VA_{m-1}}\left(\tao(\rho_{m-1},\E_m,...\E_n)\right) \\
&=&\tao(\rho_{m},\E_{m+1},...\E_n),
\end{eqnarray*}
where the last equality follows from a calculation similar to the $i=0$ case. It then follows that the result holds for $i=m$, as desired.

\underline{Item \ref{TAOFCTXTRNK49}}: We prove the equivalent statement $\tr_{\VA_{n-i}\otimes \cdots \otimes\VA_{n}}(\tao(\rho,\E_1,...,\E_{n}))=\tao(\rho,\E_{1},...,\E_{n-i-1})$ for all $i\in \{0,...,n-1\}$ by induction on $i$. For $i=0$ we have
\[
\tr_{\VA_n}\left(\tao(\rho,\E_1,...,\E_n)\right)=\tr_{\VA_n}\left(\tao(\E_n\circ \tr)\left(\tao(\rho,\E_1,...,\E_{n-1})\right)\right)=\tao(\rho,\E_1,...,\E_{n-1}),
\]
where the last equality follows from the definition of bloom map, thus the result holds for $i=0$. Now suppose the result holds for $i=m-1$ with $m-1<n-1$. Then
\begin{eqnarray*}
\tr_{\VA_{n-m}\otimes \cdots \otimes\VA_{n}}(\tao(\rho,\E_1,...,\E_{n}))&=&\tr_{\VA_{n-m}}\left(\tr_{\VA_{n-m+1}\otimes \cdots \otimes\VA_{n}}(\tao(\rho,\E_1,...,\E_{n}))\right) \\
&=&\tr_{\VA_{n-m}}\left(\tao(\rho,\E_{1},...,\E_{n-(m-1)-1})\right) \\
&=&\tr_{\VA_{n-m}}\left(\tao(\rho,\E_{1},...,\E_{n-m})\right) \\
&=&\tao(\rho,\E_{1},...,\E_{n-m-1}),
\end{eqnarray*}
where the last equality follows from a calculation similar to the $i=0$ case. It then follows that the result holds for $i=m$, as desired.
\eprf

\blem  \label{LMXHLLX767}
Let $(\E_1,...,\E_n)\in \bold{TP}(\VA_0,...,\VA_n)$ be an $n$-chain, let $0=i_1<\cdots<i_m= n$, let $\rho\in \VA_{i_1}$, and let $\aleph\in \VA_{i_1}\otimes \VA_{i_2}\otimes \cdots \otimes \VA_{i_m}$ be the element given by
\[
\aleph=\tr_{\VA_{i_{m-1}+1}\otimes \cdots \otimes \VA_{i_m-1}}\circ \cdots \circ \tr_{\VA_{i_1+1}\otimes \cdots \otimes \VA_{i_2-1}}\left(\tao(\rho,\E_{i_1+1},...,\E_{i_m})\right),
\]
where we set $\tr_{\VA_{i_{j-1}+1}\otimes \cdots \otimes \VA_{i_j-1}}=\id$  \hspace{0.5mm} if \hspace{0.5mm} $i_j-i_{j-1}=1$.
Then
\be \label{LMXHLLX76}
\aleph=\tao\left(\rho,\E_{i_2}\circ \cdots \circ \E_{i_1+1},...,\E_{i_m}\circ \cdots \circ \E_{i_{m-1}+1}\right).
\ee
\elem

\bprf
We use induction on $m$. The case $m=1$ follows from item \ref{TAOFCTXTRNK49} of Lemma~\ref{TAOFCTXTRNK747}. So assume the result holds for $m=k-1>1$, and let $\aleph_k$ be the element given by
\[
\aleph_k=\tr_{\VA_{i_{k-1}+1}\otimes \cdots \otimes \VA_{i_k-1}}\circ \cdots \circ \tr_{\VA_{i_1+1}\otimes \cdots \otimes \VA_{i_2-1}}\left(\tao(\rho,\E_{i_1+1},...,\E_{i_k})\right).
\]
The result then follows once we show
\[
\aleph_k=\tao\left(\rho,\E_{i_2}\circ \cdots \circ \E_{i_1+1},...,\E_{i_k}\circ \cdots \circ \E_{i_{k-1}+1}\right).
\]
For this, we first consider the 2-chain
\[
\VA_{i_1}\overset{\E}\lra \VA_{i_1+1}\otimes \cdots \otimes \VA_{i_2-1}\overset{\F}\lra \VA_{i_2}\otimes \VA_{i_2+1}\otimes \cdots \otimes \VA_{i_m},
\]
where $\E=\tao(\E_{i_2-1}\circ \tr)\circ \cdots \circ \tao(\E_{i_1+2})\circ \E_{i_1+1}$ and $\F=\tao(\E_{i_k}\circ \tr)\circ \cdots \circ \tao(\E_{i_2+1})\circ \E_{i_2}\circ \tr$. Then after letting $\chi=\tao(\rho,\E,\F)$, we have
\begin{eqnarray*}
\chi&=&\tao(\rho,\E,\F) \\
&=&\left(\tao(\F\circ \tr)\circ \tao(\E)\right)(\rho) \\
&=&\left(\tao(\tao(\E_{i_k}\circ \tr)\circ \cdots \circ \tao(\E_{i_2+1})\circ \E_{i_2}\circ \tr)\circ \tao\left(\tao(\E_{i_2-1}\circ \tr)\circ \cdots \circ \tao(\E_{i_1+2})\circ \E_{i_1+1}\right)\right)(\rho) \\
&=&\left(\tao(\E_{i_2}\circ \tr)\circ \cdots \circ \tao(\E_{i_1+2}\circ \tr)\circ \tao(\E_{i_1+1})\right)(\rho) \\
&=&\tao(\rho,\E_{i_1+1},...,\E_{i_k}),
\end{eqnarray*}
where the fourth equality follows from the associativity of the symmetric bloom. Then after letting $\xi=\tr_{\VA_{i_1+1}\otimes \cdots \otimes \VA_{1_2-1}}\left(\tao(\rho,\E_{i_1+1},...,\E_{i_k})\right)$, we then have
\begin{eqnarray*}
\xi&=&\tr_{\VA_{i_1+1}\otimes \cdots \otimes \VA_{1_2-1}}\left(\tao(\rho,\E_{i_1+1},...,\E_{i_k})\right) \\
&=&\tr_{\VA_{i+1}\otimes \cdots \otimes \VA_{j-1}}\left(\tao(\rho,\E,\F)\right) \\
&\overset{\eqref{CMPXALTX67}}=&\tao(\rho,\F\circ \E) \\
&=&\tao\left(\rho,\tao(\E_{i_k}\circ \tr)\circ \cdots \circ \tao(\E_{i_2+1})\circ \E_{i_2}\circ \tr\circ \tao(\E_{i_2-1}\circ \tr)\circ \cdots \circ \tao(\E_{i_1+2})\circ \E_{i_1+1}\right) \\
&=&\tao\left(\rho, \tao(\E_{i_k}\circ \tr)\circ \cdots \circ \tao(\E_{i_2+1})\circ \E_{i_2}\circ \E_{i_2-1}\circ \cdots \circ \E_{i_1+1}\right) \\
&=&\tao\left(\tao(\E_{i_k}\circ \tr)\circ \cdots \circ \tao(\E_{i_2+1})\circ \E_{i_2}\circ \E_{i_2-1}\circ \cdots \circ \E_{i_1+1}\right)(\rho) \\
&=&\left(\tao(\E_{i_k}\circ \tr)\circ \cdots \circ \tao(\E_{i_2+1}\circ \tr)\circ \tao\left(\E_{i_2}\circ \E_{i_2-1}\circ \cdots \circ \E_{i_1+1}\right)\right)(\rho) \\
&=&\tao(\rho,\E_{i_2}\circ \E_{i_2-1}\circ \cdots \circ \E_{i_1+1},\E_{i_2+1},...,\E_{i_k}),
\end{eqnarray*}
where the fifth equality follows from the fact that $\E_{i_2}\circ \tr\circ \tao(\E_{i_2-1}\circ \tr)\circ \cdots \circ \tao(\E_{i_1+2})\circ \E_{i_1+1}=\E_{i_2}\circ \E_{i_2-1}\circ \cdots \circ \E_{i_1+1}$, and the seventh equality follows from the associativity of symmetric bloom. Now let $\sigma=\tao(\rho,\E_{i_2}\circ \E_{i_2-1}\circ \cdots \circ \E_{i_1+1})$, so that
\begin{eqnarray*}
\tao(\rho,\E_{i_2}\circ \E_{i_2-1}\circ \cdots \circ \E_{i_1+1},\E_{i_2+1},...,\E_{i_k})&=&\left(\tao(\E_{i_k}\circ \tr)\circ \cdots \circ \tao\left(\E_{i_2}\circ \E_{i_2-1}\circ \cdots \circ \E_{i_1+1}\right)\right)(\rho) \\
&=&\left(\tao(\E_{i_k}\circ \tr)\circ \cdots \circ \tao\left(\E_{i_2+1}\circ \tr\right)\right)(\sigma) \\
&=&\tao(\sigma,\E_{i_2+1}\circ \tr,\E_{i_2+2},...,\E_{i_k}).
\end{eqnarray*}
We then have
\begin{eqnarray*}
\aleph_k&=&\tr_{\VA_{i_{k-1}+1}\otimes \cdots \otimes \VA_{i_k-1}}\circ \cdots \circ \tr_{\VA_{i_1+1}\otimes \cdots \otimes \VA_{i_2-1}}\left(\tao(\rho,\E_{i_1+1},...,\E_{i_k})\right) \\
&=&\tr_{\VA_{i_{k-1}+1}\otimes \cdots \otimes \VA_{i_k-1}}\circ \cdots \circ \tr_{\VA_{i_2+1}\otimes \cdots \otimes \VA_{i_3-1}}\left(\tr_{\VA_{i_1+1}\otimes \cdots \otimes \VA_{1_2-1}}\left(\tao(\rho,\E_{i_1+1},...,\E_{i_k})\right)\right) \\
&=&\tr_{\VA_{i_{k-1}+1}\otimes \cdots \otimes \VA_{i_k-1}}\circ \cdots \circ \tr_{\VA_{i_2+1}\otimes \cdots \otimes \VA_{i_3-1}}\left(\tao(\rho,\E_{i_2}\circ \E_{i_2-1}\circ \cdots \circ \E_{i_1+1},\E_{i_2+1},...,\E_{i_k})\right) \\
&=&\tr_{\VA_{i_{k-1}+1}\otimes \cdots \otimes \VA_{i_k-1}}\circ \cdots \circ \tr_{\VA_{i_2+1}\otimes \cdots \otimes \VA_{i_3-1}}\left(\tao(\sigma,\E_{i_2+1}\circ \tr,\E_{i_2+2},...,\E_{i_k})\right) \\
&\overset{\text{induction}}=&\tao\left(\sigma,\E_{i_3}\circ \cdots \circ \E_{i_2+1}\circ \tr,...,\E_{i_k}\circ \cdots \circ \E_{i_{k-1}+1}\right) \\
&=&\tao\left(\tao(\rho,\E_{i_2}\circ \E_{i_2-1}\circ \cdots \circ \E_{i_1+1}),\E_{i_3}\circ \cdots \circ \E_{i_2+1}\circ \tr,...,\E_{i_k}\circ \cdots \circ \E_{i_{k-1}+1}\right) \\
&=&\tao\left(\rho,\E_{i_2}\circ \cdots \circ \E_{i_1+1},...,\E_{i_k}\circ \cdots \circ \E_{i_{k-1}+1}\right),
\end{eqnarray*}
where the final equality follows from item \ref{TAOFCTX93123} of Theorem~\ref{TAOFCTX9312}, thus concluding the proof.
\eprf

We now prove a formula for $\tao(\rho,\E_1,...,\E_n)$ in terms of $\rho$ and the channel states $\J[\E_1]$, ..., $\J[\E_n]$.

\bnot
Let $[n]=\{1,...,n\}$ for every natural number $n\in \N$. For every subset $\{i_1,...,i_m\}\subset [n]$, the subscripts $i_j$ are chosen so that $i_1<\cdots <i_m$. In such a case, we denote the complement $[n]\setminus \{i_1,...,i_m\}$ by $\{\underline{i}_1,...,\underline{i}_{n-m}\}$, with the same convention that $\underline{i}_1<\cdots<\underline{i}_{n-m}$.
\enot

\bnot
Given an $n$-chain $(\E_1,...,\E_n)\in \bold{TP}(\VA_0,...,\VA_n)$ and an element $j\in [n]$, we let $\widetilde{\J}[\E_{j}]\in \VA_0\otimes \cdots \otimes \VA_n$ be the element given by
\[
\widetilde{\J}[\E_{j}]=
\begin{cases}
\J[\E_1]\otimes \mathds{1} \quad \hspace{0.69cm} \text{if $j=1$} \\
\mathds{1}\otimes \J[\E_j]\otimes \mathds{1} \quad \text{if $1<j<n$} \\
\mathds{1}\otimes \J[\E_n] \quad \hspace{0.64cm} \text{if $j=n$}.
\end{cases}
\]
\enot

\bt \label{MXFXMA71}
Let $(\rho,\E_1,...,\E_n)\in \mathscr{P}(\VA_0,...,\VA_n)$ be an $n$-step process.
Then
\be \label{CFX87}
\tao(\rho,\E_1,...,\E_n)=\frac{1}{2^n}\sum_{\{i_1,...,i_m\}\subset [n]} \widetilde{\J}[\E_{i_m}]\cdots \widetilde{\J}[\E_{i_1}](\rho\otimes \mathds{1})\widetilde{\J}[\E_{\underline{i}_1}]\cdots \widetilde{\J}[\E_{\underline{i}_{n-m}}]
\ee
\et

\bprf
We use induction on $n$. For $n=1$ the result holds by definition of the $\tao$-function. So now suppose the result holds for $n=k-1>1$. Then after letting $\aleph=\tao(\rho,\E_1,...,\E_k)$, we have
\begin{eqnarray*}
\aleph&=&\tao(\rho,\E_1,...,\E_k) \\
&=&(\tao(\mathscr{E}_{k}\circ \text{tr})\circ \cdots \circ \tao(\mathscr{E}_2\circ \text{tr})\circ \tao(\mathscr{E}_1))(\rho) \\
&=&\tao(\mathscr{E}_{k}\circ \text{tr})\left(\tao(\rho,\E_1,...,\E_{k-1})\right) \\
&=&\bold{Jor}\left(\tao(\rho,\E_1,...,\E_{k-1})\otimes \mathds{1},\mathscr{J}[\E_{k}\circ \tr]\right) \\
&=&\bold{Jor}\left(\tao(\rho,\E_1,...,\E_{k-1})\otimes \mathds{1},\mathds{1}\otimes \mathscr{J}[\E_{k}]\right) \\
&=&\bold{Jor}\left(\left(\frac{1}{2^{k-1}}\sum_{\{i_1,...,i_m\}\subset [k-1]} \widetilde{\J}[\E_{i_m}]\cdots \widetilde{\J}[\E_{i_1}](\rho\otimes \mathds{1})\widetilde{\J}[\E_{\underline{i}_1}]\cdots \widetilde{\J}[\E_{\underline{i}_{k-1-m}}]\right)\otimes \mathds{1},\widetilde{\J}[\E_{k}]\right) \\
&=&\frac{1}{2^k}\sum_{\{i_1,...,i_m\}\subset [k]} \widetilde{\J}[\E_{i_m}]\cdots \widetilde{\J}[\E_{i_1}](\rho\otimes \mathds{1})\widetilde{\J}[\E_{\underline{i}_1}]\cdots \widetilde{\J}[\E_{\underline{i}_{k-m}}],
\end{eqnarray*}
where the second to last equality follows from the inductive assumption. It then follows that the result holds for $n=k$, as desired.
\eprf

We now extend the normalized Jordan product to an $n$-ary operation for all $n>0$, which we will then use to derive a formula for the $\tao$-function which may be viewed as the $n$-step generalization of the 2-step formula \eqref{2STPFXM91}.   

\bd
Give a multi-matrix algebra $\VA$, the \define{extended Jordan product} is the map 
\[
\bold{Jor}:\VA\times \cdots \times \VA\lra \VA
\]
given by
\be \label{EXTNDJPX17}
\bold{Jor}(A_1,...,A_n)=\bold{Jor}\left(A_0,\bold{Jor}(A_1,\bold{Jor}(\cdots,\bold{Jor}(A_{n-1},A_n)))\right).
\ee
\ed

\bt \label{MXFXMA7187}
Let $(\rho,\E_1,...,\E_n)\in \mathscr{P}(\VA_0,...,\VA_n)$ be an $n$-step process.
Then
\be \label{CFX8771}
\tao(\rho,\E_1,...,\E_n)=\bold{Jor}\left(\rho\otimes \mathds{1},\J[\E_1]\otimes \mathds{1},...,\mathds{1}\otimes \J[\E_j]\otimes \mathds{1},...,\mathds{1}\otimes \J[\E_n]\right)
\ee
\et

\bprf
The statement follows from Theorem~\ref{MXFXMA71} together with the fact that $(\rho\otimes \mathds{1})$ commutes with $\widetilde{\J}[\E_{j}]$ for all $j\in [n]$ with $j>1$.
\eprf

To conclude this section, we show how the $\tao$-function recovers the pseudo-density matrix formalism for qubits.

\bx [The pseudo-density matrix formalism for systems of qubits] \label{PDOEXP71}
Let $\rho\in \bigotimes_{i=1}^{k}\M_{2}$ be the initial state of an $k$-qubit system, let $\{\sigma_0,\sigma_1,\sigma_2,\sigma_3\}$ be the Pauli basis of $\M_2$, and let $(\rho,\E_1,...,\E_n)$ be an $n$-step process with initial state $\rho$. In the formalism introduced by Fitzsimons, Jones and Vedral \cite{FJV15}, the pseudo-density matrix associated with $(\rho,\E_1,...,\E_n)$ is the matrix $R_n\in \bigotimes_{j=0}^{n}\M_{2}^{\otimes k}$ given by
\[
R_n=\frac{1}{2^{k(n+1)}}\sum_{i_0=1}^{4^k}\cdots \sum_{i_{n}=1}^{4^k}\left\<\left(\widetilde{\sigma}_{i_{\alpha}}\right)_{\alpha=0}^{n}\right\>\bigotimes_{\alpha=0}^{n}\widetilde{\sigma}_{i_{\alpha}},
\]
where $\widetilde{\sigma}_{i_{\alpha}}\in \{\sigma_0,...,\sigma_{3}\}^{\otimes k}$ and $\left\<\left\{\widetilde{\sigma}_{i_{\alpha}}\right\}_{\alpha=0}^{n}\right\>$ denotes the expectation value associated with the observable $\left(\widetilde{\sigma}_{i_{\alpha}}\right)_{\alpha=0}^{n}$. In \cite{LQDV}, it was recently proved that
\[
R_1=\bold{Jor}(\rho\otimes \mathds{1},\mathscr{J}[\E_1])\equiv \tao(\rho,\E_1) \quad \& \quad R_{n}=\bold{Jor}\left(R_{n-1}\otimes \mathds{1},\mathds{1}\otimes \mathscr{J}[\E_{n}]\right),
\]
and since
\begin{eqnarray*}
\tao(\rho,\E_1,...,\E_n)&=&(\tao(\mathscr{E}_{n}\circ \text{tr})\circ \cdots \circ \tao(\mathscr{E}_2\circ \text{tr})\circ \tao(\mathscr{E}_1))(\rho) \\
&=&\tao(\mathscr{E}_{n}\circ \text{tr})\left(\tao(\rho,\E_1,...,\E_{n-1})\right) \\
&=&\bold{Jor}\left(\tao(\rho,\E_1,...,\E_{n-1})\otimes \mathds{1},\mathscr{J}[\E_{n}\circ \tr]\right) \\
&=&\bold{Jor}\left(\tao(\rho,\E_1,...,\E_{n-1})\otimes \mathds{1},\mathds{1}\otimes \mathscr{J}[\E_{n}]\right), \\
\end{eqnarray*}
it follows via induction that $\tao(\rho,\E_1,...,\E_n)=R_n$ (the final equality follows from item \ref{CPXLX3} of Lemma~\ref{CPXLX731}). As such, the $\tao$-function recovers the pseudo-density matrix formalism for systems of qubits first introduced in \cite{FJV15}, and moreover, Theorems~\ref{MXFXMA71} and \ref{MXFXMA7187} yield new formulas for $R_n$.
\ex

%%%%%%%%%%%%%%%%%%%%%%%%%%%%%%%%%%%%%
\section{Extracting dynamics from a pseudo-density matrix} \label{S9}
%%%%%%%%%%%%%%%%%%%%%%%%%%%%%%%%%%%%%
In this section, we derive an explicit expression for the inverse of the $\tao$-function restricted to a suitably nice subset of $n$-step processes. In particular, given a pseudo-density matrix $\tau$, we show how to identify $\tau$ with an $n$-step process $(\rho,\E_1,...,\E_n)$ such that $\tao(\rho,\E_1,...,\E_n)=\tau$. This solves an open problem from \cite{jia2023}, where it is stated "Another interesting and closely relevant open question is, for a given PDO (pseudo-density matrix), how to find a quantum process to realize it.". 

\bnot
Given an element $\tau\in \VA_0\otimes \cdots \otimes \VA_n$ with $\tr_0(\tau)\in \VA_0$ invertible, we let $X_{\tau}\in \VA_0\otimes \cdots \otimes \VA_n$ be the unique element such that $\tau=\bold{Jor}((\tr(\tau_0)\otimes \mathds{1}),X_{\tau})$. 
\enot

\bnot
Given an $(n+1)$-tuple $(\VA_0,...,\VA_n)$ of multi-matrix algebras, we let $\mathscr{P}_{+}(\VA_0,...,\VA_n)\subset \mathscr{P}(\VA_0,...,\VA_n)$ denote the subset given by
\[
\mathscr{P}_{+}(\VA_0,...,\VA_n)=\left\{(\rho,\mathscr{E}_1,...,\mathscr{E}_n)\in \mathscr{P}(\VA_0,...,\VA_n) \hspace{1mm}|\hspace{1mm} \rho_i \hspace{1mm} \text{is invertible for $i=0,...,n-1$}\right\},
\]
and we let $\mathscr{T}_{*}(\VA_0\otimes \cdots \otimes \VA_n)\subset \mathscr{T}(\VA_0\otimes \cdots \otimes \VA_n)$ be the subset consisting of elements $\tau\in \mathscr{T}(\VA_0\otimes \cdots \otimes \VA_n)$ such that 
\begin{itemize}
\item
$\tau_{i}\in \mathscr{T}_{*}(\VA_{i-1}\otimes \VA_{i})$ for $i=1,...,n$, where $\mathscr{T}_{*}(\VA_{i-1}\otimes \VA_{i})$ is as defined by \eqref{PDMXT671}.
\item
$X_{\tau}=\bold{Jor}(X_{\tau_1}\otimes \mathds{1},...,\mathds{1}\otimes X_{\tau_n})$
\end{itemize}
where $\tau_{i}=\text{tr}_{i-1,i}(\tau)$ and $\text{tr}_{i-1,i}:\VA_0\otimes \cdots \otimes \VA_n\lra \VA_{i-1}\otimes \VA_i$ is the partial trace map.
\enot

\br
We have reason to believe that the two conditions defining $\mathscr{T}_{*}(\VA_0\otimes \cdots \otimes \VA_n)$ are in fact equivalent, but we have yet to find a proof.
\er

We now are going to prove that the $\tao$-function restricted to $\mathscr{P}_{+}(\VA_0,...,\VA_n)$ defines a bijection onto $\mathscr{T}_{*}(\VA_0\otimes \cdots \otimes \VA_n)$, and we will also give a precise formula for the inverse. Before doing so however, we first prove that the image of the $\tao$-function restricted to $\mathscr{P}_{+}(\VA_0,...,\VA_n)$ indeed lies in $\mathscr{T}_{*}(\VA_0\otimes \cdots \otimes \VA_n)$.

\bn \label{PNXPXN007}
Let $(\rho,\E_1,...,\E_n)\in \mathscr{P}_{+}(\VA_0,...,\VA_n)$. Then $\tao(\rho,\E_1,...,\E_n)\in \mathscr{T}_{*}(\VA_0\otimes \cdots \otimes \VA_n)$.
\en

\bprf
Let $(\rho,\mathscr{E}_1,...,\mathscr{E}_n)\in \mathscr{P}_{+}(\VA_0,...,\VA_n)$, let $\tau=\tao(\rho,\E_1,...,\E_n)$, and let 
\[
\tau_{i}=\text{tr}_{i-1,i}\left(\tao(\rho,\mathscr{E}_1,...,\mathscr{E}_n)\right)\in \VA_{i-1}\otimes \VA_{i}
\]
for all $i\in [n]$. Then it follows from the multi-marginal property of the $\tao$-function that $\tau_{i}=\tao(\rho_{i-1},\E_i)$, and since $\rho_{i-1}$ is invertible by the definition of $\mathscr{P}_{+}(\VA_0,...,\VA_n)$, it follows that $(\rho_{i-1},\E_i)\in \mathscr{P}_{+}(\VA_{i-1},\VA_{i})$, thus $\tau_i\in \mathscr{T}_{*}(\VA_{i-1}\otimes \VA_{i})$ for $i=1,...,n$ by Lemma~\ref{TXGDX687}, showing that $\tau$ satisfies the first condition for being an element of $\mathscr{T}_{*}(\VA_0\otimes \cdots \otimes \VA_n)$. Moreover, it follows from Theorem~\ref{EXT987189} that $\E_i=\J^{-1}(X_{\tau_i})$ for all $i\in [n]$, thus
\begin{eqnarray*}
\tau&=&\tao(\rho,\E_1,...,\E_n) \\
&=&\tao(\tr_0(\tau),\J^{-1}(X_{\tau_1}),...,\J^{-1}(X_{\tau_n})) \\
&\overset{\eqref{CFX8771}}=&\bold{Jor}\left(\tr_0(\tau)\otimes \mathds{1},\J\left(\J^{-1}(X_{\tau_1})\right)\otimes \mathds{1},...,\mathds{1}\otimes \J\left(\J^{-1}(X_{\tau_n})\right)\right) \\
&=&\bold{Jor}\left(\tr_0(\tau)\otimes \mathds{1},X_{\tau_1}\otimes \mathds{1},...,\mathds{1}\otimes X_{\tau_n}\right) \\
&=&\bold{Jor}\left((\tr_0(\tau)\otimes \mathds{1}),\bold{Jor}\left(X_{\tau_1}\otimes \mathds{1},...,\mathds{1}\otimes X_{\tau_n}\right)\right) \\
&\implies& X_{\tau}=\bold{Jor}\left(X_{\tau_1}\otimes \mathds{1},...,\mathds{1}\otimes X_{\tau_n}\right),
\end{eqnarray*}
thus $\tau$ satisfies the second condition for being an element of $\mathscr{T}_{*}(\VA_0\otimes \cdots \otimes \VA_n)$. It then follows that $\tau=\tao(\rho,\E_1,...,\E_n)\in \mathscr{T}_{*}(\VA_0\otimes \cdots \otimes \VA_n)$, as desired.

\eprf

\bt \label{MSXAX71}
Let $(\VA_0,...,\VA_n)$ be an $(n+1)$-tuple of multi-matrix algebras, let 
\[
\mathfrak{S}:\mathscr{P}_{+}(\VA_0,...,\VA_n)\lra \mathscr{T}_{*}(\VA_0\otimes \cdots \otimes \VA_n)
\]
be the restriction of the $\tao$-function to $\mathscr{P}_{+}(\VA_0,...,\VA_n)$, and let 
\[
\mathfrak{C}:\mathscr{T}_{*}(\VA_0\otimes \cdots \otimes \VA_n)\lra \mathscr{P}_{+}(\VA_0,...,\VA_n)
\]
be the map given by
\be \label{EXTRKTN87}
\mathfrak{C}(\tau)=\Big(\emph{tr}_{0}(\tau),\mathscr{J}^{-1}(X_{\tau_1}),...,\mathscr{J}^{-1}(X_{\tau_n})\Big),
\ee
where $\tau_i=\emph{tr}_{i-1,i}\left(\tau \right)$ and $X_{\tau_i}\in \VA_{i-1}\otimes \VA_{i}$ is defined via \eqref{XTAU97} for all $i\in \{1,...,n\}$.  Then $\mathfrak{S}$ is a bijection, and $\mathfrak{C}= \mathfrak{S}^{-1}$.
\et

\bprf
Let $(\rho,\mathscr{E}_1,...,\mathscr{E}_n)\in \mathscr{P}_{+}(\VA_0,...,\VA_n)$, let $\tau=\mathfrak{S}(\rho,\E_1,...,\E_1)\equiv \tao(\rho,\E_1,...,\E_n)$, and let 
\[
\tau_{i}=\text{tr}_{i-1,i}\left(\mathfrak{S}(\rho,\mathscr{E}_1,...,\mathscr{E}_n)\right)\in \VA_{i-1}\otimes \VA_{i}
\]
for all $i\in \{1,...,n\}$. Then we have seen in the proof of Proposition~\ref{PNXPXN007} that $\mathscr{J}^{-1}(X_{\tau_i})=\mathscr{E}_i$, thus
\begin{eqnarray*}
(\mathfrak{C}\circ \mathfrak{S})(\rho,\mathscr{E}_1,...,\mathscr{E}_n)&=&\mathfrak{C}\left(\mathfrak{S}(\rho,\mathscr{E}_1,...,\mathscr{E}_n)\right) \\
&=&\mathfrak{C}(\tao(\rho,\mathscr{E}_1,...,\mathscr{E}_n)) \\
&=&\left(\text{tr}_{0}(\tao(\rho,\mathscr{E}_1,...,\mathscr{E}_n)),\mathscr{J}^{-1}(X_{\tau_1}),...,\mathscr{J}^{-1}(X_{\tau_n})\right) \\
&=&(\rho,\mathscr{E}_1,...,\mathscr{E}_n).
\end{eqnarray*}
It then follows that $\mathfrak{C}\circ \mathfrak{S}=\id$, thus $\mathfrak{S}$ is injective.

Now let $\tau\in \mathscr{T}_{*}(\VA_0\otimes \cdots \otimes \VA_n)$, so that 
\[
\mathfrak{C}(\tau)=\left(\tr_{0}(\tau),\J^{-1}(X_{\tau_1}),...,\J^{-1}(X_{\tau_n})\right).
\]
From the definition of $\mathscr{T}_{*}(\VA_0\otimes \cdots \otimes \VA_n)$ we have $\tau_i\in \mathscr{T}_{*}(\VA_{i-1},\VA_i)$ for all $i\in [n]$, from which it follows that $\J^{-1}(X_{\tau_i})$ is CPTP for all $i\in [n]$ and also that $\tr_0(\tau)$ is invertible, thus $\mathfrak{C}(\tau)\in \mathscr{P}_{+}(\VA_0,...,\VA_n)$. We then have
\begin{eqnarray*}
(\mathfrak{S}\circ \mathfrak{C})(\tau)&=&\tao\left(\tr_{0}(\tau),\J^{-1}(X_{\tau_1}),...,\J^{-1}(X_{\tau_n})\right) \\
&\overset{\eqref{CFX8771}}=&\bold{Jor}\left(\tr_0(\tau)\otimes \mathds{1},\J\left(\J^{-1}(X_{\tau_1})\right)\otimes \mathds{1},...,\mathds{1}\otimes \J\left(\J^{-1}(X_{\tau_n})\right)\right) \\
&=&\bold{Jor}\left(\tr_0(\tau)\otimes \mathds{1},X_{\tau_1}\otimes \mathds{1},...,\mathds{1}\otimes X_{\tau_n}\right) \\
&=&\bold{Jor}\left((\tr_0(\tau)\otimes \mathds{1}),\bold{Jor}\left(X_{\tau_1}\otimes \mathds{1},...,\mathds{1}\otimes X_{\tau_n}\right)\right) \\
&=&\bold{Jor}\left((\tr_0(\tau)\otimes \mathds{1}),X_{\tau}\right) \\
&=&\tau,
\end{eqnarray*}
where the second to last equality follows from the fact that $\tau\in \mathscr{T}_{*}(\VA_0\otimes \cdots \otimes \VA_n)$. It then follows that $\mathfrak{S}\circ \mathfrak{C}=\id$, thus concluding the proof.
\eprf

\br
As the inverse $\mathscr{J}^{-1}:\VA\otimes \VB\lra \text{Hom}(\VA,\VB)$ of the Jamio\l kowski isomorphism admits the explicit formula
\[
\mathscr{J}^{-1}(\tau)(\rho)=\text{tr}_{\VA}((\rho\otimes \mathds{1})\tau),
\]
and since there are algorithms to compute $X_{\tau}$ with $\tau\in \mathscr{T}_{*}(\VA_0\otimes \cdots \otimes \VA_n)$, the formula \eqref{EXTRKTN87} for the inverse of the $\tao$-function is easily implementable in symbolic computing software. In particular, in Ref.~\cite{Liu_2024} computational methods are developed for to explicitly determine when an operator $\tau\in \VA_0\otimes \cdots \otimes \VA_n$ may be realized as pseudo-density matrix associated with a process $(\rho,\E_1,\ldots,\E_n)$, where $\rho$ is a state of full rank corresponding to a system of qubits and $\E_i$ is a channel where the dimension of the output always matches the dimension of the input for all $i\in \{1,\ldots,n\}$. It should then be straightforward to adapt such methods to the more general context of Theorem~\ref{MSXAX71}, which applies to arbitrary finite-dimensional systems and arbitrary channels.  
\er

%%%%%%%%BIBLIOGRAPHY%%%%%%%%%%%%
\addcontentsline{toc}{section}{\numberline{}Bibliography}
\bibliographystyle{quantum}
\bibliography{QDX}

\begin{thebibliography}{10}

\bibitem{MKWSKI}
Hermann Minkowski.
\newblock ``Espace et temps''.
\newblock \href{https://dx.doi.org/10.24033/asens.613}{Ann. Sci. \'{E}cole Norm. Sup. (3) {\bf 26}, 499--517}~(1909).

\bibitem{Wheeler_1989}
John~Archibald Wheeler.
\newblock ``{Information, physics, quantum: The search for links}''.
\newblock In {3rd International Symposium on Foundations of Quantum Mechanics in Light}.
\newblock ~(1989).

\bibitem{Atiyah88}
Michael~F Atiyah.
\newblock ``Topological quantum field theory''.
\newblock \href{https://dx.doi.org/10.1007/BF02698547}{Publications Math{\'e}matiques de l'IH{\'E}S {\bf 68}, 175--186}~(1988).

\bibitem{Ba06}
John~C. {Baez}.
\newblock ``Quantum quandaries: A category-theoretic perspective''.
\newblock In Steven French, Dean Rickles, and Juha Saatsi, editors, Structural Foundations of Quantum Gravity.
\newblock \href{https://dx.doi.org/10.1093/acprof:oso/9780199269693.003.0008}{Pages 240--265}.
\newblock Oxford U. Press~(2006).
\newblock  \href{http://arxiv.org/abs/quant-ph/0404040}{arXiv:quant-ph/0404040}.

\bibitem{HHPBS17}
Dominic {Horsman}, Chris {Heunen}, Matthew~F. {Pusey}, Jonathan {Barrett}, and Robert~W. {Spekkens}.
\newblock ``{Can a quantum state over time resemble a quantum state at a single time?}''.
\newblock \href{https://dx.doi.org/10.1098/rspa.2017.0395}{Proc. R. Soc. A {\bf 473}, 20170395}~(2017).
\newblock  \href{http://arxiv.org/abs/1607.03637}{arXiv:1607.03637}.

\bibitem{Le06}
Matthew~S. Leifer.
\newblock ``Quantum dynamics as an analog of conditional probability''.
\newblock \href{https://dx.doi.org/10.1103/PhysRevA.74.042310}{Phys. Rev. A {\bf 74}, 042310}~(2006).
\newblock  \href{http://arxiv.org/abs/0606022}{arXiv:0606022}.

\bibitem{Le07}
Matthew~S. {Leifer}.
\newblock ``{Conditional Density Operators and the Subjectivity of Quantum Operations}''.
\newblock In Guillaume {Adenier}, Chrisopher {Fuchs}, and Andrei~Yu {Khrennikov}, editors, Foundations of Probability and Physics - 4.
\newblock \href{https://dx.doi.org/10.1063/1.2713456}{Volume 889 of American Institute of Physics Conference Series, pages 172--186}.
\newblock ~(2007).
\newblock  \href{http://arxiv.org/abs/quant-ph/0611233}{arXiv:quant-ph/0611233}.

\bibitem{LeSp13}
Matthew~S. Leifer and Robert~W. Spekkens.
\newblock ``Towards a formulation of quantum theory as a causally neutral theory of {B}ayesian inference''.
\newblock \href{https://dx.doi.org/10.1103/PhysRevA.88.052130}{Phys. Rev. A {\bf 88}, 052130}~(2013).
\newblock  \href{http://arxiv.org/abs/1107.5849}{arXiv:1107.5849}.

\bibitem{Cotler2018}
Jordan Cotler, Chao-Ming Jian, Xiao-Liang Qi, and Frank Wilczek.
\newblock ``Superdensity operators for spacetime quantum mechanics''.
\newblock \href{https://dx.doi.org/10.1007/jhep09(2018)093}{Journal of High Energy Physics{\bf 9}}~(2018).

\bibitem{OCB12}
Ognyan Oreshkov, Fabio Costa, and {\v C}aslav Brukner.
\newblock ``Quantum correlations with no causal order''.
\newblock \href{https://dx.doi.org/10.1038/ncomms2076}{Nat. Comm. {\bf 3}, 1092}~(2012).
\newblock  \href{http://arxiv.org/abs/1105.4464}{arXiv:1105.4464}.

\bibitem{Ohya1983}
M.~Ohya.
\newblock ``Note on quantum probability''.
\newblock \href{https://dx.doi.org/}{Lett. Nuovo Cimento (2) {\bf 38}, 402--404}~(1983).

\bibitem{Ts22}
Mankei Tsang.
\newblock ``Generalized conditional expectations for quantum retrodiction and smoothing''.
\newblock \href{https://dx.doi.org/10.1103/PhysRevA.105.042213}{Phys. Rev. A {\bf 105}, 042213}~(2022).
\newblock  \href{http://arxiv.org/abs/1912.02711}{arXiv:1912.02711}.

\bibitem{Woot87}
William~K. Wootters.
\newblock ``A {W}igner-function formulation of finite-state quantum mechanics''.
\newblock \href{https://dx.doi.org/10.1016/0003-4916(87)90176-X}{Ann. Physics {\bf 176}, 1--21}~(1987).

\bibitem{MaCh22}
Takashi Matsuoka and Dariusz Chru{\'s}ci{\'n}ski.
\newblock ``Compound state, its conditionality and quantum mutual information''.
\newblock In International Conference on Quantum Probability \& Related Topics.
\newblock \href{https://dx.doi.org/10.1007/978-3-031-06170-7_7}{Pages 135--150}.
\newblock Springer, Cham~(2022).

\bibitem{GuoZ1}
Zhiqiang Huang and Xiao-Kan Guo.
\newblock ``{L}egget-{G}arg inequalities for multitime processes''~(2022).
\newblock  \href{http://arxiv.org/abs/2211.13396}{arXiv:2211.13396}.

\bibitem{Jia_2024}
Zhian Jia and Dagomir Kaszlikowski.
\newblock ``The spatiotemporal doubled‐density operator: A unified framework for analyzing spatial and temporal quantum processes''.
\newblock \href{https://dx.doi.org/10.1002/qute.202400102}{Advanced Quantum Technologies{\bf 7}}~(2024).

\bibitem{FuPa22}
James Fullwood and Arthur~J. Parzygnat.
\newblock ``On quantum states over time''.
\newblock \href{https://dx.doi.org/10.1098/rspa.2022.0104}{Proc. R. Soc. A{\bf 478}}~(2022).
\newblock  \href{http://arxiv.org/abs/2202.03607}{arXiv:2202.03607}.

\bibitem{LiNg23}
Seok~Hyung Lie and Nelly H.~Y. Ng.
\newblock ``Quantum state over time is unique''.
\newblock \href{https://dx.doi.org/10.1103/physrevresearch.6.033144}{Physical Review Research{\bf 6}}~(2024).

\bibitem{PFBC23}
Arthur Parzygnat, James Fullwood, Francesco Buscemi, and Giulio Chiribella.
\newblock ``Virtual quantum broadcasting''.
\newblock \href{https://dx.doi.org/10.1103/PhysRevLett.132.110203}{Phys. Rev. Lett. {\bf 132}, 110203}~(2024).
\newblock  \href{http://arxiv.org/abs/2310.13049}{arXiv:2310.13049}.

\bibitem{PaRuBayes}
Arthur~J. Parzygnat and Benjamin~P. Russo.
\newblock ``A non-commutative {B}ayes' theorem''.
\newblock \href{https://dx.doi.org/https://doi.org/10.1016/j.laa.2022.02.030}{Linear Algebra Its Appl. {\bf 644}, 28--94}~(2022).
\newblock  \href{http://arxiv.org/abs/2005.03886}{arXiv:2005.03886}.

\bibitem{PaRu19}
Arthur Parzygnat and Benjamin Russo.
\newblock ``Non-commutative disintegrations: Existence and uniqueness in finite dimensions''.
\newblock \href{https://dx.doi.org/10.4171/jncg/493}{Journal of Noncommutative Geometry {\bf 17}, 899–955}~(2023).

\bibitem{FuPa22a}
Arthur~J. Parzygnat and James Fullwood.
\newblock ``{From time-reversal symmetry to quantum Bayes' rules}''.
\newblock \href{https://dx.doi.org/10.1103/PRXQuantum.4.020334}{PRX Quantum {\bf 4}, 020334}~(2023).
\newblock  \href{http://arxiv.org/abs/quant-ph/2212.08088}{arXiv:quant-ph/2212.08088}.

\bibitem{PaFu24}
Arthur~J. Parzygnat and James Fullwood.
\newblock ``Time-symmetric correlations for open quantum systems''~(2024).
\newblock  \href{http://arxiv.org/abs/2407.11123}{arXiv:2407.11123}.

\bibitem{BHKK_2024}
Gabriele Bressanini, Farhan Hanif, Hyukjoon Kwon, and M.~S. Kim.
\newblock ``Quantum observables over time for information recovery''~(2024).
\newblock  \href{http://arxiv.org/abs/2412.11659}{arXiv:2412.11659}.

\bibitem{FJV15}
Joseph~F. Fitzsimons, Jonathan~A. Jones, and Vlatko Vedral.
\newblock ``Quantum correlations which imply causation''.
\newblock \href{https://dx.doi.org/10.1038/srep18281}{Sci. Rep. {\bf 5}, 18281}~(2015).
\newblock  \href{http://arxiv.org/abs/1302.2731}{arXiv:1302.2731}.

\bibitem{Marletto_2021}
Chiara Marletto, Vlatko Vedral, Salvatore Virzì, Alessio Avella, Fabrizio Piacentini, Marco Gramegna, Ivo~Pietro Degiovanni, and Marco Genovese.
\newblock ``Temporal teleportation with pseudo-density operators: How dynamics emerges from temporal entanglement''.
\newblock \href{https://dx.doi.org/10.1126/sciadv.abe4742}{Science Advances{\bf 7}}~(2021).

\bibitem{ZPTGVF18}
Zhikuan Zhao, Robert Pisarczyk, Jayne Thompson, Mile Gu, Vlatko Vedral, and Joseph~F. Fitzsimons.
\newblock ``Geometry of quantum correlations in space-time''.
\newblock \href{https://dx.doi.org/10.1103/PhysRevA.98.052312}{Phys. Rev. A {\bf 98}, 052312}~(2018).
\newblock  \href{http://arxiv.org/abs/1711.05955}{arXiv:1711.05955}.

\bibitem{Pisar19}
Robert Pisarczyk, Zhikuan Zhao, Yingkai Ouyang, Vlatko Vedral, and Joseph~F. Fitzsimons.
\newblock ``Causal limit on quantum communication''.
\newblock \href{https://dx.doi.org/}{Phys. Rev. Lett.{\bf 123}}~(2019).

\bibitem{Marletto_2020}
Chiara Marletto, Vlatko Vedral, Salvatore {Virz{\`\i}}, Enrico Rebufello, Alessio Avella, Fabrizio Piacentini, Marco Gramegna, Ivo~Pietro Degiovanni, and Marco Genovese.
\newblock ``Non-monogamy of spatio-temporal correlations and the black hole information loss paradox''.
\newblock \href{https://dx.doi.org/10.3390/e22020228}{Entropy {\bf 22}, 228}~(2020).

\bibitem{Marletto_2019}
Chiara Marletto, Vlatko Vedral, Salvatore {Virz{\`\i}}, Enrico Rebufello, Alessio Avella, Fabrizio Piacentini, Marco Gramegna, Ivo~Pietro Degiovanni, and Marco Genovese.
\newblock ``Theoretical description and experimental simulation of quantum entanglement near open time-like curves via pseudo-density operators''.
\newblock \href{https://dx.doi.org/10.1038/s41467-018-08100-1}{Nat. Commun.{\bf 10}}~(2019).

\bibitem{Liu_2024}
Xiangjing Liu, Qian Chen, and Oscar Dahlsten.
\newblock ``Inferring the arrow of time in quantum spatiotemporal correlations''.
\newblock \href{https://dx.doi.org/10.1103/physreva.109.032219}{Phys. Rev. A{\bf 109}}~(2024).

\bibitem{Liu_2024X}
Xiangjing Liu, Zhian Jia, Yixian Qiu, Fei Li, and Oscar Dahlsten.
\newblock ``Unification of spatiotemporal quantum formalisms: mapping between process and pseudo-density matrices via multiple-time states''.
\newblock \href{https://dx.doi.org/10.1088/1367-2630/ad264c}{New Journal of Physics {\bf 26}, 033008}~(2024).

\bibitem{LQDV}
Xiangjing Liu, Yixian Qiu, Oscar Dahlsten, and Vlatko Vedral.
\newblock ``Quantum causal inference with extremely light touch''.
\newblock \href{https://dx.doi.org/10.1038/s41534-024-00956-0}{npj Quantum Information{\bf 11}}~(2025).

\bibitem{FuPa24}
James Fullwood and Arthur~J. Parzygnat.
\newblock ``Operator representation of spatiotemporal quantum correlations''~(2024).
\newblock  \href{http://arxiv.org/abs/2405.17555}{arXiv:2405.17555}.

\bibitem{song23}
Minjeong Song, Varun Narasimhachar, Bartosz Regula, Thomas~J. Elliott, and Mile Gu.
\newblock ``Causal classification of spatiotemporal quantum correlations''.
\newblock \href{https://dx.doi.org/10.1103/PhysRevLett.133.110202}{Phys. Rev. Lett. {\bf 133}, 110202}~(2024).
\newblock  \href{http://arxiv.org/abs/2306.09336}{arXiv:2306.09336}.

\bibitem{jia2023}
Zhian Jia, Minjeong Song, and Dagomir Kaszlikowski.
\newblock ``Quantum space-time marginal problem: global causal structure from local causal information''.
\newblock \href{https://dx.doi.org/10.1088/1367-2630/ad1416}{New J. Phys. {\bf 25}, 123038}~(2023).

\bibitem{FZ_24}
James Fullwood, Zhen Wu, Arthur~J. Parzygnat, and Vlatko Vedral.
\newblock ``Quantum mutual information in time''~(2024).
\newblock  \href{http://arxiv.org/abs/2410.02137}{arXiv:2410.02137}.

\bibitem{Utagi_2021}
Shrikant Utagi.
\newblock ``Quantum causal correlations and non-markovianity of quantum evolution''.
\newblock \href{https://dx.doi.org/10.1016/j.physleta.2020.126983}{Physics Letters A {\bf 386}, 126983}~(2021).

\bibitem{Fa01}
Douglas~R. Farenick.
\newblock ``Algebras of linear transformations''.
\newblock \href{https://dx.doi.org/10.1007/978-1-4613-0097-7}{Pages xiv+238}.
\newblock Universitext. Springer-Verlag, New York. ~(2001).

\bibitem{Jam72}
A.~Jamio\l~kowski.
\newblock ``Linear transformations which preserve trace and positive semidefiniteness of operators''.
\newblock \href{https://dx.doi.org/10.1016/0034-4877(72)90011-0}{Rep. Mathematical Phys. {\bf 3}, 275--278}~(1972).

\bibitem{fritz2023free}
Tobias Fritz and Wendong Liang.
\newblock ``Free gs-monoidal categories and free markov categories''.
\newblock \href{https://dx.doi.org/10.1007/s10485-023-09717-0}{Applied Categorical Structures{\bf 31}}~(2023).

\bibitem{Datta}
Biswa~Nath Datta.
\newblock ``Numerical methods for linear control systems''.
\newblock \href{https://dx.doi.org/}{Elsevier, Inc., Amsterdam, Netherlands}. ~(2004).

\end{thebibliography}

\end{document}